\documentclass[aps,twocolumn,nofootinbib,showpacs,prd,10pt,superscriptaddress]{revtex4-2}
\usepackage{graphicx}
\usepackage[english]{babel}
\usepackage[T1]{fontenc}
\usepackage{mathrsfs}
\usepackage[tbtags]{amsmath}
\usepackage{amssymb}
\usepackage{amsxtra}
\usepackage{amsopn}
\usepackage{latexsym}
\usepackage[mathcal]{eucal}
\usepackage{mathtools}
\usepackage{slashed}
\usepackage{xcolor}
\usepackage[hyperfootnotes=false,bookmarks=false]{hyperref}

\newcommand{\BE}{\begin{equation}}
\newcommand{\EE}{\end{equation}}
\newcommand{\BA}{\begin{align}}
\newcommand{\EA}{\end{align}}

\newcommand{\mc}{\mathcal}
\newcommand{\psibar}{\overline{\psi}}
\newcommand{\cbar}{\overline{c}}

\begin{document}

\title{QCD phase diagram from the gluon propagator at finite temperature and density}

\author{Giorgio Comitini}
\email{giorgio.comitini@dfa.unict.it}
\affiliation{Dipartimento di Fisica e Astronomia ``E. Majorana'', Universit\`a di Catania, Via S. Sofia 64, I-95123 Catania, Italy}
\affiliation{INFN Sezione di Catania, Via S. Sofia 64, I-95123 Catania, Italy}

\author{Fabio Siringo}
\email{fabio.siringo@ct.infn.it}
\affiliation{Dipartimento di Fisica e Astronomia ``E. Majorana'', Universit\`a di Catania, Via S. Sofia 64, I-95123 Catania, Italy}
\affiliation{INFN Sezione di Catania, Via S. Sofia 64, I-95123 Catania, Italy}

\date{\today}

\begin{abstract}
    The screened massive expansion of full QCD is used in conjunction with a model for infrared quark masses to compute the Landau-gauge gluon propagator at finite temperature and baryonic density. Analytic expressions up to a one-dimensional momentum integral are provided for the propagator, and its behavior is studied at zero Matsubara frequency with respect to temperature, chemical potential, and the parameters of the expansion. The phase diagram of QCD is explored under the assumption that the deconfinement temperature can be identified as the position of the maximum of the longitudinal gluon propagator at zero Matsubara frequency and fixed spatial momentum.
\end{abstract}



\maketitle
\section{Introduction}

Over the last fifteen years, the discovery of dynamical mass generation in the gluon sector of low-energy QCD \cite{LSWP98a,LSWP98b,BBLW00,BBLW01,SIMS05,CM08b,BIMS09,ISI09,BMM10,BLLM12,OS12,BBCO15,DOS16,BHLP04,BHLP07,IMSS07,SO10,ABBC12} has led to the formulation of massive perturbative techniques which proved to be effective in describing the infrared regime of the strong interactions. Such techniques were first developed in the context of vacuum pure Yang-Mills theory \cite{DSVV08,DGSVV08,DOV10,DSV11,CFGM15,CFGM16,CDFG16,CDPF17,CFPS17,CDGP18,MPPS19,DFPR19,CDDS24,TW10,TW11,PTW13,RSTW17,HK19,GPRT19,BPRW20,BPR22,SIR16a,SIR17c,SIR17a,SC18,SIR19a,CS20,SIR19b,SC22b,SC22a,SIR23,SC23}, and then extended to full QCD \cite{PTW14,PTW15,PRST17,HK20,PRST21a,BGPR21,FP22,PRSW23,SIR16b,SIR17b,CRBS21} and to non-zero temperatures and chemical potentials \cite{RSTW14,RST15,RSTW15a,RSTW15b,RSTW16,RSTT17,MRS18a,MRS18b,SK19,MRS20,KS21,vERST22,vER22,CDJP15,DV23,DvERV22,SIR17d,CS18,SC21}. Among these, the so-called screened massive expansion \cite{SIR16a,SIR16b,SIR17a,SIR17b,SIR17c,SIR17d,CS18,SC18,SIR19a,SIR19b,CS20,SC21,CRBS21,SC22a,SC22b,SIR23,SC23} was recently used to study the pure Yang-Mills Landau-gauge Euclidean gluon propagator at finite temperature $T$ and zero Matsubara frequency $\omega$ \cite{SC21}. The behavior of the latter was found to agree with the lattice already at one loop and for the simplest choice of parameters: keeping the parameters fixed at their $T=0$ value yields a three-dimensionally transverse propagator that strictly decreases with temperature, and a longitudinal one that first increases with temperature at low $T$'s, and then decreases at large $T$'s after having attained a maximum. On the lattice, at $\omega=0$, pure Yang-Mills theory's longitudinal gluon propagator attains its maximum at the critical temperature of the deconfinement transition \cite{Fischer2010,SOBC14}, $T_{c}\approx 270$~MeV for SU(3), which in the absence of quarks is a (weakly) first-order phase transition \cite{ABIM12,SOBC14}. The entity and abruptness of the increase around $T_{c}$ has lead to the suggestion -- see e.g. \cite{Fischer2010} -- that the longitudinal propagator computed at fixed (usually vanishing) spatial momentum may be used as an order parameter for the deconfiment transition. A natural question then arises -- namely, whether a similar behavior also holds in full QCD and to what extent one may obtain information on the phase diagram of QCD by studying how the gluon propagator changes with temperature.

Understanding the phase diagram of QCD is one of the main challenges of modern research in the strong interactions \cite{GUE21}. At physical quark masses and vanishing chemical potential, lattice simulations predict the existence of a crossover between a low-temperature confined phase, in which the degrees of freedom of strongly interacting matter are hadrons, and a high-temperature deconfined phase -- the quark-gluon plasma -- in which the degrees of freedom are quarks and gluons \cite{Aoki2009,Borsanyi2010,Bazavov2012,Bhattacharya2014}. The pseudocritical temperature of the crossover is usually defined as the position of the peak of chiral or quark-number susceptibilities, or as the inflection point of the Polyakov loop \cite{Aoki2006a,Aoki2006b,Steinbrecher2019}, and is found to lie between 155 and 175 MeV depending on its definition. At non-vanishing chemical potential $\mu\neq 0$, lattice simulations suffer from the infamous sign problem. In order to make predictions about the finite-density regime of QCD, one resorts to extrapolations that make use of imaginary chemical potentials or of Taylor expansions around $\mu=0$ \cite{dFP02,AEHK02,DL03,ADEH05,GG08,KKLM11,CCP14,BdFDP14,BDMM15,BBFG15,CCP16,BDNS18,BFGK20}, or to restricting the scope of the calculation to domains where the sign problem does not arise, like quark-gluon matter at finite isospin density \cite{KS02,KS04,dFSW07,DOS12,BES18,BNRT21} and two-color QCD \cite{NAK84,HKLM99,HKS06,SKU09,HKS10,BM12,CGHS13,BCF13,BIKM16,BBI18,HBMS18,BHMS19,BGHS20,BBNR20,IIL20,BR21}. Studies carried out with these methods have shown that the pseudocritical temperature of the deconfinement crossover decreases with chemical potential; moreover, they hint to the existence of a second-order critical endpoint, after which the crossover is expected to become a true phase transition of the first order that stretches down to $T_{c}=0$ for some critical $\mu=\mu_{c}$. Beyond lattice QCD, several semi-analytic methods, like those based on the Schwinger-Dyson equations and on the functional Renormalization Group \cite{RS00,YCL06,CYC08,CYL09,FW09,FLM11,MBW13,FL13,FLW14,WFL15,GL16,FIS19,BLP20,HSC20,FPR20,BFPR20,GP20,GP21}, and effective approaches such as the Nambu-Jona-Lasinio model \cite{VW91,KLE92,HK94,RWB05,BFG05,FZL08,CGI08,FUK08,FS17}, are available to probe the theory at $\mu\neq0$. At high densities $\mu>\mu_{c}$ and low temperatures, these predict the existence of color-superconducting phases of matter known as two-color superconductivity (2SC) and color-flavor locking (CFL) \cite{ARW98,RSS98,ARW99}.

Thanks to its analytical nature, the screened massive expansion can be extended to non-vanishing chemical potentials in a straightforward way by using the formalism of thermal field theory at finite density. The gluon propagator can then be computed in full QCD as a function of both temperature and chemical potential at any given order in perturbation theory (at least in principle), in any covariant gauge and for any given quark configuration. In the present paper we will show that, when a suitable model is employed for the quark masses in the confined phase, the one-loop zero-frequency Landau-gauge propagator of full QCD at $\mu=0$ turns out to be no different than that of pure Yang-Mills theory: since confined quarks have masses of the order of the QCD scale $\Lambda_{\text{QCD}}$ due to chiral symmetry violation, at $\mu=0$ their presence only has the effect of taming -- without completely neutralizing -- the non-monotonic behavior of the longitudinal propagator. If one assumes that a parallelism can be drawn between pure Yang-Mills theory and full QCD, then this feature translates to a decrease of the (now pseudo-) critical deconfinement temperature $T_{c}$ -- defined as the temperature $T_{\text{max}}$ at which the longitudinal propagator is maximum -- with respect to the pure Yang-Mills case. Going ahead, we will show that, as (baryonic) chemical potential is turned on, the critical temperature starts to decrease due to the quarks' finite density effects setting in. Eventually, $T_{c}$ becomes zero for some $\mu=\mu_{c}$, which at one loop is equal to the value of the lightest infrared effective quark mass. Studying the temperature $T_{\text{max}}$ as a function of chemical potential will allow us to map out the phase diagram of QCD in the $T$-$\mu$ plane, at least to the extent to which the parallelism with pure Yang-Mills theory can be assumed to hold -- that is, within the region involved in the deconfinement transition. Beyond this region, various interesting behaviors could emerge; we will discuss those suggested by our model.

This paper is organized as follows. In Sec.~II we will describe the setup of the screened massive expansion and review some of the finite-temperature results obtained in pure Yang-Mills theory. In Sec.~III we will formulate our model for full QCD at finite temperature and baryonic density and provide expressions for the Landau-gauge one-loop gluon propagator which are analytic up to a one-dimensional momentum integral. Then, at zero Matsubara frequency, we will study its dependence on temperature, chemical potential, and on the free parameters that enter the screened massive expansion, including the gluon and quark masses. In Sec.~IV we will compute the temperature $T_{\text{max}}$ as a function of chemical potential and use it to obtain information on the phase diagram of QCD. In Sec.~V we will discuss our results and present our conclusions.

\section{The screened massive expansion of pure Yang-Mills theory}

\subsection{General setup: zero temperature}

The screened massive expansion is defined by a shift of the expansion point of the QCD perturbative series, carried out after gauge fixing in such a way that the transverse gluons propagate as massive at tree level. Denoting the renormalized Euclidean pure Yang-Mills Faddeev-Popov Lagrangian\footnote{For simplicity, we will keep the usual set of field and coupling renormalization counterterms implicit.} with $\mathcal{L}_{\text{YM}}$,
\begin{align}\label{fplag}
    \mathcal{L}_{\text{YM}} & =\frac{1}{4}\,F_{\mu\nu}^{a}F^{a\,\mu\nu}+\frac{1}{2\xi}\,(\partial\cdot A^{a})^{2}+\cbar^{a}\partial\cdot D c^{a}=                                       \\
    \notag                  & =\frac{1}{2}\,\partial_{\mu}A_{\nu}^{a}\left(\partial^{\mu}A^{a\,\nu}-(1-\frac{1}{\xi})\partial^{\nu}A^{a\,\mu}\right)+\overline{c}^{a}\partial^{2}c^{a}+ \\
    \notag                  & \quad+gf^{a}_{bc}\,\partial^{\mu}A^{a\,\nu}A_{\mu}^{b}A_{\nu}^{c}+\frac{g^{2}}{4}f^{a}_{bc}f^{a}_{de}A_{\mu}^{b}A_{\nu}^{c}A^{d\,\mu}A^{e\,\nu}+          \\
    \notag                  & \quad+gf^{a}_{bc}\,\partial^{\mu}\overline{c}^{a}c^{b}A_{\mu}^{c}\ ,
\end{align}
where $F_{\mu\nu}^{a}=\partial_{\mu}A_{\nu}^{a}-\partial_{\nu}A_{\mu}^{a}+gf^{a}_{bc}A_{\mu}^{b}A_{\nu}^{c}$ is the gluonic field-strength tensor, $D_{\mu}^{ab}=\delta^{ab}\partial_{\mu}-f^{abc}A_{\mu}^{c}$ is the SU($N$) covariant derivative acting on the adjoint representation, $f_{abc}$ are the SU($N$) structure constants, $g$ is the strong coupling and $\xi$ is the gauge parameter, one defines a massive zero-order Lagrangian $\mathcal{L}_{\text{YM},m}$ by introducing a mass term $\delta\mc{L}_{m}$ for the transverse gluons:
\begin{equation}\label{fpkin}
    \mathcal{L}_{\text{YM},m}=\mathcal{L}_{\text{YM},0}+\delta\mc{L}_{m}=\mathcal{L}_{\text{YM},0}+\frac{1}{2}\,m^{2}\,A_{\mu}^{a}t^{\mu\nu}A_{\nu}^{a}\ ,
\end{equation}
where $\mathcal{L}_{\text{YM},0}$ is the ordinary (massless) kinetic term of the Faddeev-Popov Lagrangian,
\begin{equation}
    \mathcal{L}_{\text{YM},0}=-\frac{1}{2}\,A_{\mu}^{a}\,\left(t^{\mu\nu}+\frac{1}{\xi}\ell^{\mu\nu}\right)\partial^{2} A_{\nu}^{a}+\overline{c}^{a}\partial^{2}c^{a}\ ,
\end{equation}
while $t^{\mu\nu}$ and $\ell^{\mu\nu}$ are, respectively, the (4-dimensionally) transverse and longitudinal projectors: in momentum space,
\begin{equation}
    t_{\mu\nu}(p)=\delta_{\mu\nu}-\frac{p_{\mu}p_{\nu}}{p^{2}}\ ,\qquad \ell_{\mu\nu}(p)=\frac{p_{\mu}p_{\nu}}{p^{2}}\ .
\end{equation}
The zero-order Euclidean gluon propagator $\Delta_{m,ab}^{\mu\nu}(p)$ derived from $\mathcal{L}_{\text{YM},m}$ reads
\begin{equation}\label{glupropzero}
    \Delta_{m,ab}^{\mu\nu}(p)=\delta_{ab}\left(\frac{t^{\mu\nu}(p)}{p^{2}+m^{2}}+\xi\,\frac{\ell^{\mu\nu}(p)}{p^{2}}\right)
\end{equation}
and describes transverse gluons propagating with a tree-level mass $m$, and massless longitudinal gluons. The zero-order ghost propagator $\mc{G}_{0}(p)$,
\begin{equation}
    \mc{G}_{0}(p)=\frac{1}{-p^{2}}\ ,
\end{equation}
on the other hand, remains massless.

Having added a transverse gluon mass term to $\mathcal{L}_{\text{YM},m}$, the same term must be subtracted from the interaction part of the Lagrangian in order for the total action to remain unchanged. Thus, in the screened massive expansion, the interaction Lagrangian $\mathcal{L}_{\text{I}}$ must be defined as
\begin{equation}\label{fpint}
    \mc{L}_{\text{I}}=\mc{L}_{\text{YM}}-\mc{L}_{\text{YM},m}=\mc{L}_{3}+\mc{L}_{4}+\mc{L}_{gh}-\delta\mc{L}_{m}\ ,
\end{equation}
where $\mc{L}_{3,4,gh}$ are, respectively, the 3-gluon, 4-gluon and ghost-gluon terms in Eq.~\eqref{fplag}. $-\delta\mc{L}_{m}$ above translates to a new two-gluon interaction vertex $-\delta\Gamma$ in the Feynman rules of the theory, proportional to the gluon mass parameter squared and transverse in momentum space,
\begin{equation}\label{glu2vertex}
    -\delta\Gamma^{\mu\nu}(p)=-m^{2}t^{\mu\nu}(p)\ ,
\end{equation}
which will be referred to as the (gluon) mass counterterm\footnote{Not to be confused with a renormalization counterterm.}.

In order to compute quantities of physical interest within the screened massive expansion, one makes use of the ordinary pQCD techniques, replacing the massless gluon propagator by Eq.~\eqref{glupropzero} and including the two-gluon vertex of Eq.~\eqref{glu2vertex} in the Feynman rules. Since the mass counterterm is not proportional to the coupling, one must decide how many mass counterterms to include at any fixed loop order in perturbation theory. Most of the studies undertaken to date have adopted the criterion by which the number of counterterms is no less than needed to cancel the divergences proportional to the gluon mass parameter, while saturating the maximum number of vertices that appear in the loops.

\begin{figure}[h]
    \centering
    \includegraphics[width=0.15\textwidth]{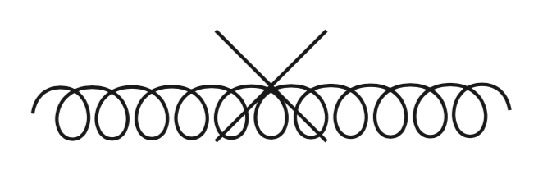}
    \vskip 5pt
    \includegraphics[width=0.45\textwidth]{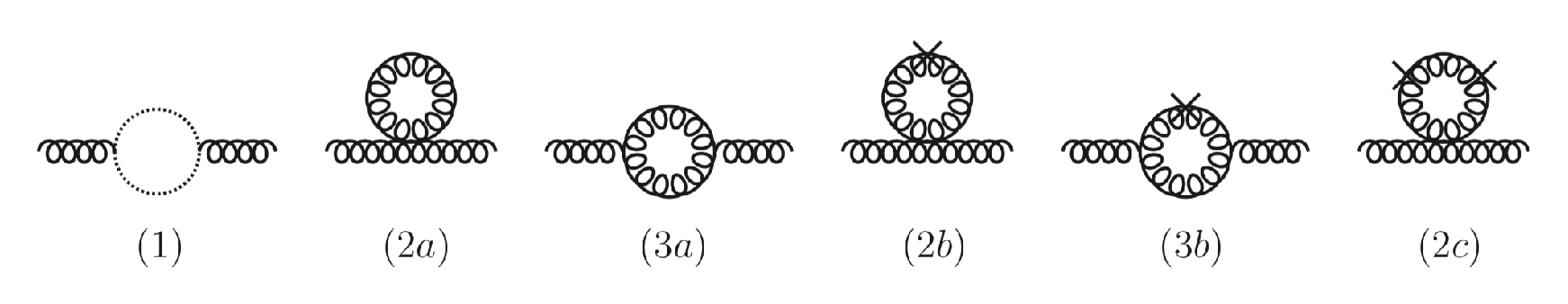}
    \caption{1PI diagrams with no more than three vertices used to compute the one-loop gluon polarization in the screened massive expansion of pure Yang-Mills theory. Top: tree-level gluon mass counterterm. Bottom: loop diagrams.}
    \label{glu1ldiags}
\end{figure}

Fig.~\ref{glu1ldiags} lists the diagrams included in the one-loop one-particle-irreducible (1PI) gluon polarization according to such a criterion. Diagrams $1$, $2a$ and $3a$ are the ordinary one-loop pQCD diagrams, albeit computed with a massive rather than massless gluon propagator. Diagrams $2a$ and $3a$ contain divergences proportional to the gluon mass parameter squared which are not renormalizable because of the absence of a gluon mass (and thus of a gluon mass renormalization counterterm) in the original Faddeev-Popov Lagrangian. Nonetheless, diagrams $2b$ and $3b$, which contain one insertion of the gluon mass counterterm $-\delta\Gamma$, can be shown \cite{SIR16a} to provide opposite diverging terms which, when summed to $2a$ and $3a$, make the sum finite. Thus diagrams $2b$ and $3b$ are included in the calculation to cure the spurious mass divergences and, since diagram $3b$ has three vertices, all the remaining diagrams with no more than one loop and no more than three vertices are also included in the polarization. These are the tree-level gluon mass counterterm at the top of Fig.~\ref{glu1ldiags}, and diagram $2c$.

Neglecting the diagonal color structure and denoting the momentum-space dressed gluon propagator with $\Delta_{\mu\nu}(p)$, after resumming the 1PI polarization diagrams $\Pi_{\mu\nu}(p)$ to obtain the full one-loop polarization, we can write $\Delta_{\mu\nu}(p)$ as
\begin{equation}
    \Delta_{\mu\nu}^{-1}(p)=(p^{2}+m^{2})\,t_{\mu\nu}(p)+\frac{p^{2}}{\xi}\,\ell_{\mu\nu}(p)-\Pi_{\mu\nu}(p)\ .
\end{equation}
While to any finite loop order $\Pi_{\mu\nu}(p)$ may contain longitudinal contributions as a consequence of having introduced a gluon mass at tree level, we know that the screened massive expansion is perturbatively equivalent to pQCD \cite{SIR16a}, and that the exact polarization is transverse. Thus we may assume that any resummation needed to make the longitudinal part of the polarization vanish has been performed, and simply set $\Pi_{\mu\nu}(p)=\Pi(p)\,t_{\mu\nu}(p)$. The polarization function $\Pi(p)$, modulo field renormalization counterterms, is easily seen to be equal to
\begin{equation}
    \Pi(p)=m^{2}+\Pi_{\text{loop}}(p)\ ,
\end{equation}
where the first term in the equation comes from the tree-level mass counterterm in Fig.~\ref{glu1ldiags}, while $\Pi_{\text{loop}}(p)$ is given by diagrams $1$, $2a$-$2c$ and $3a$-$3b$. Explicit expressions for $\Pi_{\text{loop}}(p)$ are provided in Ref.~\cite{SIR16a}.

With the definitions above, the zero-temperature one-loop dressed Euclidean gluon propagator computed in a generic covariant gauge within the screened mass expansion can be expressed as
\begin{equation}\label{ymvacgluprop}
    \Delta_{\mu\nu}(p)=\frac{t_{\mu\nu}(p)}{p^{2}-\Pi_{\text{loop}}(p)}+\frac{\xi}{p^{2}}\ \ell_{\mu\nu}(p)\ .
\end{equation}
In this equation, the tree-level gluon mass counterterm has canceled out the tree-level mass contained in the zero-order propagator. It follows that any mass generated for the dressed gluon propagator will come from the interaction loops. In Refs.\cite{SIR16a,SIR16b,SC18,CS20}, the prediction provided by Eq.~\eqref{ymvacgluprop} was shown to be in excellent agreement with the lattice data.

\subsection{The gluon propagator at finite temperature}

In order to extend the theory to finite temperatures, one restricts the Euclidean Faddeev-Popov action to the imaginary time interval $\tau\in[0,\beta]$, where $\beta=1/T$ is the inverse temperature of the system, and imposes periodic boundary conditions in imaginary time for the gluon and ghost field configurations. It follows that, at $T\neq0$, the SO(4) rotational symmetry of the Euclidean theory is broken down to spatial SO(3). Moreover, the time-components $p^{4}=\omega$ of the Euclidean momenta take on the discrete values $\omega=\omega_{n}=2\pi n T$ ($n\in\mathbb{Z}$) known as Matsubara frequencies; four-dimensional momentum integrals are then replaced by sums over the Matsubara index $n$ and three-dimensional integrals,
\begin{equation}
    \int\frac{d^{4}p}{(2\pi)^{4}}\to  T\sum_{n}\int\frac{d^{3}p}{(2\pi)^{3}}\ .
\end{equation}
The gluon propagator can no longer be expressed in terms of a single unknown scalar function (its four-dimensionally transverse component), but must instead be split into a three-dimensionally transverse term, a three-dimensionally longitudinal term, and the usual four-dimensionally longitudinal gauge term,
\begin{align}
           & \Delta_{\mu\nu}(\omega,{\bf p})=                                                                                                                  \\
    \notag & =\Delta_{T}(\omega,|{\bf p}|)\ \mc{P}_{\mu\nu}^{T}(p)+\Delta_{L}(\omega,|{\bf p}|)\ \mc{P}_{\mu\nu}^{L}(p)+\frac{\xi}{p^{2}}\,\ell_{\mu\nu}(p)\ .
\end{align}
In the above equation, $\Delta_{T,L}(\omega,{\bf p})$ are the three-dimensionally transverse and longitudinal projections of the propagator, which will henceforth be referred to simply as the transverse and the longitudinal gluon propagator, and $\mc{P}^{T,L}_{\mu\nu}(p)$ are transverse and longitudinal projectors defined as
\begin{align}
    \mc{P}^{T}_{\mu\nu}(p)       & =\left(1-\delta_{\mu4}\right)\left(1-\delta_{\nu4}\right)\left(\delta_{\mu\nu}-\frac{p_{\mu}p_{\nu}}{{\bf p}^{2}}\right)\ , \\
    \notag\mc{P}^{L}_{\mu\nu}(p) & =t_{\mu\nu}(p)-\mc{P}^{T}_{\mu\nu}(p)\ .
\end{align}

The Landau-gauge pure Yang-Mills gluon propagator was computed at finite temperature to one loop using the screened massive expansion in Refs.~\cite{SC21}. There $\Delta_{T,L}(\omega,{\bf p})$ were expressed as
\begin{equation}\label{ymglupropTraw}
    \Delta_{T,L}^{-1}(\omega,{\bf p})=p^{2}+\delta Z_{A}\,p^{2}-Ng^{2}\Pi_{T,L}^{(1)}(\omega,{\bf p})\ ,
\end{equation}
where $p^{2}=\omega^{2}+{\bf p}^2$, $\delta Z_{A}$ is the gluon field-strength renormalization constant and $\Pi^{(1)}_{T,L}(\omega,{\bf p})$ -- modulo a factor of $Ng^{2}$ -- are the (non-renormalized) finite-temperature transverse and longitudinal components of the gluon polarization obtained from the loop diagrams in Fig.~\ref{glu1ldiags}. These functions can be split as \cite{SC21}
\begin{equation}
    \Pi^{(1)}_{T,L}(\omega,{\bf p})=[\Pi^{(1)}_{T,L}(\omega,{\bf p})]_{\text{V}}+[\Pi^{(1)}_{T,L}(\omega,{\bf p})]_{\text{Th}}\ ,
\end{equation}
where
\begin{equation}
    [\Pi^{(1)}_{T,L}(\omega,{\bf p})]_{\text{V}}=\lim_{T\to 0}\Pi^{(1)}_{T,L}(\omega,{\bf p})\ ,
\end{equation}
the (divergent) vacuum polarization, is equal for both components and was first computed in \cite{SIR16a}, whereas $[\Pi^{(1)}_{T,L}(\omega,{\bf p})]_{\text{Th}}$ are the thermal ($T\neq0$) contribution to the polarization. Finally, $\Delta_{T,L}(p)$ can be put in the form
\begin{equation}\label{ymglupropadim}
    p^{2}\Delta_{T,L}(p)=\frac{z_{\pi}}{\pi_{1}(s)+\pi_{0}+[\pi_{T,L}(p)]_{\text{Th}}}\ ,
\end{equation}
where $s=p^{2}/m^{2}$,
\begin{align}
    \pi_{1}(s)                       & =-\frac{16\pi^{2}}{3}\frac{[\Pi^{(1)}_{T,L}(\omega,{\bf p})]_{\text{V,ren.}}}{p^{2}}\ , \\
    \notag[\pi_{T,L}(p)]_{\text{Th}} & =-\frac{16\pi^{2}}{3}\frac{[\Pi^{(1)}_{T,L}(\omega,{\bf p})]_{\text{Th}}}{p^{2}}\ ,
\end{align}
and the coupling constant was reabsorbed into the definition of the adimensional constants $z_{\pi}$ and $\pi_{0}$. The latter, together with the dimensionful value of the gluon mass parameter $m^{2}$, are the free parameters of the expansion. In particular, $\pi_{0}$ plays the role of an additive renormalization constant and depends on the choice of the finite terms in the gluon field renormalization constant $\delta Z_{A}$ in Eq.~\eqref{ymglupropTraw}. Explicit expressions for the functions $\pi_{1}(s)$ and $[\pi_{T,L}(p)]_{\text{Th}}$ can be found in Ref.~\cite{SC21}.

In momentum-subtraction renormalization schemes, all three free parameters $z_{\pi}$, $\pi_{0}$ and $m^{2}$ can depend on the temperature: $m^{2}$, being a mass scale introduced in the Lagrangian in order to account for dynamical mass generation in the gluon sector, can and should be fixed independently at each temperature, whereas the dependence of $z_{\pi}$ and $\pi_{0}$ on $T$ is inherited by that of the coupling and of the relations needed to enforce the renormalization conditions for the propagators. In what follows, the latter will be chosen to be\footnote{Since we will only present results at zero Matsubara frequencies, the actual renormalization momentum will be defined as $p^{2}=\omega^{2}+{\bf p}^{2}={\bf p}^{2}=\mu_{0}^{2}$.}
\begin{equation}\label{rencond}
    \Delta_{T,L}(\omega,{\bf p};T)|_{p^{2}=\mu_{0}^{2}}=\frac{1}{\mu_{0}^{2}}\ ,
\end{equation}
with $\mu_{0}=4$~GeV.

\begin{figure}[h]
    \includegraphics[width=0.3\textwidth,angle=270]{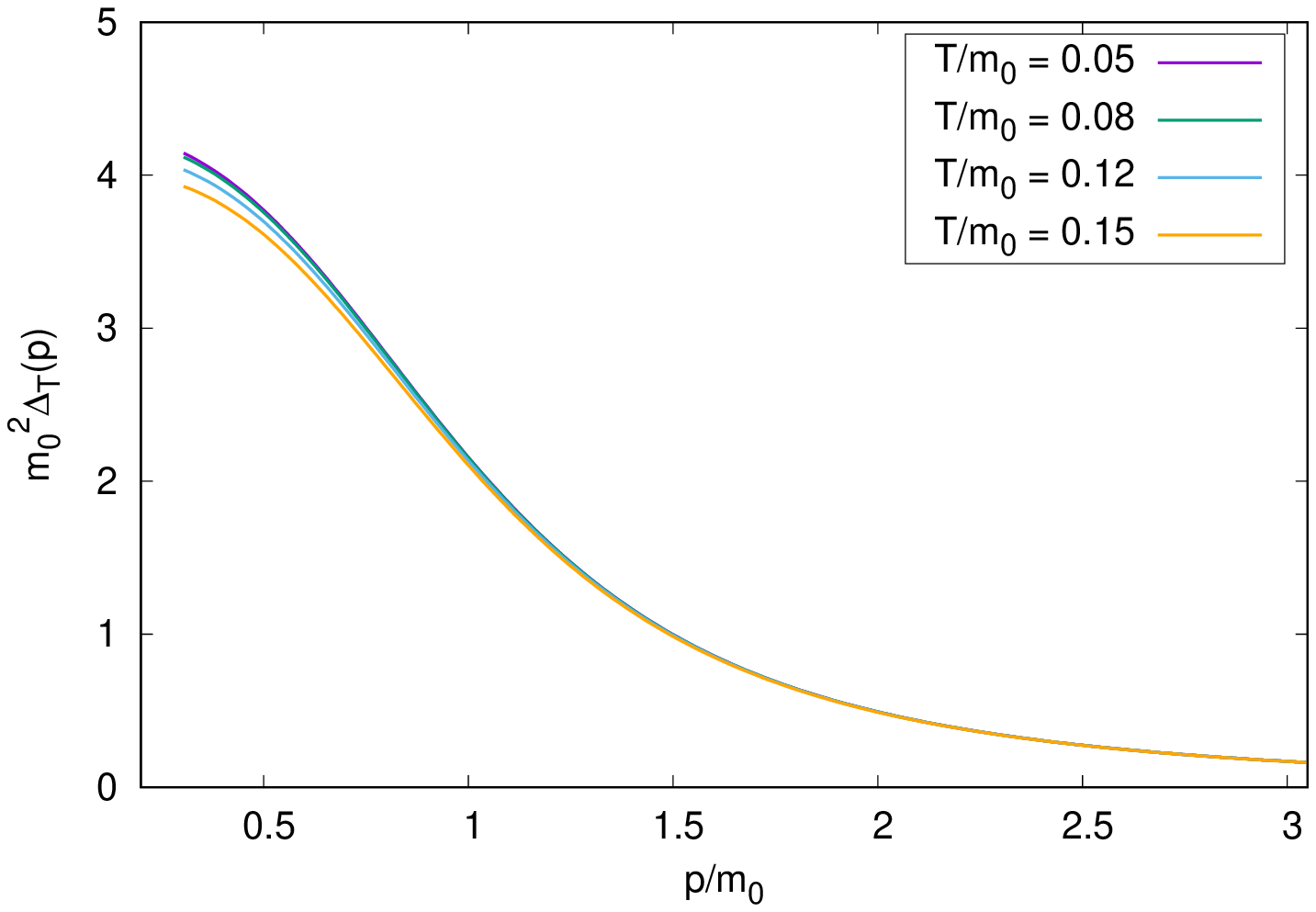}
    \vskip 5pt
    \includegraphics[width=0.3\textwidth,angle=270]{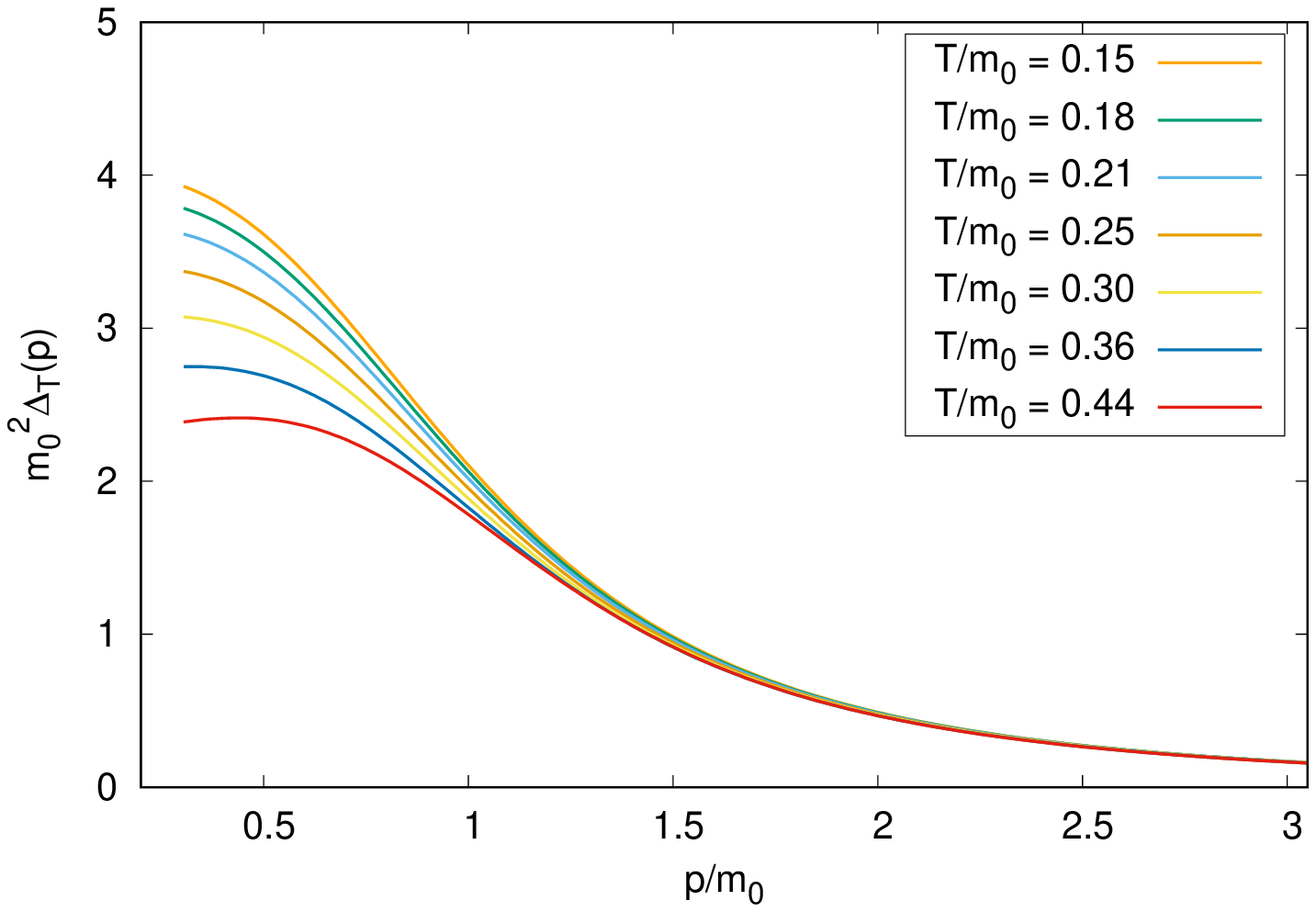}
    \caption{Transverse component of the Landau-gauge Euclidean gluon propagator at zero Matsubara frequency ($\omega_{n}=0$) as a function of spatial momentum, for different values of the temperature $T$, renormalized at $\mu_{0}=6.098\,m_{0}=4$~GeV. $\pi_{0}(T)=-0.876$, $m(T)=m_{0}=0.656$~GeV. Top: $T\leq0.15\,m_{0}$. Bottom: $T\geq0.15\,m_{0}$.}
    \label{ymfixedtransprop}
\end{figure}

\begin{figure}[h]
    \includegraphics[width=0.3\textwidth,angle=270]{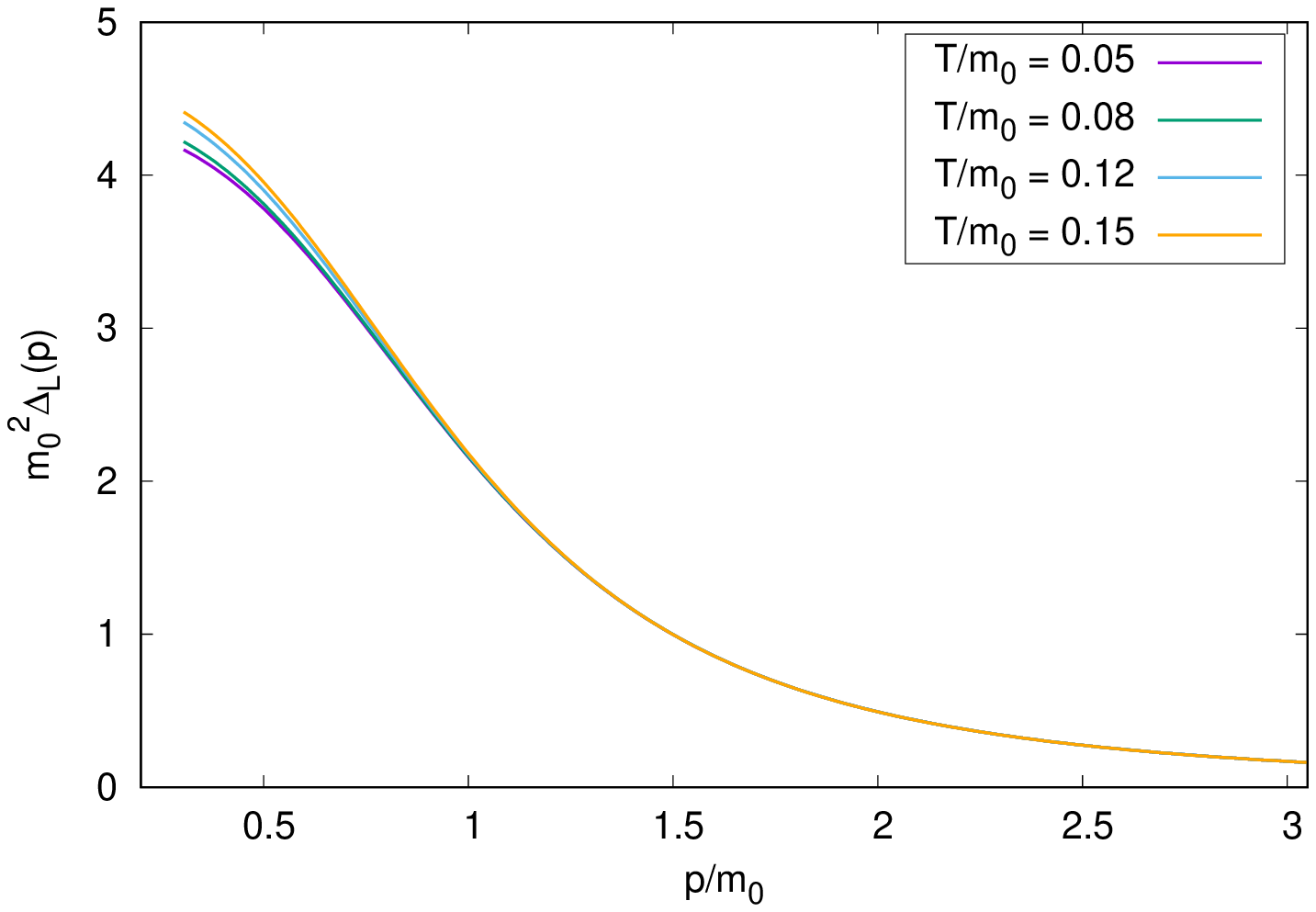}
    \vskip 5pt
    \includegraphics[width=0.3\textwidth,angle=270]{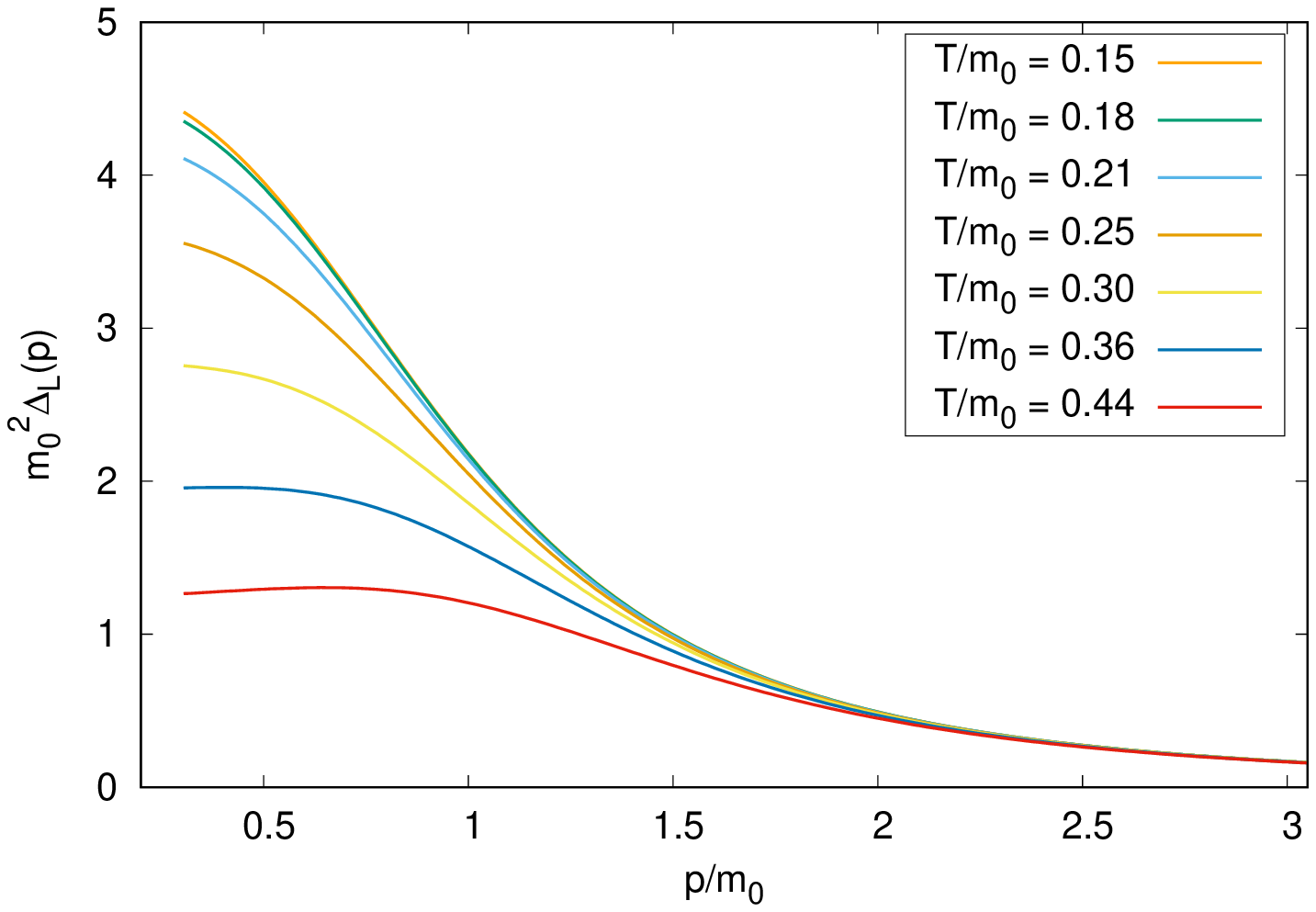}
    \caption{Longitudinal component of the Landau-gauge Euclidean gluon propagator at zero Matsubara frequency ($\omega_{n}=0$) as a function of spatial momentum, for different values of the temperature $T$, renormalized at $\mu_{0}=6.098\,m_{0}=4$~GeV. $\pi_{0}(T)=-0.876$, $m(T)=m_{0}=0.656$~GeV. Top: $T\leq0.15\,m_{0}$. Bottom: $T\geq0.15\,m_{0}$.}
    \label{ymfixedlongprop}
\end{figure}

The behavior of the Landau-gauge transverse and longitudinal gluon propagators at zero Matsubara frequency, as a function of spatial momentum and at fixed, non-zero temperature, was studied in Ref.~\cite{SC21}. As a first approximation, one can choose the free parameters of the expansion to be equal to those obtained by a fit of the lattice data at $T=0$. Here, since $z_{\pi}$ is fixed by renormalization and $\pi_{0}$ can be obtained by optimizing the expansion as in Ref.~\cite{SC18}, the only value that needs to be extracted from the lattice is that of the gluon mass parameter $m$. The resulting propagators are displayed in Figs.~\ref{ymfixedtransprop} and \ref{ymfixedlongprop}. One finds that, at fixed spatial momentum, the transverse propagator monotonically decreases with the temperature, while the longitudinal one first increases when $T$ is below a critical temperature $T_{c}$, then decreases for $T>T_{c}$. Within said approximation, $T_{c}$ was found to be equal to about $0.15\, m_{0}\approx 100$~MeV, where $m_{0}\approx656$~MeV is the value of the gluon mass parameter at $T=0$ obtained by fixing the energy scale from the lattice data of \cite{DOS16}. We shall comment on this result in a moment.

\begin{figure}[h]
    \includegraphics[width=0.3\textwidth,angle=270]{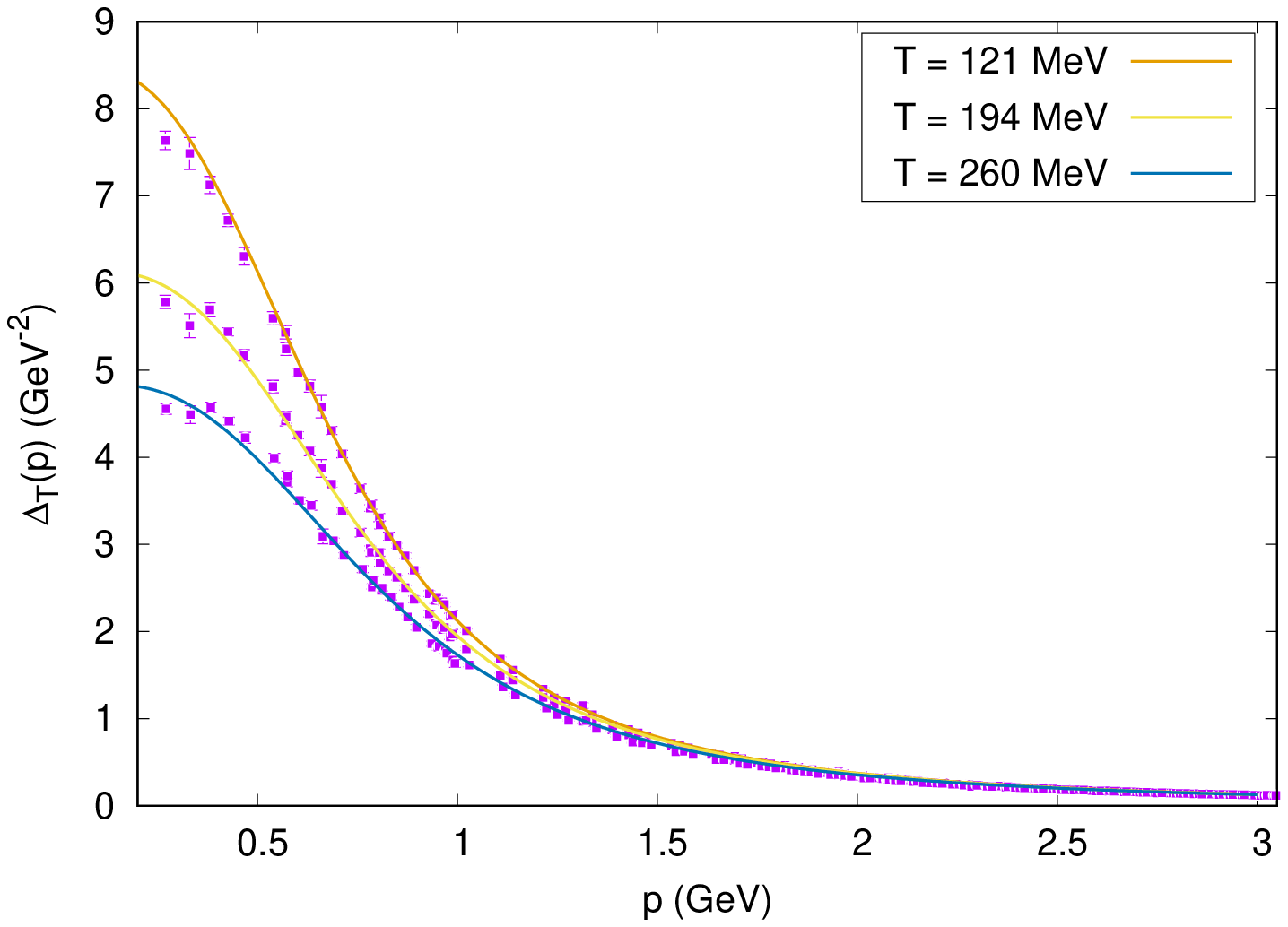}
    \vskip 5pt
    \includegraphics[width=0.3\textwidth,angle=270]{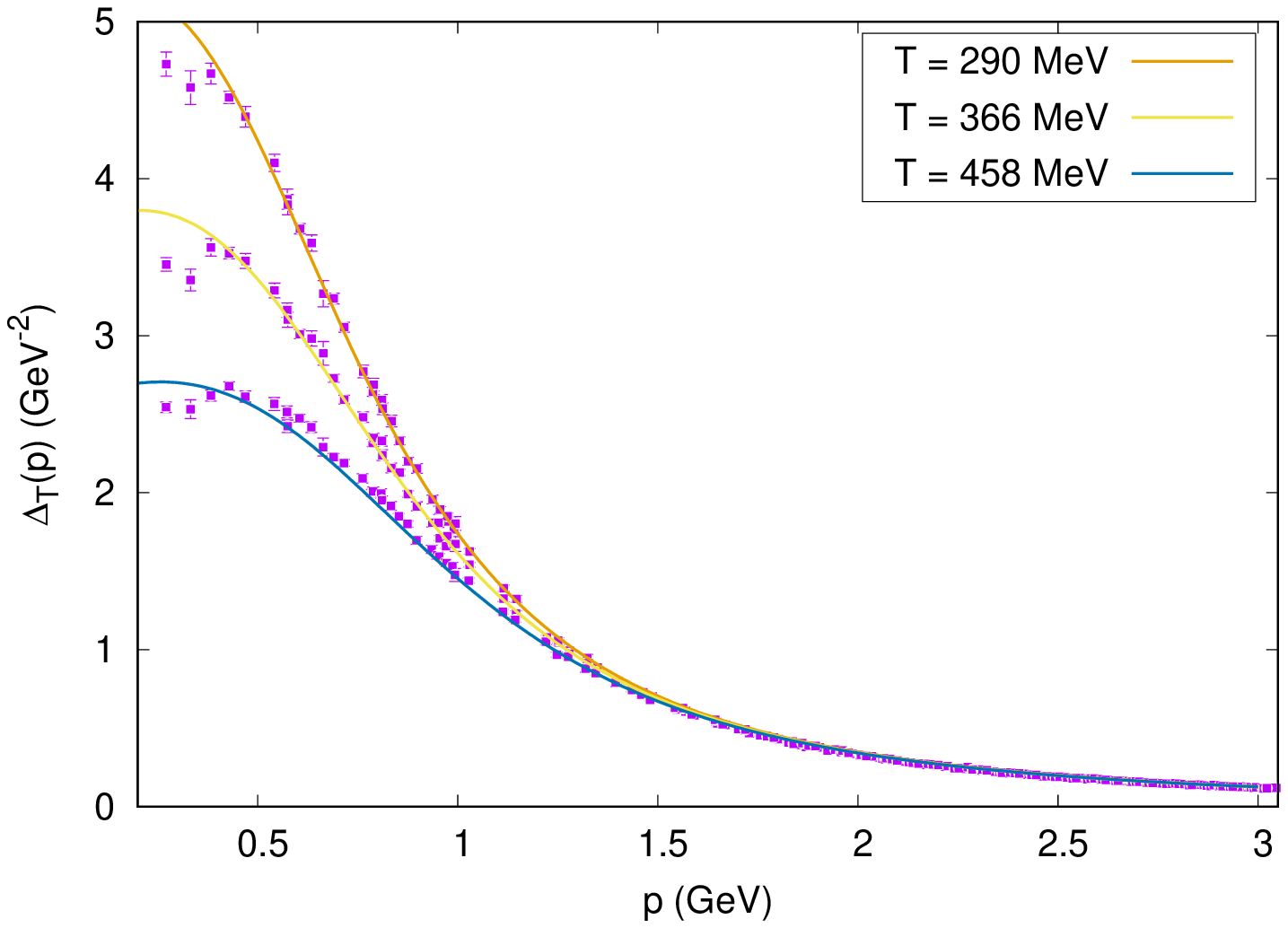}
    \caption{Transverse component of the Landau-gauge Euclidean gluon propagator at zero Matsubara frequency ($\omega_{n}=0$) as a function of spatial momentum, for different values of the temperature $T$, renormalized at $\mu_{0}=4$~GeV. The values of $\pi_{0}(T)$ and of $m(T)$ used in the figures are reported in Tab.~\ref{ymlatfitparams}. Top: $T<T_{c}$. Bottom: $T>T_{c}$. $T_{c}\approx270$~MeV. Lattice data from Ref.~\cite{SOBC14}.}
    \label{ymlattransprop}
\end{figure}

\begin{figure}[h]
    \includegraphics[width=0.3\textwidth,angle=270]{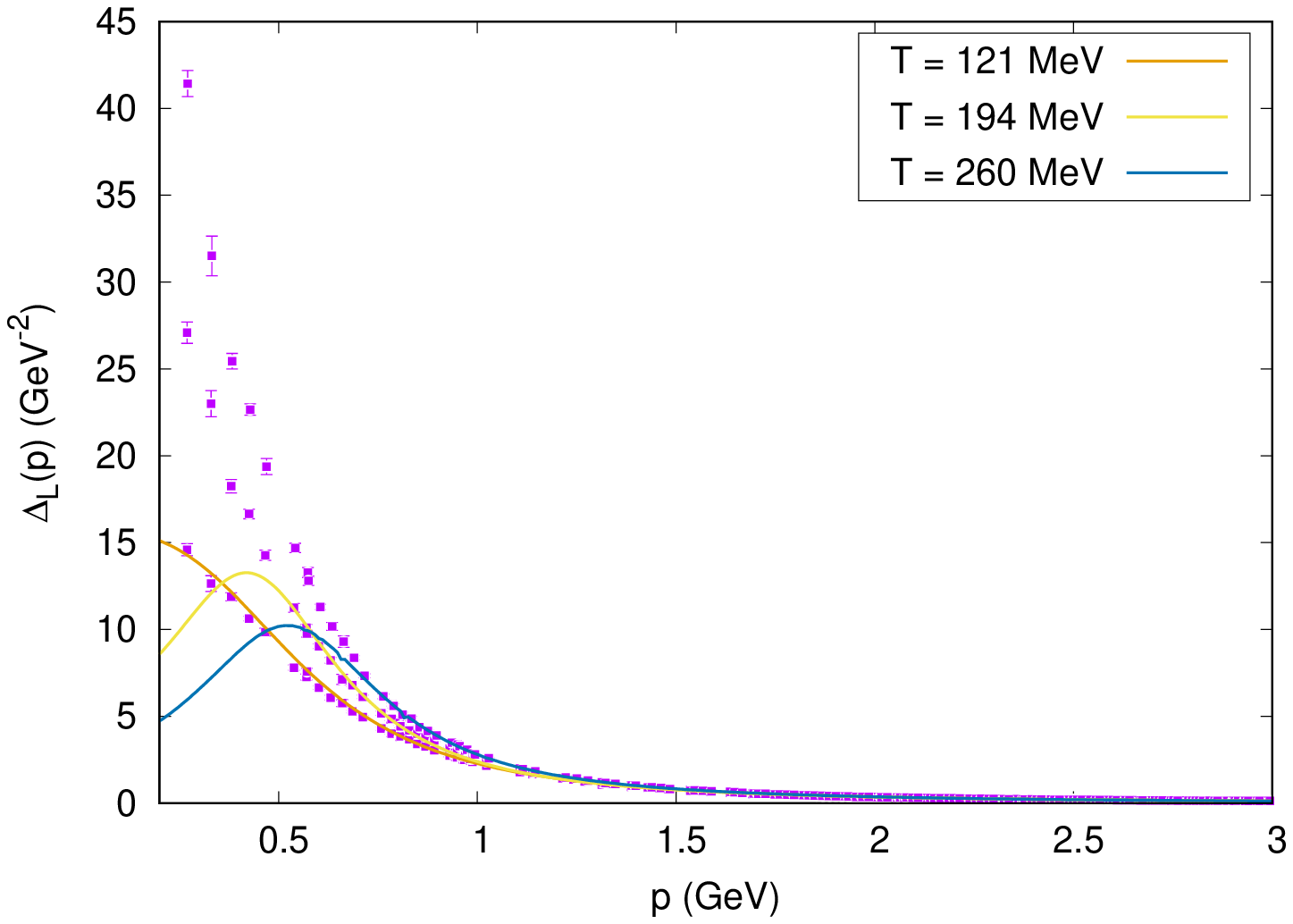}
    \vskip 5pt
    \includegraphics[width=0.3\textwidth,angle=270]{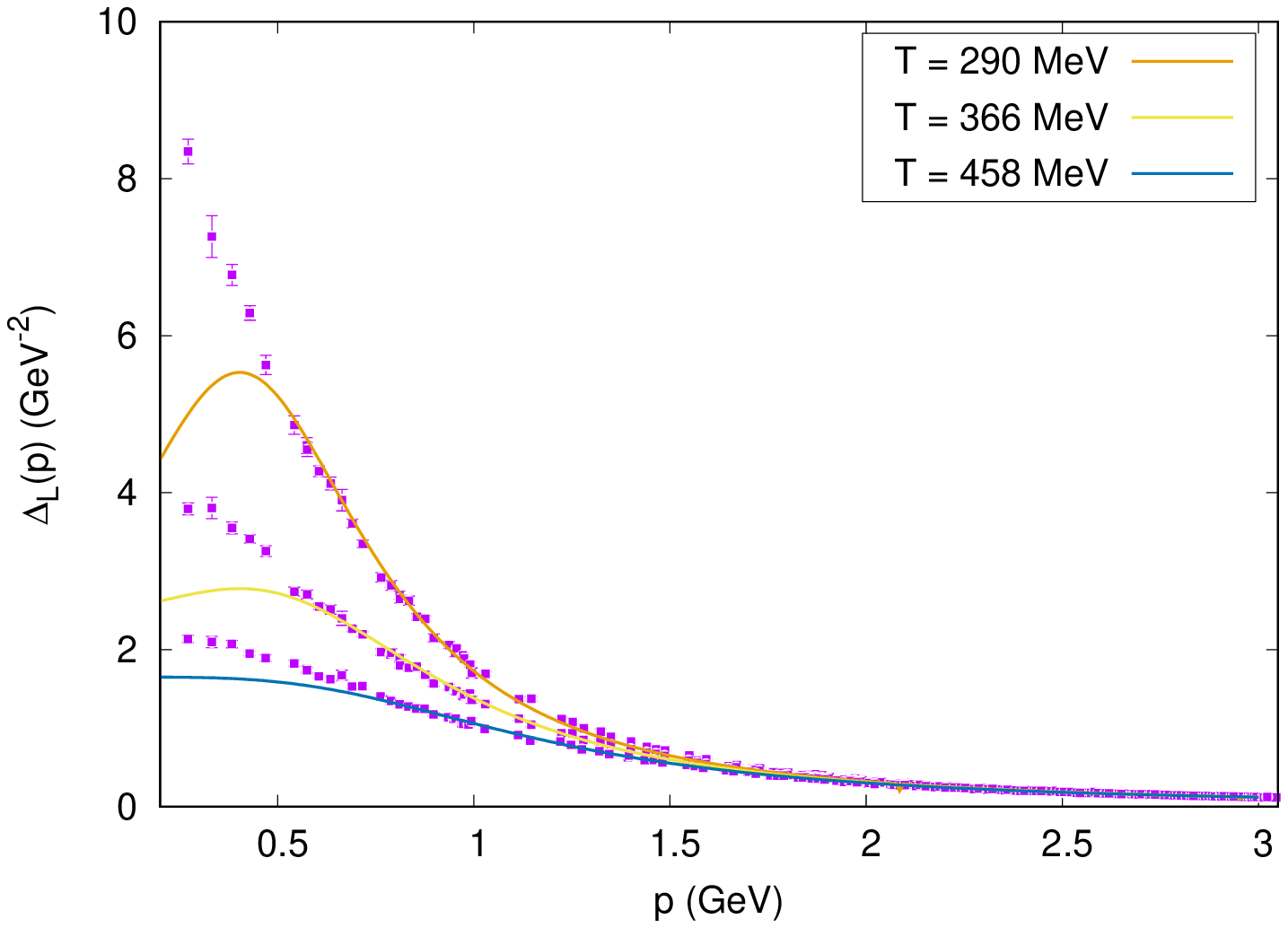}
    \caption{Longitudinal component of the Landau-gauge Euclidean gluon propagator at zero Matsubara frequency ($\omega_{n}=0$) as a function of spatial momentum, for different values of the temperature $T$, renormalized at $\mu_{0}=4$~GeV. The values of $\pi_{0}(T)$ and of $m(T)$ used in the figures are reported in Tab.~\ref{ymlatfitparams}. Top: $T<T_{c}$. Bottom: $T>T_{c}$. $T_{c}\approx270$~MeV. Lattice data from Ref.~\cite{SOBC14}.}
    \label{ymlatlongprop}
\end{figure}

As anticipated in the Introduction, the change in the behavior of the longitudinal component of the gluon propagator hints to the existence of two phases in which the gluons have different dynamical properties. This is made explicit by the lattice calculations -- Figs.~\ref{ymlattransprop} and \ref{ymlatlongprop} -- which show, first of all, that the change is in fact much more pronounced than observed to one-loop in the screened massive expansion by fixing the parameters at $T=0$, and second of all that the temperature $T_{\text{max}}$ at which the longitudinal propagator is maximum at fixed momentum is $T_{c}\approx270$~MeV -- that is, it is in fact the critical temperature of the first-order phase transition of pure Yang-Mills theory. The simple approximation adopted in Figs.~\ref{ymfixedtransprop} and \ref{ymfixedlongprop}, being a one-loop truncation and not letting the parameters vary with the temperature, is unable to quantitatively reproduce either the sudden increase of the longitudinal propagator to values which are as much as three times larger than at lower temperatures, or the value of the critical temperature itself. Nevertheless, the approximation is still able to capture the qualitative behavior of the longitudinal propagator and, in particular, the existence of a transition -- interpreted as a point across which such a behavior changes.

Insofar as the dependence of the parameters on the temperature is concerned, in Ref.~\cite{SC21} we found that the one-loop screened massive expansion of pure Yang-Mills theory can only reproduce the finite-temperature lattice data of Ref.~\cite{SOBC14} if $m^{2}$ and $\pi_{0}$ -- $z_{\pi}$ having been fixed by renormalizing the propagators at $\mu_{0}$ as in Eq.~\eqref{rencond} -- are fitted independently as functions of the temperature for the two components of the propagator. One major reason for this is the wildly different behavior of the two components as computed on the lattice, together with the fact that the ordinary setup of the expansion at finite temperature forces the gluon mass parameter to be the same for both the (three-dimensionally) transverse and longitudinal zero-order gluon propagators. It was argued that defining \textit{ab initio} a zero-order propagator of the form
\begin{equation}
    \Delta^{-1}_{\mu\nu}(p)=(p^2+m_{T}^{2})\ \mc{P}^{T}_{\mu\nu}(p)+(p^2+m_{L}^{2})\ \mc{P}^{L}_{\mu\nu}(p)+\frac{p^{2}}{\xi}\ \ell_{\mu\nu}(p)
\end{equation}
with $m_{L}(T)\neq m_{T}(T)$ for $T\neq0$ could put our findings on a more rigorous basis, but no further research was carried out in this direction.

While preparing this paper we discovered an error in the code used for fitting the propagators in Ref.~\cite{SC21}, which we amend in what follows. Tab.~\ref{ymlatfitparams} reports the corrected values of the parameters obtained from a fit of the lattice data, while Figs.~\ref{ymlattransprop} and \ref{ymlatlongprop} display the corresponding plots. As one can see by comparison with Figs.~4 and 5 in Ref.~\cite{SC21}, the corrected parameters do not yield substantially different propagators: the one-loop functions computed at finite temperature using the screened massive expansion are still in good agreement with the lattice data in the transverse sector, whereas they manage to reproduce well the longitudinal sector at low temperatures or at momenta larger than about $0.5$-$0.75$~GeV. The fact that, at $T=458$~MeV, the longitudinal propagator seems to match the lattice data better\footnote{We draw the reader's attention to the fact that in Ref.~\cite{SC21} we were simply unable to obtain a longitudinal fit for this value of the temperature.} than at all lower temperatures except for the lowest ($T=121$~MeV) points to the possibility that the region where the screened massive expansion fails to reproduce the lattice's longitudinal sector is in fact $p\lesssim0.5$-$0.75$~GeV and $T$ around $T_{c}$, instead of large $T$. This possibility can only be confirmed by extending the study to higher temperatures and assessing whether the agreement with the lattice data gets better as it seems here, instead of worse as appeared to be the case in Ref.~\cite{SC21}.

\begin{table}
    \begin{tabular}{|c|c|c|}
        \hline
        $T$ (MeV) & $m(T)$ (MeV) (trans., long.) & $\pi_{0}(T)$ (trans., long.) \\
        \hline
        121       & 675, 525                     & -0.83, -0.93                 \\
        194       & 725, 175                     & -0.78, -2.18                 \\
        260       & 775, 10*                     & -0.58, -6.58                 \\
        290       & 725, 100                     & -0.48, -1.88                 \\
        366       & 800, 350                     & -0.38, 0.32                  \\
        458       & 900, 650                     & -0.28, 0.92                  \\
        \hline
    \end{tabular}
    \caption{Values of the parameters used to obtain Figs.~\ref{ymlattransprop} and \ref{ymlatlongprop}. The starred value is the lowest that could be reached by our numerical routines without incurring in non-converging integrals. See the text for more details.}
    \label{ymlatfitparams}
\end{table}

As for the value of the critical temperature itself, if we define $T_{c}$ as the point at which the longitudinal propagator changes behavior with respect to $T$ at fixed momentum, then the screened massive expansion trivially agrees with the lattice finding, $T_{c}\approx270$~MeV, given that at large enough momenta it is able to reproduce the lattice propagators for all temperatures. Of course, to reach this conclusion, we must disregard the low-momentum region, where the longitudinal propagator does not match the lattice in the first place.

\begin{figure}[h]
    \includegraphics[width=0.3\textwidth,angle=270]{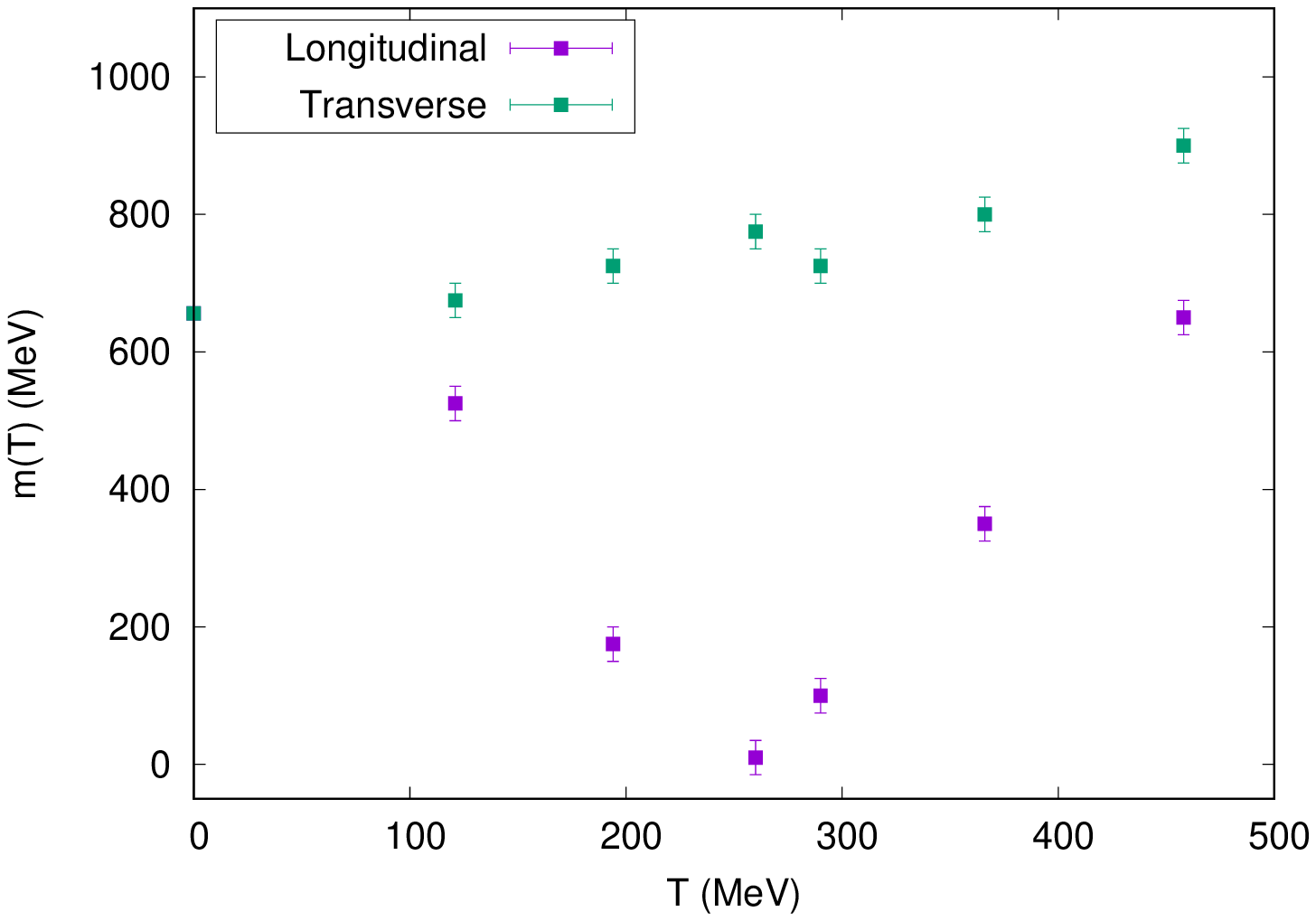}
    \vskip 5pt
    \hskip 12pt\includegraphics[width=0.278\textwidth,angle=270]{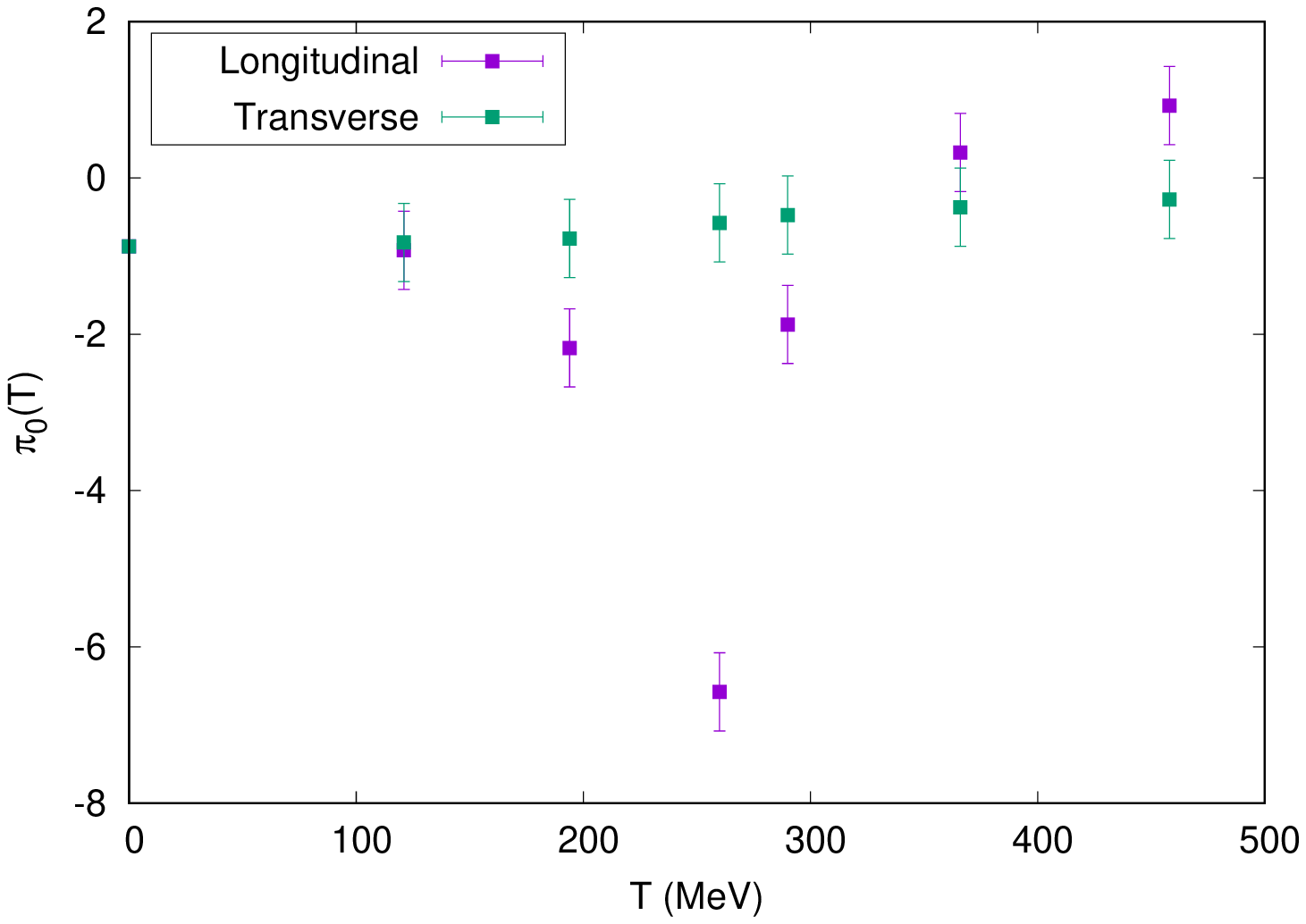}
    \caption{Temperature dependence of the free parameters $m(T)$ and $\pi_{0}(T)$ as reported in Tab.~\ref{ymlatfitparams}. At $T=0$, $m=656$~MeV and $\pi_{0}=-0.876$ as in Ref.~\cite{SC18}.}
    \label{ymlatparams}
\end{figure}

One relevant difference between the old fit parameters and the corrected ones lies in the dependence of the gluon mass parameter $m(T)$ on the temperature, Fig.~\ref{ymlatparams}. At variance with our previous determination, which saw the transverse propagator's $m(T)$ decrease with the temperature below $T_{c}$ to then remain essentially constant above $T_{c}$, the parameter is found to be a strictly increasing\footnote{Except for a small decrease around $T\approx T_{c}$ which is within errors.} function of the temperature, as one would expect from a mass that only gets corrected by thermal effects. The longitudinal $m(T)$, on the other hand, instead of being a decreasing function of temperature, is found to have a non-monotonous behavior that closely follows that of $\Delta_{L}^{-1}$: it decreases to a very small value\footnote{Due to numerical limitations we were unable to push the calculation to values lower than $m(T)=10$~MeV.} as $T$ approaches $T_{c}$, to then start increasing linearly with $T$ for $T>T_{c}$. This outcome is in qualitative agreement with the mass scales reported in \cite{SOBC14}.

\section{Full QCD: the gluon propagator at finite temperature and density}

\subsection{Modeling dynamical quarks across the phase transition}

The exact Lagrangian of full QCD $\mc{L}_{\text{QCD}}$ is obtained by adding dynamical terms for each of the quark flavors $f$ to the Faddeev-Popov Lagrangian $\mc{L}_{\text{YM}}$,
\begin{equation}\label{qcdexactlag}
    \mc{L}_{\text{QCD}}=\mc{L}_{\text{YM}}+\sum_{f}\ \psibar_{f}(\slashed{D}+m_{f})\psi_{f}\ ,
\end{equation}
where the $m_{f}$'s are the quark current masses and $\slashed{D}=\gamma^{\mu}D_{\mu}=\gamma^{\mu}(\partial_{\mu}-igA_{\mu}^{a}T_{a})$ -- the $T_{a}$'s being the generators of the fundamental representation of SU($N$) -- is the gauge covariant derivative acting on such a representation. While Eq.~\eqref{qcdexactlag} provides an accurate description of the ultraviolet regime of the strong interactions, when used in the ordinary perturbative setting it fails to reproduce the infrared phenomenon of low-momentum enhancement of the quark masses triggered by chiral symmetry violation\footnote{Or, to be more precise, if the current masses do not vanish, of what is left of the approximate chiral symmetry.}: at low momenta, even quarks with vanishing or small current masses are known to acquire a dynamically generated mass of $\approx 350$-$450$~MeV \cite{KBLW05} due to the formation of non-perturbative condensates which pQCD is unable to account for.

In Refs.~\cite{CRBS21} it was shown that adding and subtracting an extra mass term for quarks in the Lagrangian like we did in Eqs.~\eqref{fpkin} and \eqref{fpint} for the gluons, while treating the newly introduced quark masses as free parameters of the expansion, yields Euclidean quark mass functions which agree well with the lattice data for various current masses. Such a reformulation of QCD perturbation theory can be proven to be equivalent to ordinary pQCD just like the one introduced in Sec.~II above, and is able to describe dynamical mass generation in the quark sector.

In the present paper we adopt a simpler effective approach. Rather than shifting the expansion point of the quark perturbative series, we replace the quark current masses with the effective mass with which they propagate in the infrared. In other words, we model the system with the Lagrangian
\begin{equation}\label{qcdmodlag}
    \mc{L}_{\text{QCD}}=\mc{L}_{\text{YM}}+\sum_{f}\ \psibar_{f}(\slashed{D}+M_{f})\psi_{f}\ ,
\end{equation}
where the $M_{f}$'s are the masses dynamically generated at $T=\mu=0$. The approximation given by Eq.~\eqref{qcdmodlag} provides a better description of the non-perturbative regime of QCD at small momenta and low temperatures $T$ and chemical potentials $\mu$. As for larger temperatures and chemical potentials, as $T$ and $\mu$ increase beyond the confined phase, chiral symmetry is restored and the effective quark masses decrease. In regions of the phase diagram where the decrease is slow, the latter is unlikely to have a large impact on the radiative corrections to the gluon propagator, and can thus be neglected. On the other hand, an abrupt decrease of the effective masses to values closer to the current masses may actually have an appreciable effect on the propagator. We will come back to this aspect later on, when we will discuss the phase diagram of QCD in Sec.~IV.

By expanding Eq.~\eqref{qcdmodlag} around massive gluons and massless ghosts -- like we did in Eqs.~\eqref{fpkin} and \eqref{fpint} -- and quarks of masses $M_{f}$, one obtains a perturbative series for full QCD whose Feynman rules are those of pQCD plus the gluon mass counterterm in Eq.~\eqref{glu2vertex}, with the zero-order gluon propagator given by Eq.~\eqref{glupropzero} and the zero-order quark propagator given by
\begin{equation}
    S_{f,0}(p)=\frac{1}{i\slashed{p}+M_{f}}\ .
\end{equation}

At a finite temperature $T$, restricting the imaginary-time integrals to $\tau\in[0,1/T]$ and imposing anti-periodic boundary conditions for the quark fields causes the time component $p^{4}$ of the quarks' momenta to take on values $p^{4}=\omega_{n}=(2n+1)\pi T$, $n\in \mathbb{Z}$. Just like in the bosonic case, these fermionic frequencies will appear as the summands of finite-$T$ momentum-space integrals. In order to account for a non-zero (baryonic) chemical potential $\mu$\footnote{Note that we will be using a microscopic definition of the chemical potential $\mu$ -- i.e. one based on Eq.~\eqref{qcdchempotsubst}. The baryonic chemical potential $\mu_{B}$ is $\mu_{B}=3\mu$.}, one then adds the number-density operator $N_{q}=\sum_{f}\psibar_{f}\gamma^{4}\psi_{f}$ to the Euclidean QCD Lagrangian as
\begin{equation}\label{qcdchempotsubst}
    \mc{L}_{\text{QCD}}\to\mc{L}_{\text{QCD}}-\mu N_{q}\ .
\end{equation}
Being quadratic in the quark fields, the operator can be included into the kinetic part of the Lagrangian, and it causes the zero-order Euclidean quark propagator to take on the form
\begin{equation}
    S_{f,0}(p)=\frac{1}{i\slashed{p}+M_{f}-\mu\gamma^{4}}=\frac{1}{i\slashed{\widetilde{p}}+M_{f}}\ ,
\end{equation}
where $\widetilde{p}^{4}=p^{4}+i\mu$, $\widetilde{p}^{i}=p^{i}$.\\

\begin{figure}[h]
    \hskip 12pt\includegraphics[width=0.30\textwidth]{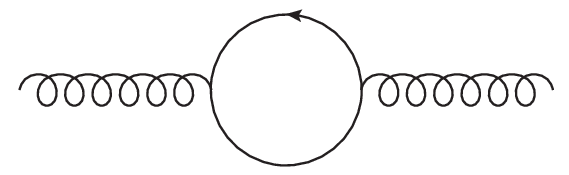}
    \caption{Full-QCD one-loop quark polarization diagram.}
    \label{qcdquarkloop}
\end{figure}

At one loop, by modeling the quark masses as discussed in the present section, the gluon propagator can be evaluated within the screened massive expansion of full QCD by adding one quark loop (Fig.~\ref{qcdquarkloop}) for each quark flavor to the Yang-Mills 1PI polarization of Fig.~\ref{glu1ldiags}. One then obtains the components of the propagator in the form
\begin{align}\label{qcdglupropTraw}
    &\Delta_{T,L}^{-1}(\omega,{\bf p})=\\
    \notag&=p^{2}+\delta Z_{A}\,p^{2}-Ng^{2}\Pi_{T,L}^{(1)}(\omega,{\bf p})-g^{2}\sum_{f}\Pi_{T,L}^{(f)}(\omega,{\bf p})\ ,
\end{align}
where $\Pi_{T,L}^{(f)}(\omega,{\bf p})$, modulo a factor of $g^{2}$, is the (transverse or longitudinal component of the) contribution to the gluon polarization due to the quark loop of flavor $f$.

\subsection{The one-loop gluon propagator: calculation}

The one-loop polarization functions $\Pi_{T,L}^{(f)}(\omega,{\bf p})$ in Eq.~\eqref{qcdglupropTraw} can be evaluated at finite temperature and density by making use of the fermionic frequency sum formula ($\omega_{n}=(2n+1)\pi T$, $n\in\mathbb{Z}$)
\begin{align}
    T\sum_{n}f(\omega_{n}+i\mu)&=-\frac{1}{2\pi}\oint_{C}\,dz\ \frac{f(z)}{e^{i\beta(z-i\mu)+1}}\ ,
\end{align}
where $C$ is a contour that encircles the $\text{Im}(z)=\mu$ line of the complex plane counterclockwise. By manipulating the integrand on the right-hand side and deforming the contour $C$, one can easily show that the fermionic sum can be split as \cite{KG23}
\begin{align}\label{fermisums}
    &T\sum_{n}f(\omega_{n}+i\mu)=\\
    \notag&=\int_{-\infty}^{+\infty}\frac{dp^{4}}{2\pi}\ f(p^{4})-\oint_{C'}\frac{dz}{2\pi}\ f(z)+\\
    \notag&\quad-\int_{-\infty+i\mu+i\epsilon}^{+\infty+i\mu+i\epsilon}\frac{dz}{2\pi}\frac{f(z)}{e^{-i\beta (z-i\mu)}+1}+\\
    \notag&\quad-\int_{-\infty+i\mu-i\epsilon}^{+\infty+i\mu-i\epsilon}\frac{dz}{2\pi}\frac{f(z)}{e^{i\beta (z-i\mu)}+1}\ ,
\end{align}
where $C'$ is a second counterclockwise contour, delimiting the region between the real axis and the $\text{Im}(z)=\mu$ line. The second term in the above equation vanishes for $\mu=0$, whereas the third and the fourth one vanish for $T=0$. Since the first term does not depend on temperature and chemical potential, it then follows that said term makes up for the vacuum contribution to the sum. The second term, on the other hand, only depends on chemical potential (i.e. it does not depend on temperature) and thus represents a pure finite-density contribution. Finally, the last two terms depend on both temperature and chemical potential.

Because of Eq.~\eqref{fermisums}, $\Pi_{T,L}^{(f)}(\omega,{\bf p})$ can be split as
\begin{equation}
    \Pi_{T,L}^{(f)}(\omega,{\bf p})=[\Pi_{T,L}^{(f)}(\omega,{\bf p})]_{\text{V}}+[\Pi_{T,L}^{(f)}(\omega,{\bf p})]_{\text{M}}\ ,
\end{equation}
where $[\Pi_{T,L}^{(f)}(\omega,{\bf p})]_{\text{V}}$ is the pure vacuum contribution -- and as such is the same for both components --, whereas $[\Pi_{T,L}^{(f)}(\omega,{\bf p})]_{\text{M}}$ is the medium contribution, i.e. that due to finite-temperature and density effects. These functions are well known in the literature as they are equal to analogous electron loop contributions to the photon polarization functions with $e^{2}$ -- the electromagnetic coupling -- replaced by $g^{2}/2$ and the electron mass replaced by $M_{f}$. They read
\begin{widetext}
\begin{align}
    [\Pi_{T,L}^{(f)}(\omega,{\bf p})]_{\text{V}}&=-\frac{p^{2}}{24\pi^{2}}\left\{\frac{2}{\epsilon}+\ln\frac{\overline{\mu}^{2}}{M_{f}^{2}}+\frac{s-2}{s}\sqrt{\frac{4+s}{s}}\ln\left(\frac{\sqrt{s+4}-\sqrt{s}}{\sqrt{s+4}+\sqrt{s}}\right)-\frac{4}{s}+\frac{5}{3}\right\}_{s=p^{2}/M_{f}^{2}}\ ,\label{quarkvacpol}\\
    [\Pi_{T}^{(f)}(\omega,{\bf p})]_{\text{M}}&=-\frac{1}{4\pi^{2}}\,\mathfrak{R}e\int_{0}^{+\infty}\frac{dq\, q^{2}}{\varepsilon_{q}}\ n_{F}(\varepsilon_{q})\left\{\frac{{\bf p}^{2}-\omega^{2}}{{\bf p}^{2}}-\frac{4q^{2}{\bf p}^{2}+4\omega^{2}\varepsilon_{q}^{2}+p^{2}({\bf p}^{2}-\omega^{2}-4i\omega\varepsilon_{q})}{4q|{\bf p}|^{3}}\ln\left(\frac{R_{+}}{R_{-}}\right)\right\}\ ,\label{quarktranspol}\\
    [\Pi_{L}^{(f)}(\omega,{\bf p})]_{\text{M}}&=-\frac{1}{2\pi^{2}}\ \frac{p^{2}}{{\bf p}^{2}}\ \mathfrak{R}e\int_{0}^{+\infty}\frac{dq\,q^{2}}{\varepsilon_{q}}\ n_{F}(\varepsilon_{q})\left\{1-\frac{p^{2}-4\varepsilon_{q}^{2}+4i\omega\varepsilon_{q}}{4q|{\bf p}|}\ln\left(\frac{R_{+}}{R_{-}}\right)\right\}\ ,\label{quarklongpol}
\end{align}
\end{widetext}
where $p^{2}=\omega^{2}+{\bf p}^2$, $\epsilon=4-d$, $\overline{\mu}$ is the scale introduced by dimensional regularization, and in the second and third equations
\begin{equation}
    R_{\pm}=p^{2}\pm 2q|{\bf p}|+2i\omega\varepsilon_{q}\ ,\qquad\varepsilon_{q}=\sqrt{q^{2}+M_{f}^{2}}\ ,
\end{equation}
\begin{equation}
    n_{F}(\varepsilon)=\frac{1}{e^{\beta(\varepsilon-\mu)}+1}+\frac{1}{e^{\beta(\varepsilon+\mu)}+1}\ .
\end{equation}
The $\mathfrak{R}e$ operator is defined as
\begin{equation}
    \mathfrak{R}e\left\{f(\omega)\right\}=\frac{f(\omega)+f(-\omega)}{2}\ ,
\end{equation}
so that, as long as $\omega$ is a Matsubara frequency, it acts in Eqs.~\eqref{quarktranspol}-\eqref{quarklongpol} as the real part operator.

If we now define adimensional functions $\pi_{f}(s)$ and $[\pi_{T,L}^{(f)}(p)]_{\text{M}}$ as
\begin{align}
    \pi_{f}(s)                       & =-\frac{16\pi^{2}}{3N}\frac{[\Pi^{(f)}_{T,L}(\omega,{\bf p})]_{\text{V,ren.}}}{p^{2}}\ , \\
    \notag[\pi_{T,L}^{(f)}(p)]_{\text{M}} & =-\frac{16\pi^{2}}{3N}\frac{[\Pi^{(1)}_{T,L}(\omega,{\bf p})]_{\text{M}}}{p^{2}}\ ,
\end{align}
then, going back to Eq.~\eqref{ymglupropadim}, we can express the gluon propagator in the form
\begin{align}\label{qcdglupropadim}
    &p^{2}\Delta_{T,L}(p)=\\
    \notag&=\frac{z_{\pi}}{\pi_{1}(s)+\pi_{0}+[\pi_{T,L}(p)]_{\text{Th}}+\sum_{f}\left(\pi_{f}(s)+[\pi_{T,L}^{(f)}(p)]_{\text{M}}\right)}\ .
\end{align}
We fix the renormalization of $[\Pi^{(f)}_{T,L}(\omega,{\bf p})]_{\text{V}}$ so that
\begin{equation}\label{pifvac}
\pi_{f}(s)=\frac{2}{9N}\, \left[\ln\frac{m^{2}}{M_{f}^{2}}+f_q\left(s\,\frac{m^{2}}{M_{f}^{2}}\right)\right]\ ,
\end{equation}
where $s=p^{2}/m^{2}$,
\begin{equation}\label{quarkfq}
    f_{q}(x)=\frac{x-2}{x}\sqrt{\frac{4+x}{x}}\ln\left(\frac{\sqrt{x+4}-\sqrt{x}}{\sqrt{x+4}+\sqrt{x}}\right)-\frac{4}{x}+\frac{5}{3}\ ,
\end{equation}
and the logarithm of $M_{f}$ is kept in Eq.~\eqref{pifvac} instead of being absorbed into the gluon field-strength renormalization constant in order to avoid an artificial divergence of the form $\ln M_{f}$ that would otherwise arise in the $M_{f}\to 0$ limit. As we will see later on, the choice of a finite additive constant in Eqs.~\eqref{pifvac}-\eqref{quarkfq}, being equivalent to a choice of $\pi_{0}$ in Eq.~\eqref{qcdglupropadim}, has little to no effect on the qualitative behavior of the gluon propagator.

\subsection{The one-loop gluon propagator: results}

In order to obtain physically meaningful predictions from Eq.~\eqref{qcdglupropadim}, we assume our Lagrangian Eq.~\eqref{qcdmodlag} to describe the infrared dynamics of $n_{f}=2+1$ light quark fields with current masses $m_\text{1}\approx2$-4~MeV and $m_{2}\approx90$~MeV, that is, the dynamics of the up, down and strange quarks. Thanks to lattice calculations we know that, in the infrared, these quarks will propagate with masses around $350$-$450$~MeV \cite{KBLW05}, with the heavier quark's mass being somewhat larger than that of the lighter ones. For the purpose of distinguishing between the two in our analysis, we will choose the effective infrared masses $M_{1}$ and $M_{2}$ to be $M_{1}=350$~MeV, $M_{2}=450$~MeV, although in reality the difference may be smaller. As for the gluon mass parameter $m$ and additive renormalization constant $\pi_{0}$ in Eq.~\eqref{qcdglupropadim}, we fix these from a fit of the optimized one-loop screened expansion to the zero-temperature pure Yang-Mills lattice data as in Sec.~ IIB; namely, we set $m=m_{0}=656$~MeV and $\pi_{0}=-0.876$ \cite{SC18} for all $(T,\mu)$. Although these parameters should depend on temperature -- and, in the case of finite-density full QCD, also on chemical potential --, the pure Yang-Mills results presented in Sec.~II suggest that taking them to be constants will yield propagators with the correct qualitative behavior. In the next section we will explore other parameter choices and show that, from a qualitative perspective, the results of the present section are general enough not to depend on them.

\begin{figure*}[t]
    \includegraphics[width=0.34\textwidth,angle=270]{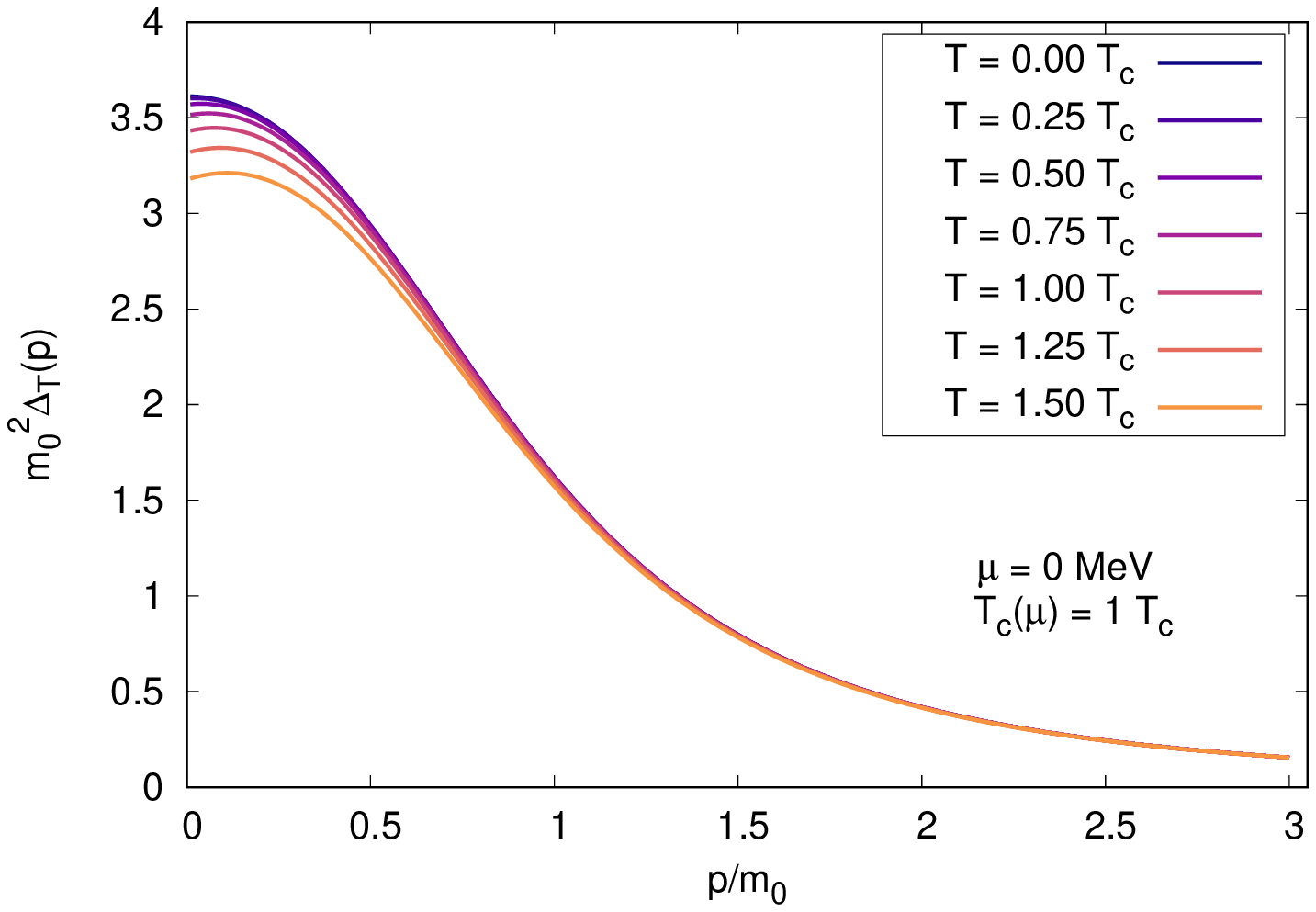}\hspace{5pt}%
    \includegraphics[width=0.34\textwidth,angle=270]{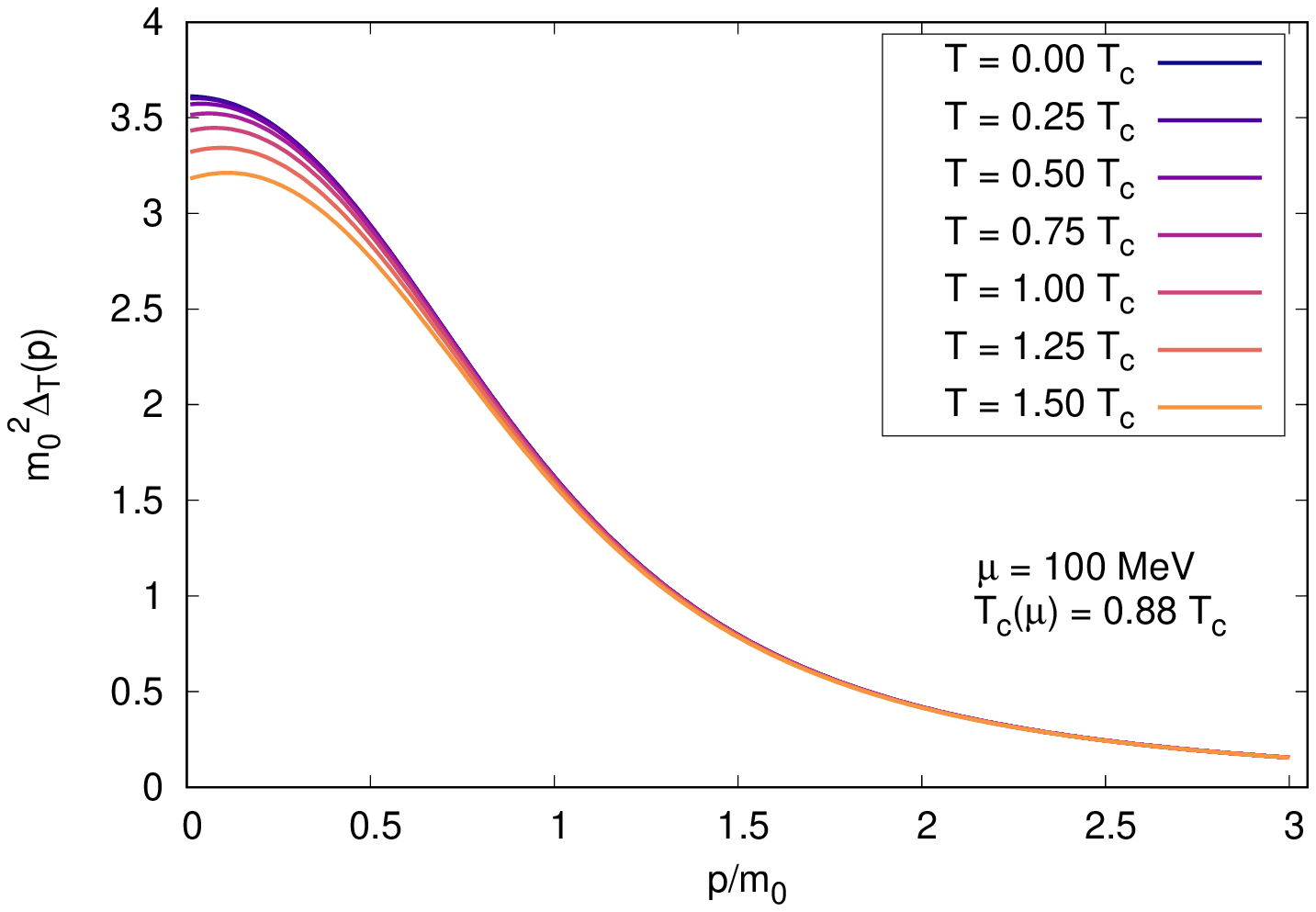}
    \includegraphics[width=0.34\textwidth,angle=270]{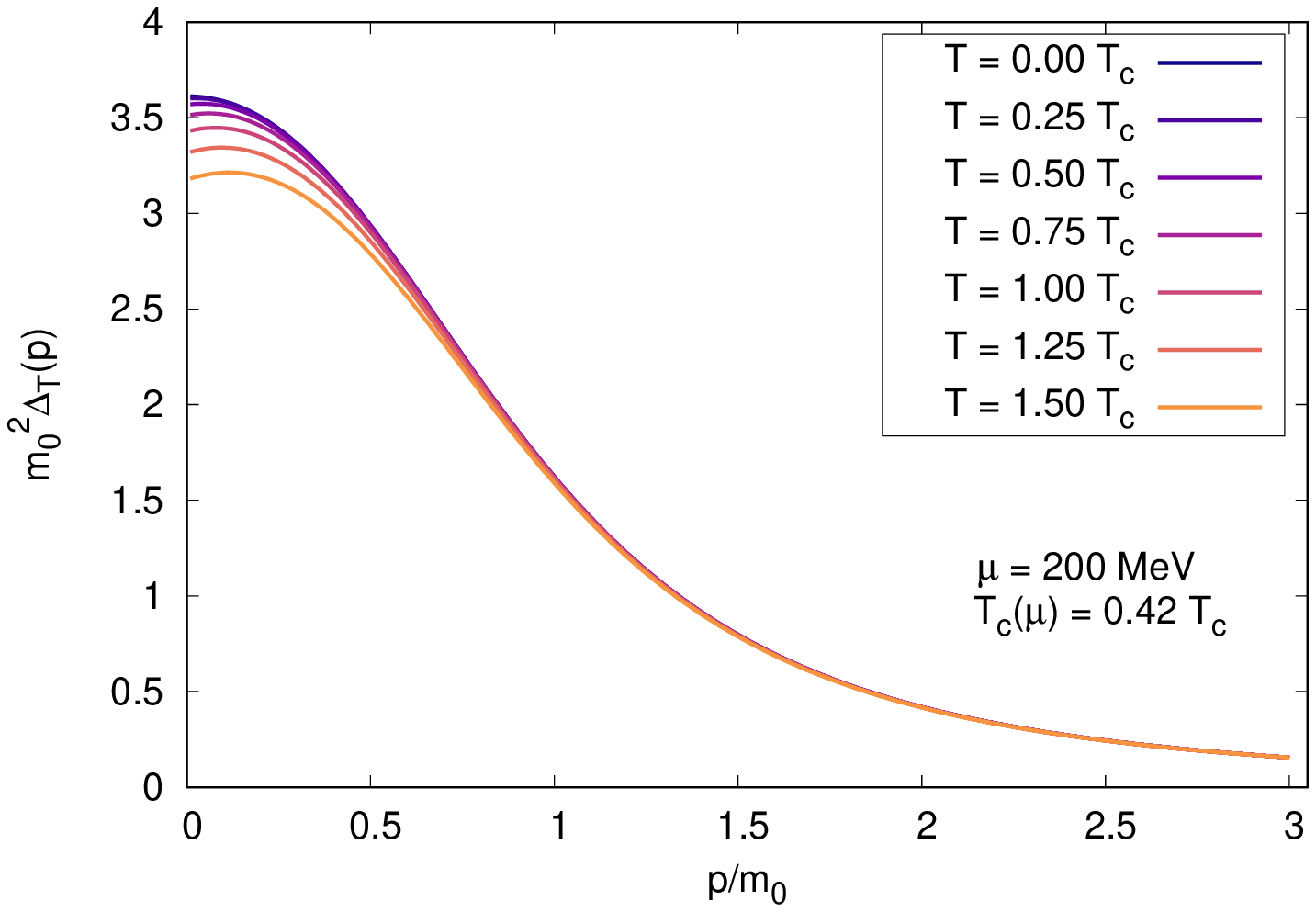}\hspace{5pt}%
    \includegraphics[width=0.34\textwidth,angle=270]{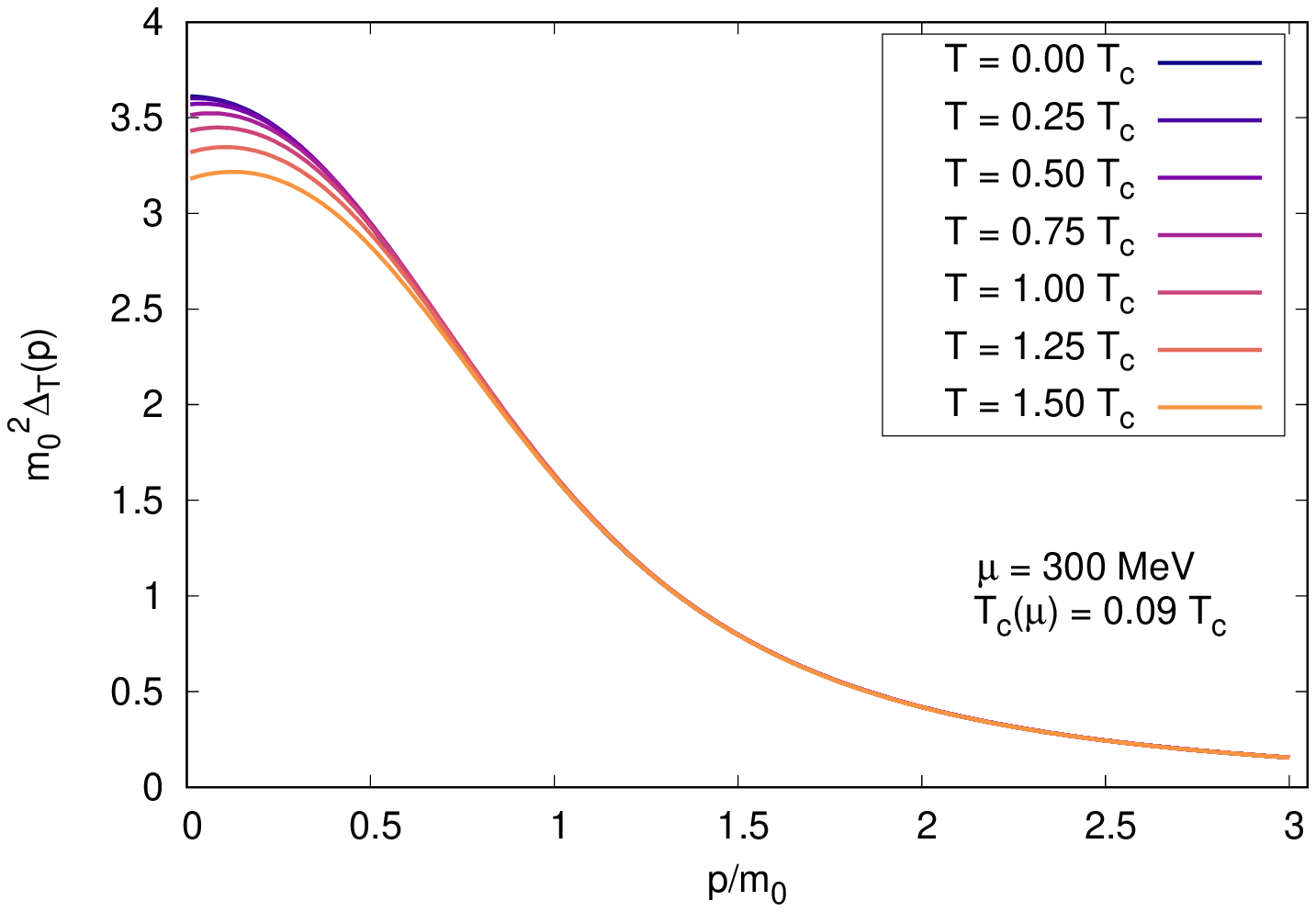}
    \includegraphics[width=0.34\textwidth,angle=270]{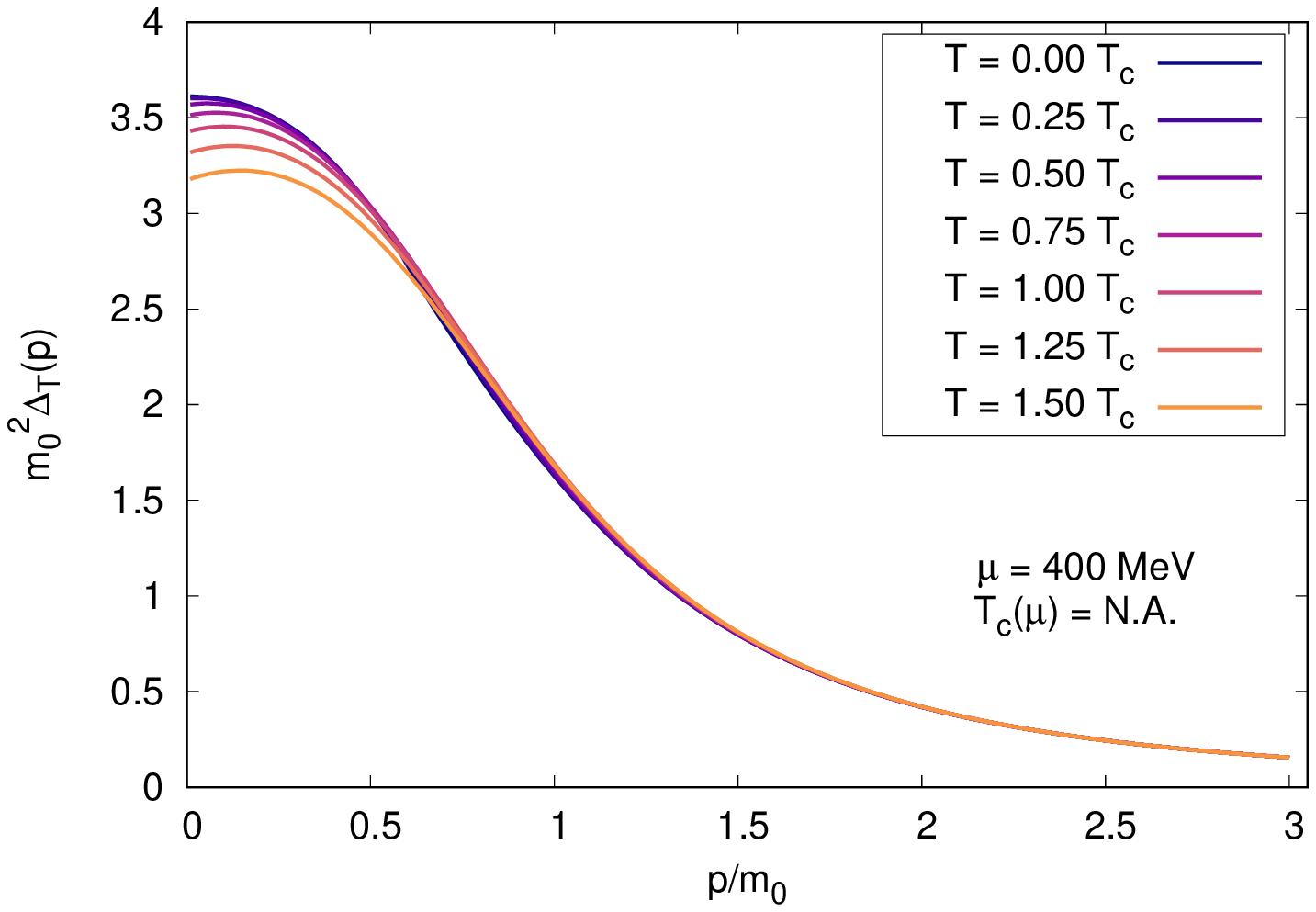}\hspace{5pt}%
    \includegraphics[width=0.34\textwidth,angle=270]{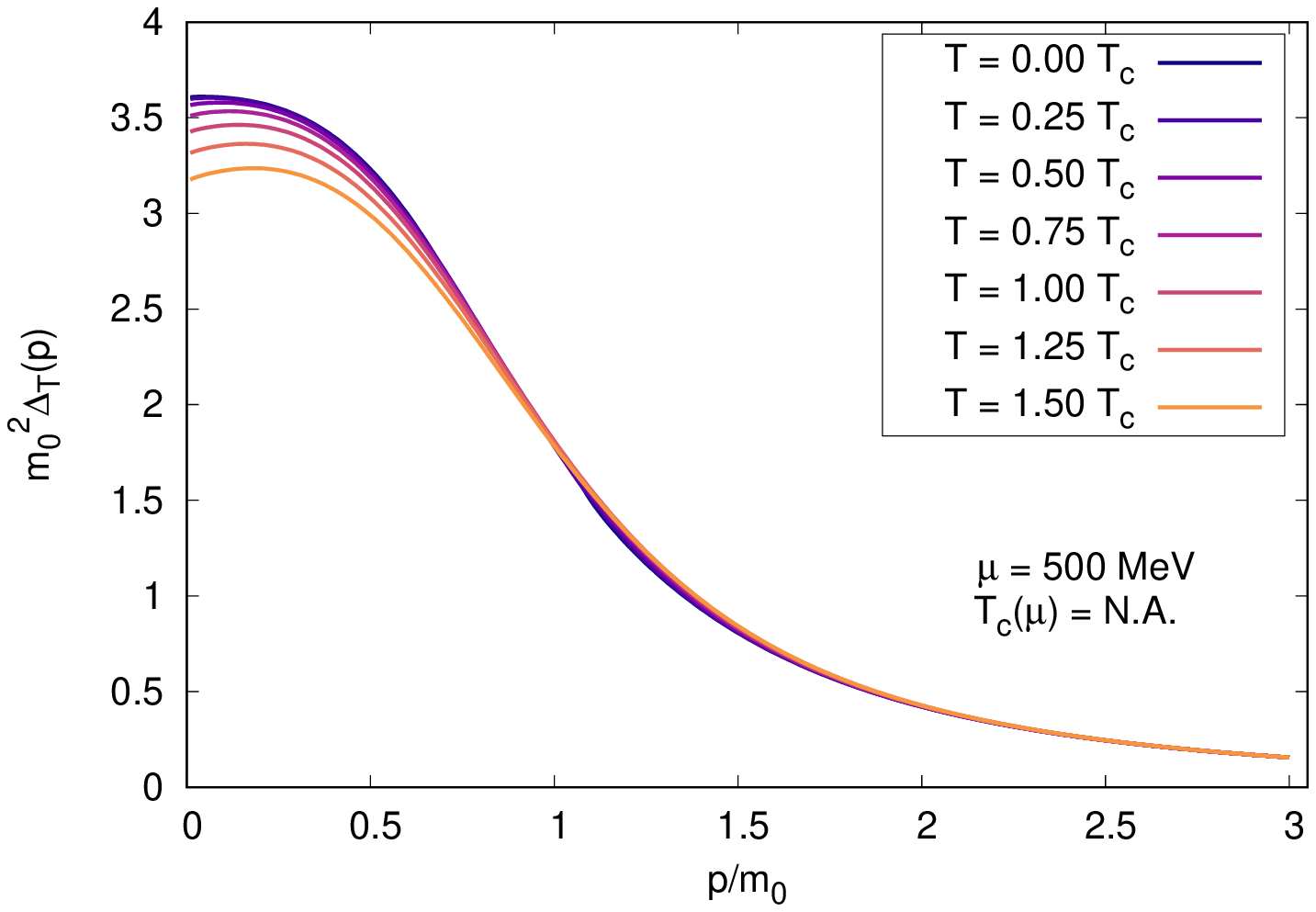}
    \caption{Transverse gluon propagator as a function of spatial momentum at finite $(T,\mu)$. Zero Matsubara frequency ($\omega_{n}=0$). $n_{f}=2+1$ with effective infrared quark masses $350,450$~MeV. Gluon mass parameter $m_{0}=656$~MeV. $T_{c}=T_{c}(\mu=0)$.}
    \label{qcdtransprop}
\end{figure*}

\begin{figure*}[t]
    \includegraphics[width=0.34\textwidth,angle=270]{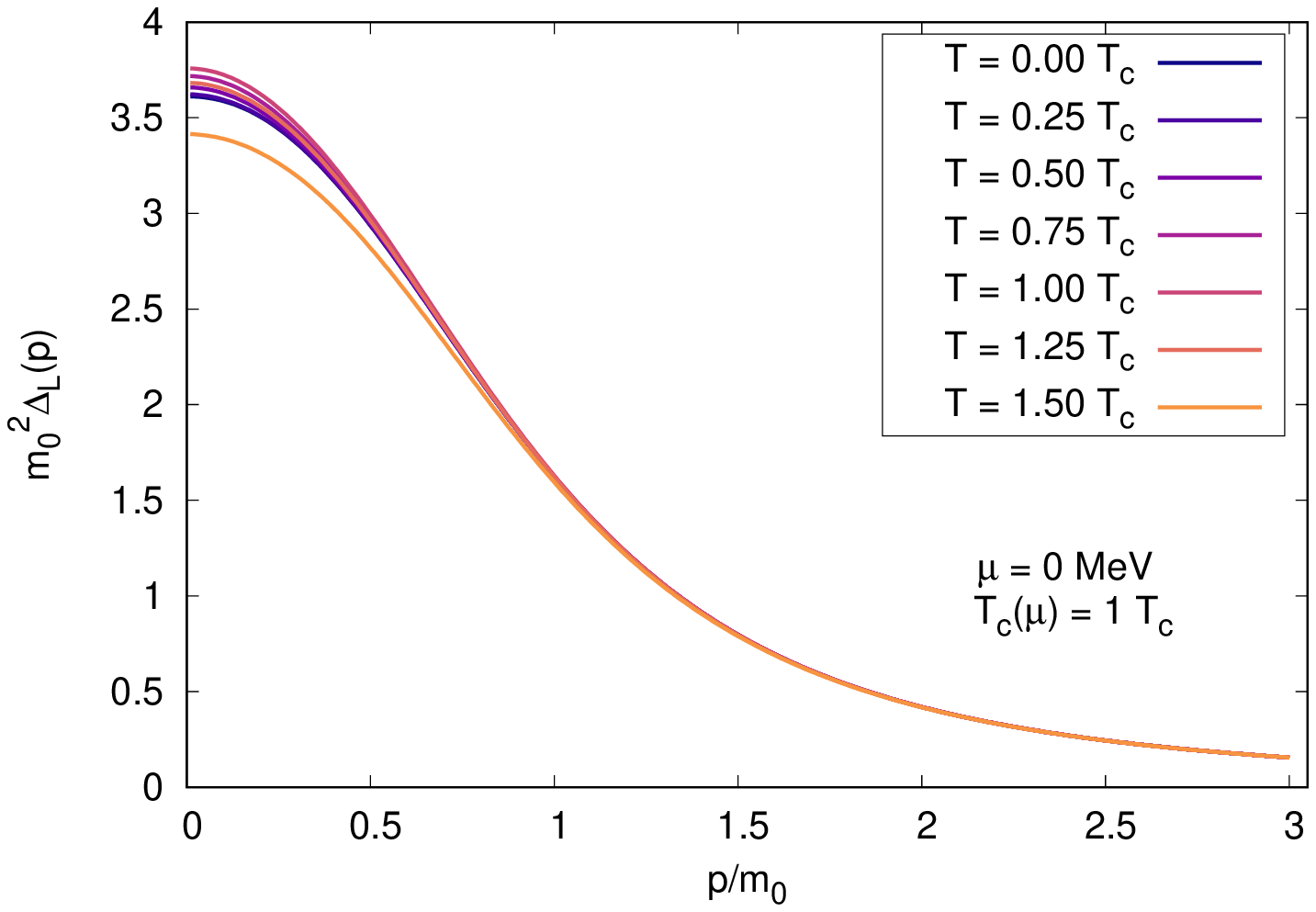}\hspace{5pt}%
    \includegraphics[width=0.34\textwidth,angle=270]{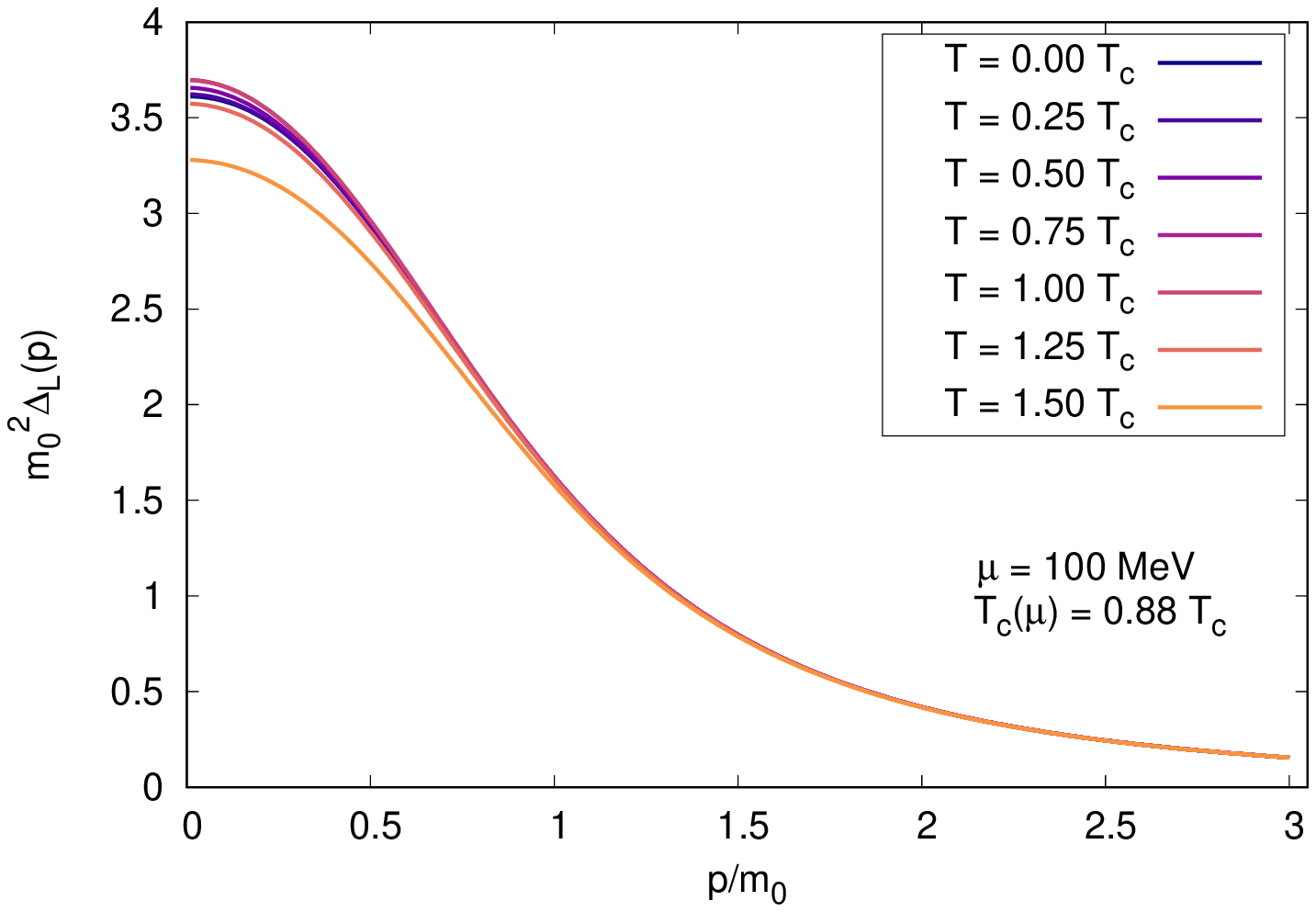}
    \includegraphics[width=0.34\textwidth,angle=270]{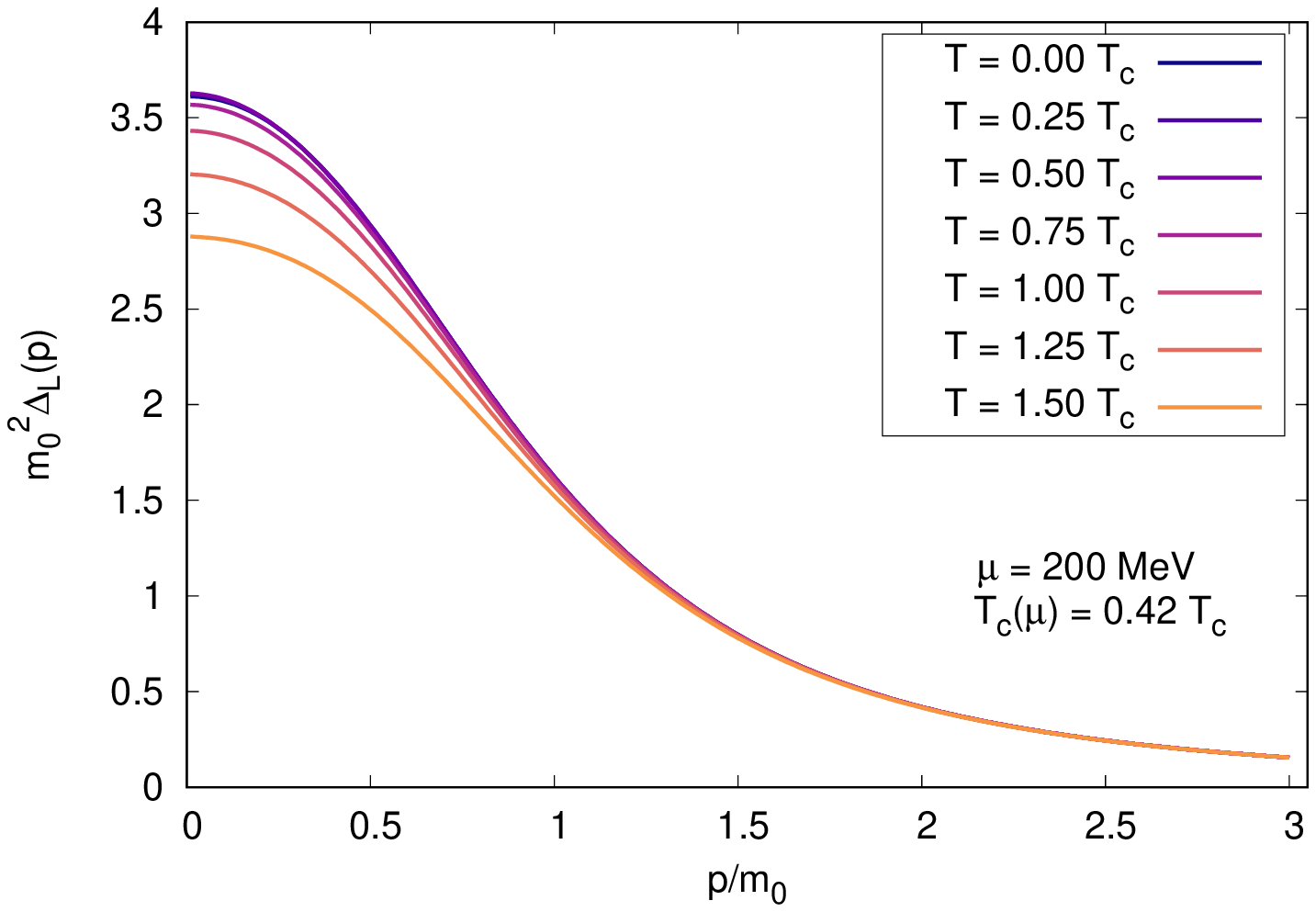}\hspace{5pt}%
    \includegraphics[width=0.34\textwidth,angle=270]{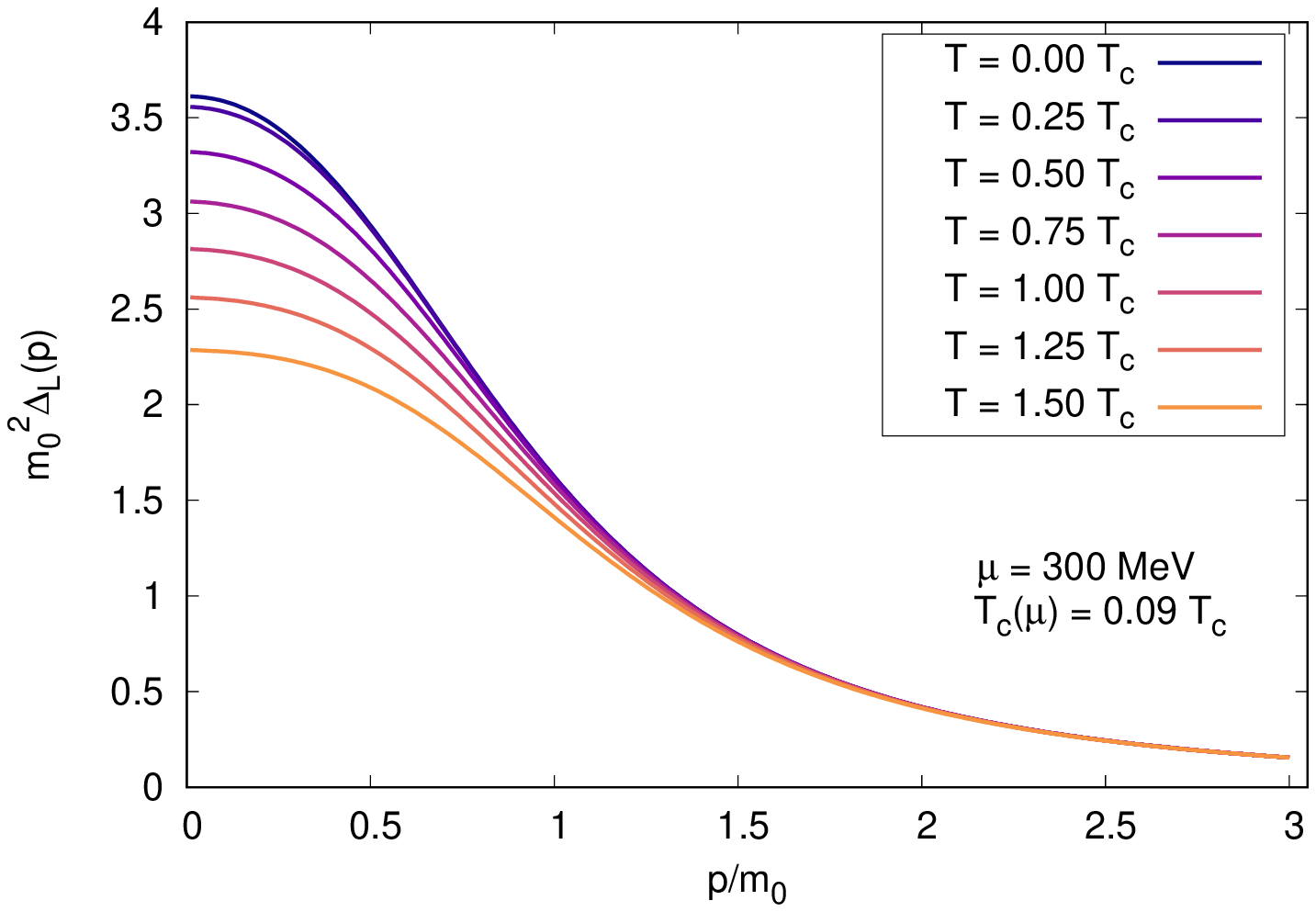}
    \includegraphics[width=0.34\textwidth,angle=270]{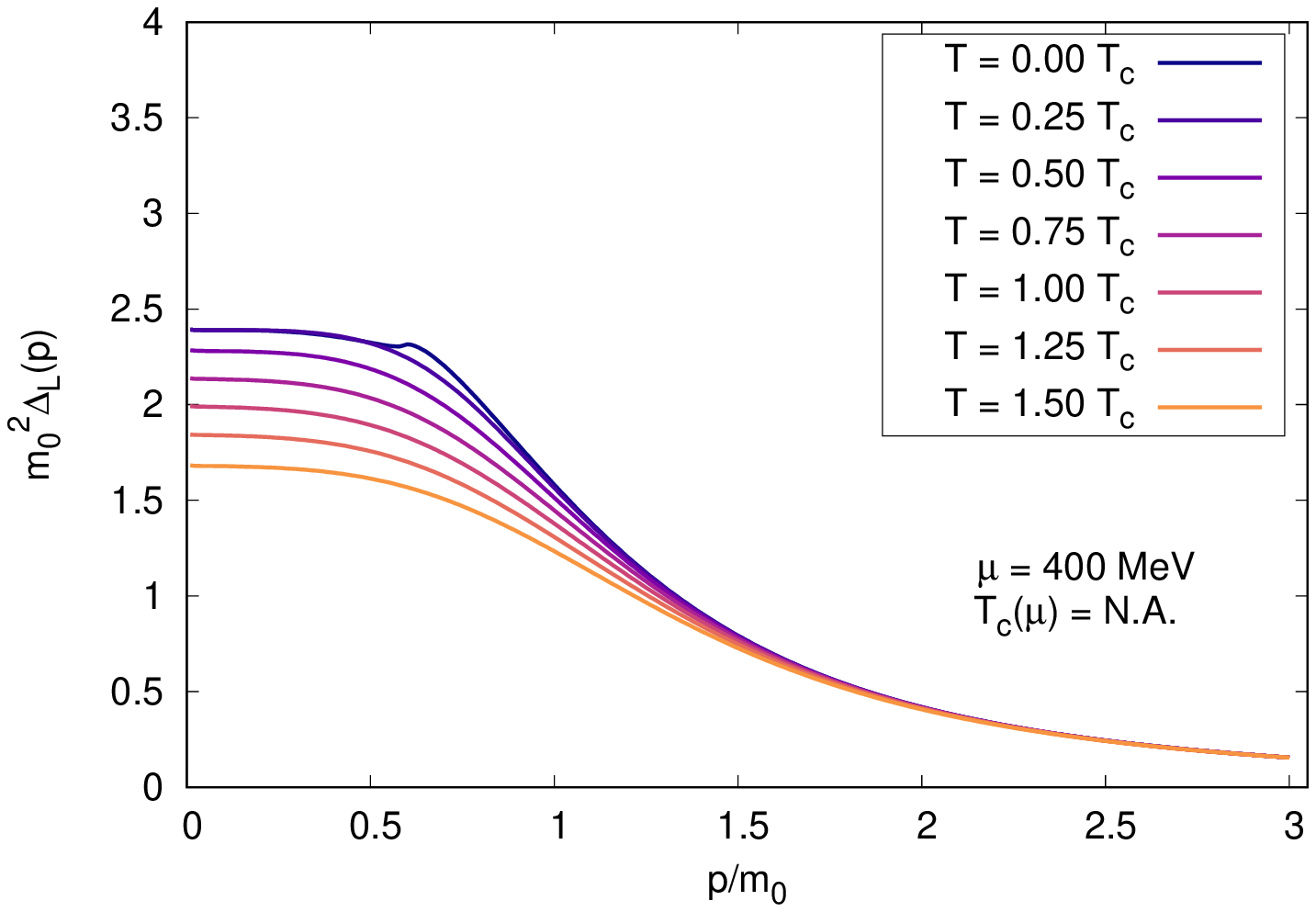}\hspace{5pt}%
    \includegraphics[width=0.34\textwidth,angle=270]{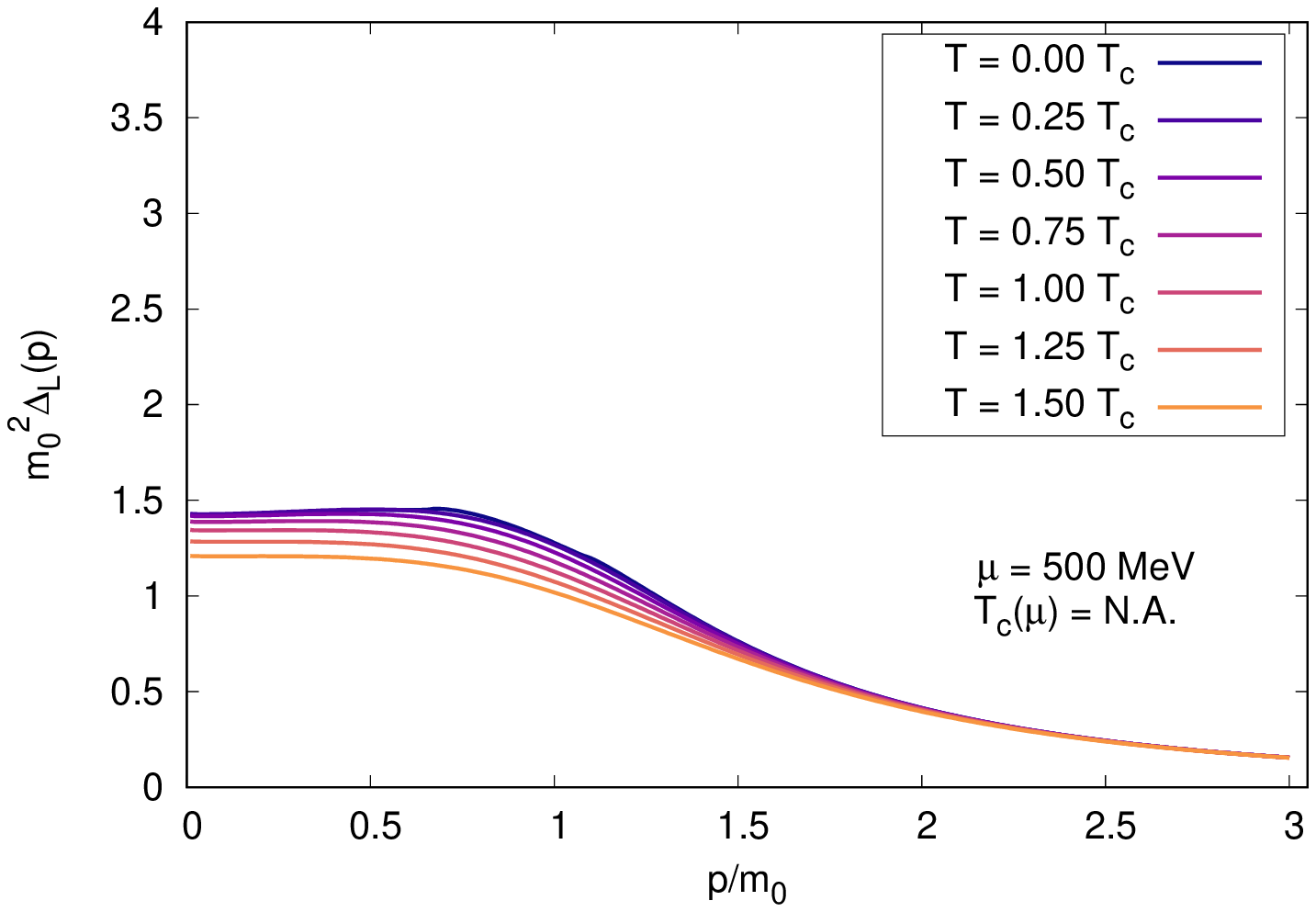}
    \caption{Longitudinal gluon propagator as a function of spatial momentum at finite $(T,\mu)$. Zero Matsubara frequency ($\omega_{n}=0$). Parameters and notation as in Fig.~\ref{qcdtransprop}.}
    \label{qcdlongprop}
\end{figure*}

Figs.~\ref{qcdtransprop} and \ref{qcdlongprop} display the transverse and the longitudinal component of the gluon propagator computed within the screened massive expansion of full QCD at zero Matsubara frequency ($\omega_{n}=0$), as functions of spatial momentum and at finite temperature and chemical potential, for the matter content and parameters described in the previous paragraph. For chemical potentials $\mu < M_{1} < M_{2}$, the transverse propagator is a decreasing function of both momentum and temperature, and its dependence on chemical potential at fixed $(T,|{\bf p}|)$ is practically negligible. While obvious for very small temperatures being a manifestation of the silver blaze property\footnote{At very small temperatures the Fermi distributions in Eqs.~\eqref{quarktranspol}-\eqref{quarklongpol} essentially reduce to a $\Theta\left(|\mu|-\sqrt{q^{2}+M_{f}^{2}}\right)$ step function: the (non-vacuum) quark contribution to the gluon polarization and any dependence on chemical potential that comes with it are strongly suppressed for $T\approx 0$ and $|\mu|<M_{1}<M_{2}$.}, this kind of dependence on $\mu$ is not trivial at larger $T$'s, as we will see shortly when we will describe the longitudinal propagator. For $\mu > M_{1}$ the same behavior with respect to $(T,|{\bf p}|)$ is also observed at low momenta, whereas at intermediate momenta -- below the renormalization scale of 4~GeV, corresponding to $|{\bf p}|/m_{0}\approx6.1$ -- we see the low-temperature curves slightly dropping below the high-temperature ones. This effect is somewhat enhanced for $\mu > M_{2}$ (bottom-right plot of Fig.~\ref{qcdtransprop}). The dependence on chemical potential becomes more prominent for $\mu > M_{1}$ and, although it remains quite unimportant overall while $\mu<m$, it is already clear from the bottom-right plot in Fig.~\ref{qcdtransprop} ($\mu=500$~MeV) that the transverse propagators tend to expand toward larger values of the momentum as $\mu$ increases toward and beyond $m$. This observation will be confirmed in Sec.~IIID.

For chemical potentials $\mu < M_{1}$, just like in pure Yang-Mills theory, the longitudinal gluon propagator -- Fig.~\ref{qcdlongprop} -- displays a distinct non-monotonic behavior with respect to temperature: at every fixed momentum, the propagator first increases with $T$, then drops to lower values. At $\mu=0$, the temperature at which the propagator attains its maximum is $T_{\text{max}}(\mu=0)\approx 0.117\, m_{0}\approx 77$~MeV, marked with $T_{c}=T_{c}(0)$ in the figure. As soon as $\mu$ crosses the $M_{1}$ threshold, the longitudinal propagator becomes a decreasing function of temperature for all but the smallest momenta. In the low-momentum region, the approximation we are employing -- namely, that of quark masses which remain constant at all temperatures and chemical potentials -- predicts a small suppression of the lowest-temperature ($T\approx 0$) curves below those with slightly larger $T$'s. As it is clear upon comparing the $\mu=400$~MeV ($M_{1}<\mu<M_{2}$) and $\mu=500$~MeV ($M_{1}<M_{2}<\mu$) curves, this is due to $T\approx0$ near-discontinuities in the longitudinal propagator with respect to momentum, one of which exists at fixed $\mu$ for each quark becoming active (i.e. whose mass threshold is being crossed by the chemical potential). In Sec.~IV we will argue that this feature may be an artifact of the approximation, and present results that do not display such a behavior.

At variance with the transverse one, the longitudinal propagator has a large dependence on chemical potential already for $\mu<m$: if for $\mu<M_{1}$ most of the dependence translates to the already noted behavioral change from non-monotonic to decreasing with temperature, for $\mu>M_{1}$ increasing the chemical potential strongly suppresses the propagator. We have checked that this trend continues up to larger chemical potentials than displayed in Fig.~\ref{qcdlongprop}. Overall, the behavior of both the propagators at $T=0$ and finite $\mu$ agrees with the one reported in \cite{BBNR20,BR21,BNRT21} for two-color QCD and QCD at finite isospin density, at least for values of chemical potential much smaller than the inverse lattice spacing, where the lattice data is more reliable.

\subsection{The one-loop gluon propagator: stability}

In the previous section we studied the zero-frequency full QCD gluon propagator as a function of momentum, temperature and chemical potential by fixing the values of the free parameters $m$ and $\pi_{0}$ from pre-existing zero-temperature Yang-Mills results, and $M_{1}$ and $M_{2}$ (the lighter and heavier quark masses) on the basis of our expectations for the effective masses with which the $n_{f}=2+1$ light quarks propagate in the infrared -- with a somewhat exaggerated $\Delta M=M_{2}-M_{1}$ in order to distinguish the effects of the two on the propagator. Since ultimately, in the absence of a comparison with the phenomenology or with the lattice, these choices are arbitrary, it is essential to determine if and how the qualitative behavior of the propagators we described in Sec.~IIIC depends on them.

\begin{figure*}[t]
    \includegraphics[width=0.34\textwidth,angle=270]{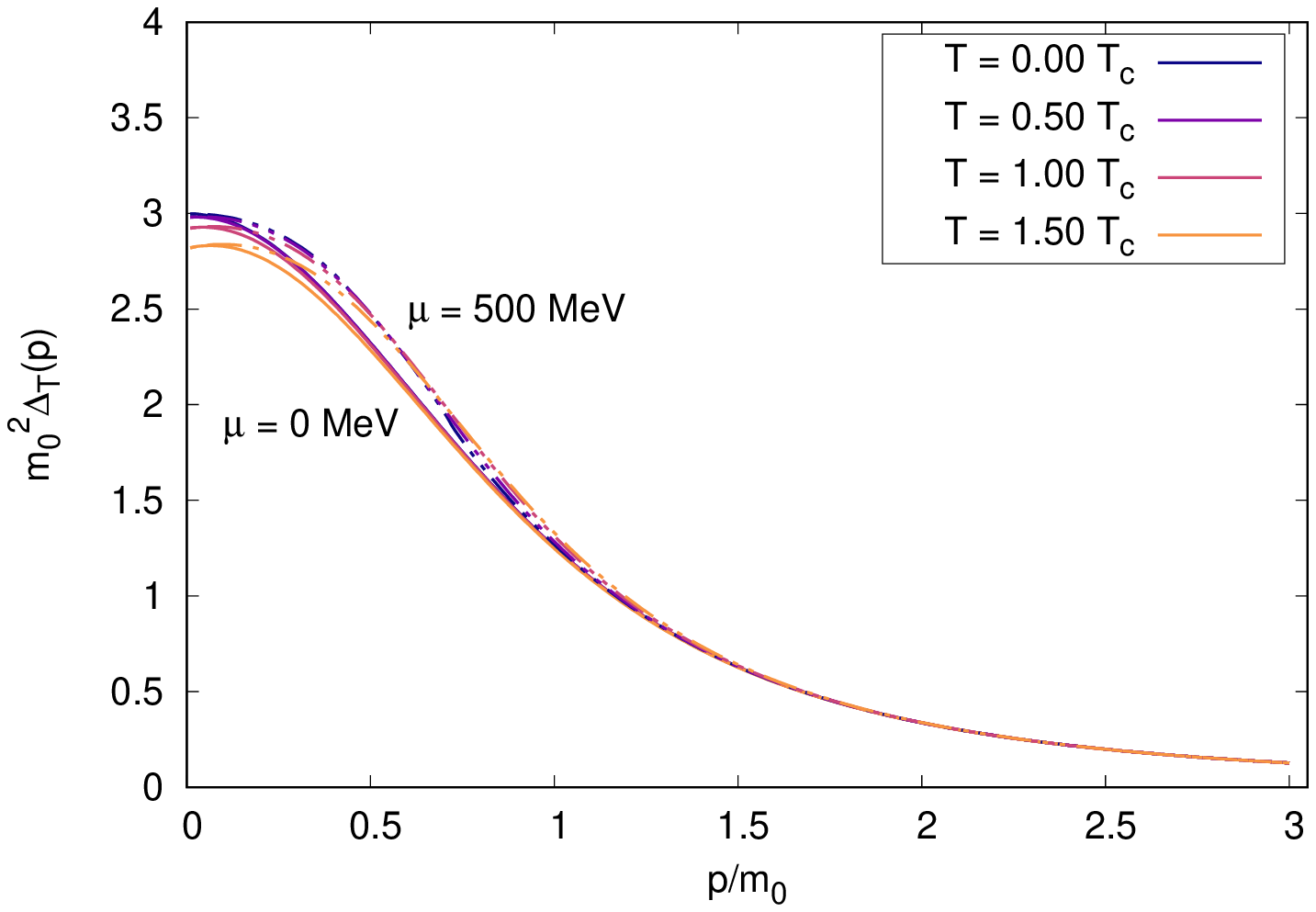}\hspace{5pt}%
    \includegraphics[width=0.34\textwidth,angle=270]{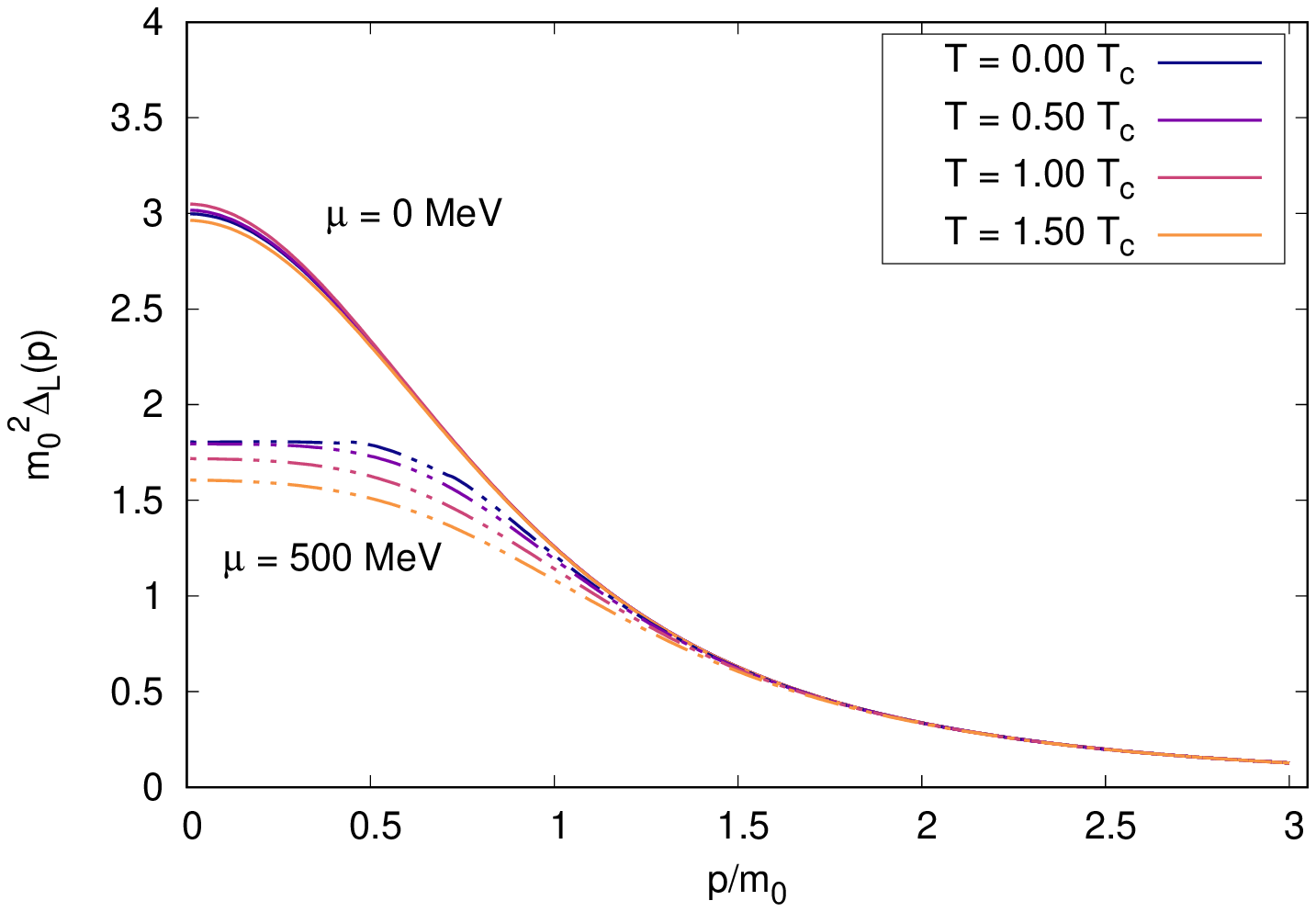}
    \includegraphics[width=0.34\textwidth,angle=270]{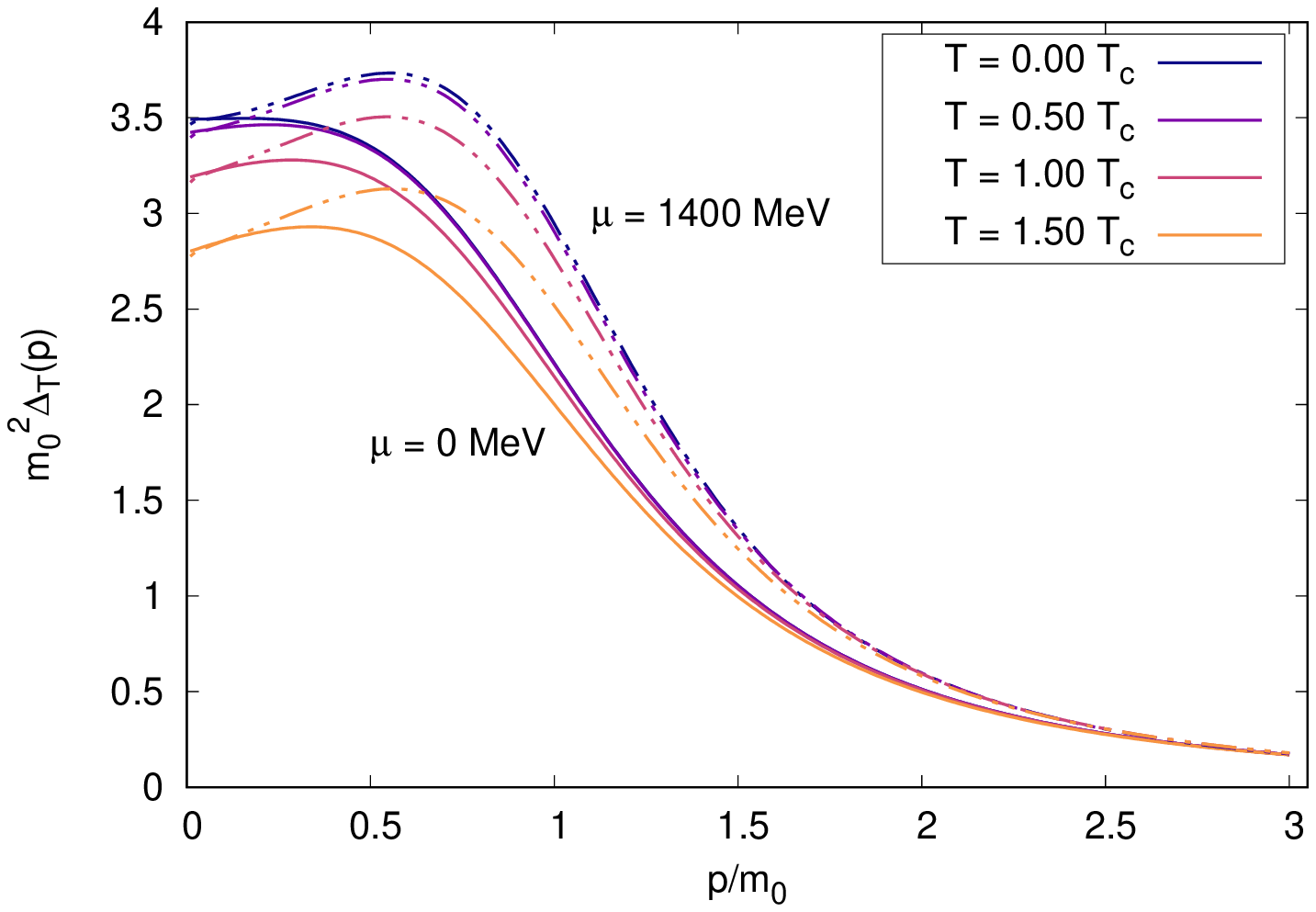}\hspace{5pt}%
    \includegraphics[width=0.34\textwidth,angle=270]{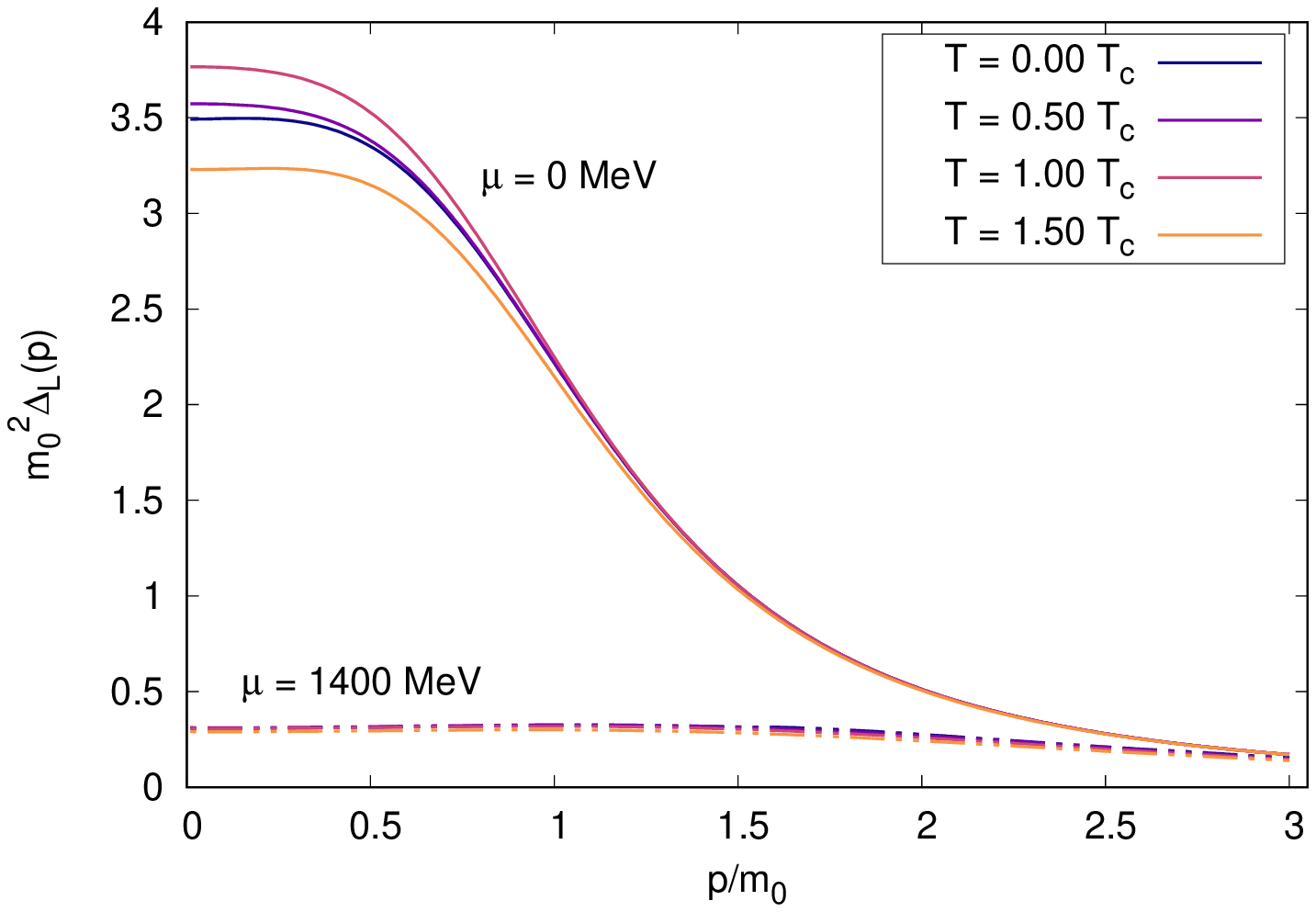}
    \caption{Transverse (left) and longitudinal (right) gluon propagator as a function of spatial momentum at finite $(T,\mu)$. Zero Matsubara frequency ($\omega_{n}=0$). $n_{f}=2+1$ with effective infrared quark masses $350,450$~MeV (top) and $1100,1300$~MeV (bottom). Gluon mass parameter $m_{0}=1000$~MeV (top) and $m_{0}=656$~MeV (bottom).}
    \label{qcdpropstabilitymasses}
\end{figure*}

Let us start from the dependence of $\Delta_{T,L}$ on the gluon and quark masses. In Fig.~\ref{qcdpropstabilitymasses} we display the same plots as Figs.~\ref{qcdtransprop} and \ref{qcdlongprop} for fewer values of the chemical potential -- namely, $\mu=0$ (solid line) and one value of $\mu > M_{2} > M_{1}$ (broken line) -- and of the temperature. The plots on the top row were obtained for $(M_{1},M_{2})=(350,450)$~MeV and $m=1000$~MeV, whereas those at the bottom were obtained for $(M_{1},M_{2})=(1100,1300)$~MeV and $m=656$~MeV. As we can see, from a qualitative perspective there is no meaningful difference between the mass configurations in Fig.~\ref{qcdpropstabilitymasses} and that of Sec.~IIIC\footnote{Although for $\mu > M_{2}$ the transverse propagator on the bottom left seems to lack any intermediate-momentum curve crossing, we have determined that the corresponding configuration does in fact display such a crossing, only at somewhat larger momenta and on a smaller scale that makes it harder to see in the figure.}. The evolution from lower to higher chemical potentials, not displayed in the figure, follows the pattern we have discussed in the previous section. The transverse propagator is only weakly dependent on $\mu$ for $\mu < M_{1}$ but acquires a stronger dependence for $\mu > M_{1}$. The longitudinal propagator has a non-monotonic behavior with temperature for $\mu < M_{1}$ and a decreasing one for $\mu > M_{1}$ -- with the exception of small momenta, due to $T\approx 0$ near-discontinuities. Quantitatively, for $\mu > M_{1}$, heavier quark masses correspond to a stronger suppression of the longitudinal propagators, and to transverse propagators which tend to acquire a maximum that expands to higher momenta as $\mu$ is increased. This last feature was only barely visible in the $\mu=400,500$~MeV plots presented in Sec.~IIIC, and we confirm it here.

Two other mass configurations were tested, $(M_{1}, M_{2})=(350,450)$~MeV, $m=200$~MeV and $(M_{1}, M_{2})=(150,225)$~MeV, $m=656$~MeV. Depending on whether $m<M_{1},M_{2}$ or $m>M_{1},M_{2}$, we obtained results which are entirely redundant with those displayed in Fig.~\ref{qcdpropstabilitymasses}, so we will not report them explicitly.

As for the dependence of $\Delta_{T,L}$ on the constant $\pi_{0}$ -- see Eq.~\eqref{qcdglupropadim} --, we start by observing that, since the renormalization scale $\mu_{0}=4$~GeV in Eq.~\eqref{rencond} is much larger than the temperatures and chemical potentials that interest us here, the multiplicative factors $z_{\pi}$, which are determined by the normalization condition $[p^{2}\Delta_{T,L}(\omega,{\bf p})]_{p^{2}=\mu^{2}_{0}}=1$, are independent of $(T,\mu)$ and of the considered projection to a very good approximation\footnote{In Sec.~IIIC, $z_{\pi}$ varied from $2.256$ to $2.267$ -- that is, to less than $0.5\%$ -- across all temperatures, chemical potentials and both the components.}. If we take $\pi_{0}$ to be a constant with respect to temperature like we did for the entirety of Sec.~III, we then find that
\begin{equation}
    \frac{\partial\Delta_{T,L}}{\partial T}\propto \frac{\partial}{\partial T}\left([\pi_{T,L}(p)]_{\text{Th}}+\sum_{f}\ [\pi_{T,L}^{(f)}(p)]_{\text{M}}\right)\ ,
\end{equation}
implying that the temperature at which $\Delta_{L}$ attains its maximum at fixed momentum and chemical potential, if it exists, is practically insensitive to a change in $\pi_{0}$. Moreover, since in the denominator of Eq.~\eqref{qcdglupropadim} $p^{2}\pi_{0}$ vanishes for $\omega=0$ as $|{\bf p}|\to 0$, whereas -- again due to the renormalization conditions -- $\pi_{0}$ enters $z_{\pi}$ additively, in the limit of vanishing momentum the propagators depend on $\pi_{0}$ as
\begin{equation}
    \lim_{|{\bf p}|\to0}\ \Delta_{T,L}(\omega=0, {\bf p})=\frac{\pi_{0}+\kappa}{Q^{2}}\ ,
\end{equation}
where $\kappa$ and $Q$ are constants of mass dimension $0$ and $1$ respectively. In particular, in the limit of vanishing momentum, the $\omega=0$ propagators increase with $\pi_{0}$.

\begin{figure*}[t]
    \includegraphics[width=0.34\textwidth,angle=270]{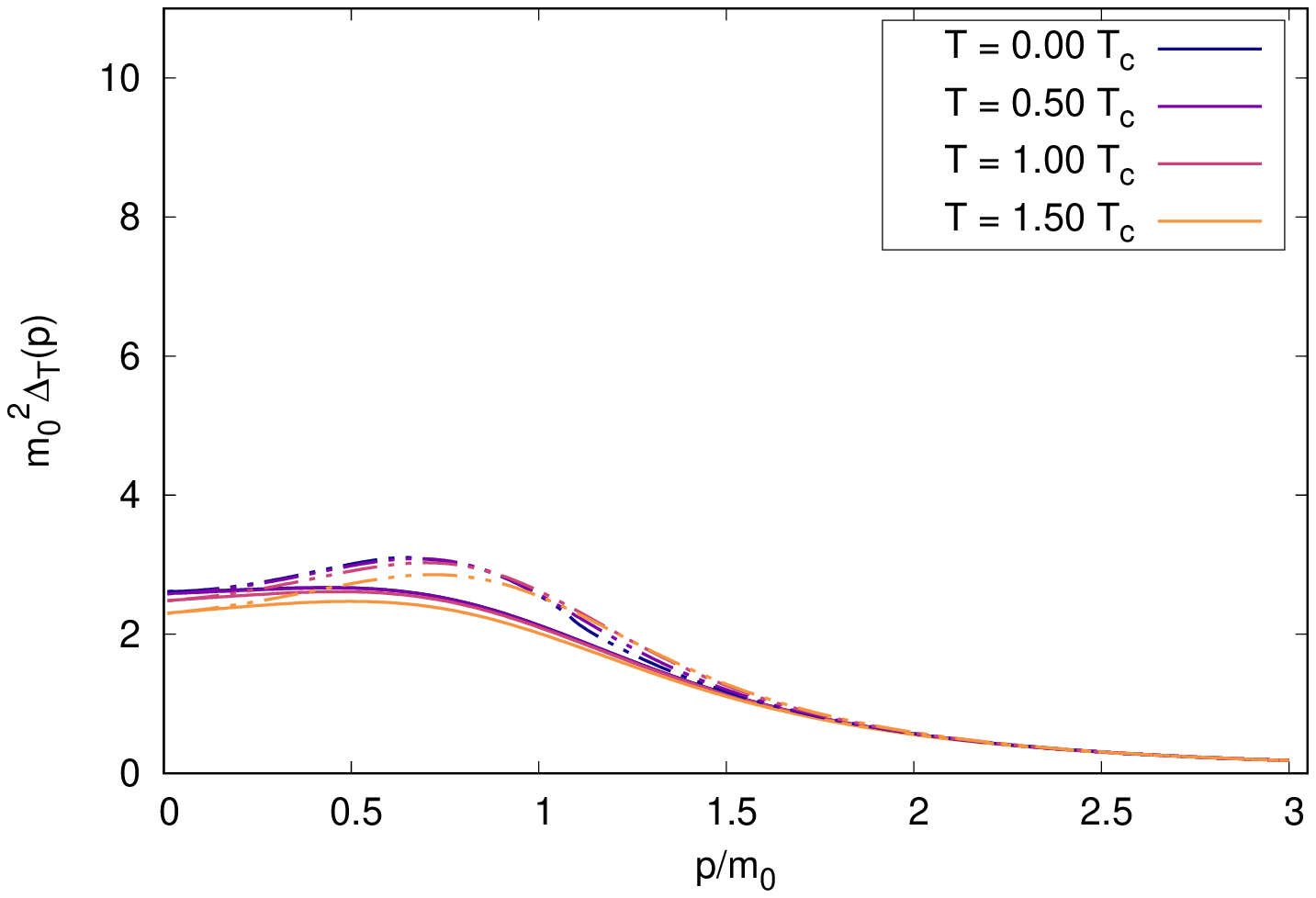}\hspace{5pt}%
    \includegraphics[width=0.34\textwidth,angle=270]{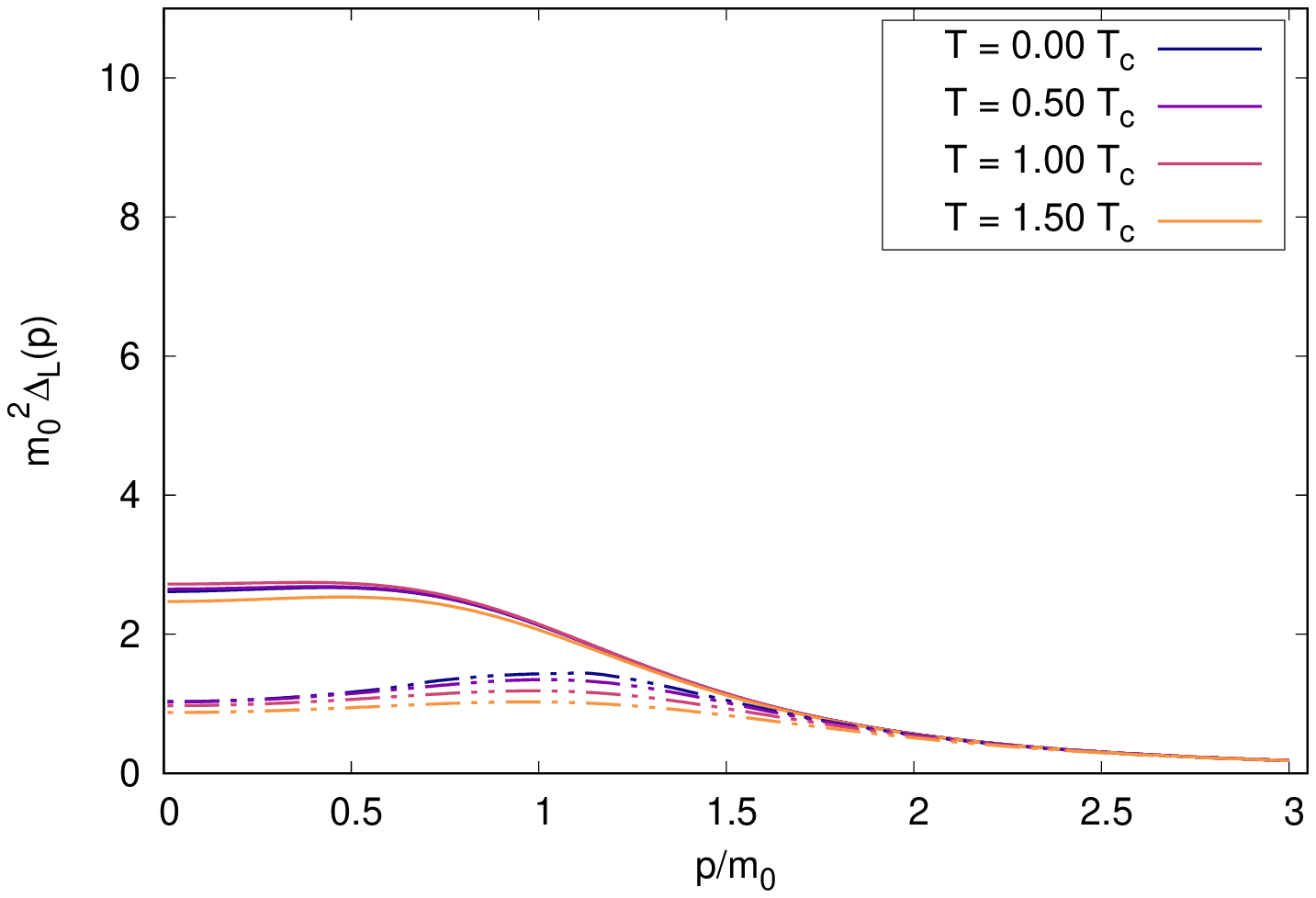}
    \includegraphics[width=0.34\textwidth,angle=270]{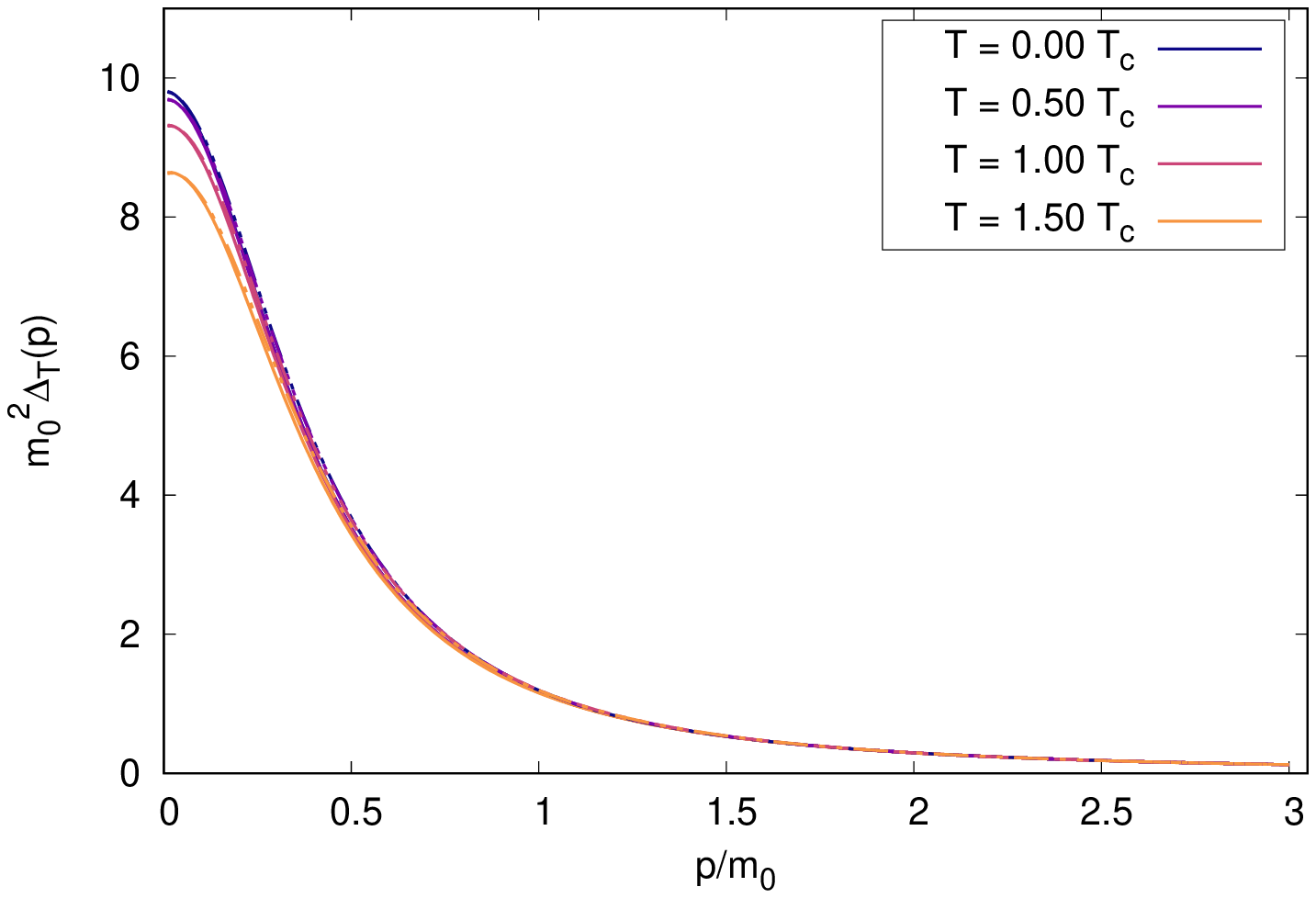}\hspace{5pt}%
    \includegraphics[width=0.34\textwidth,angle=270]{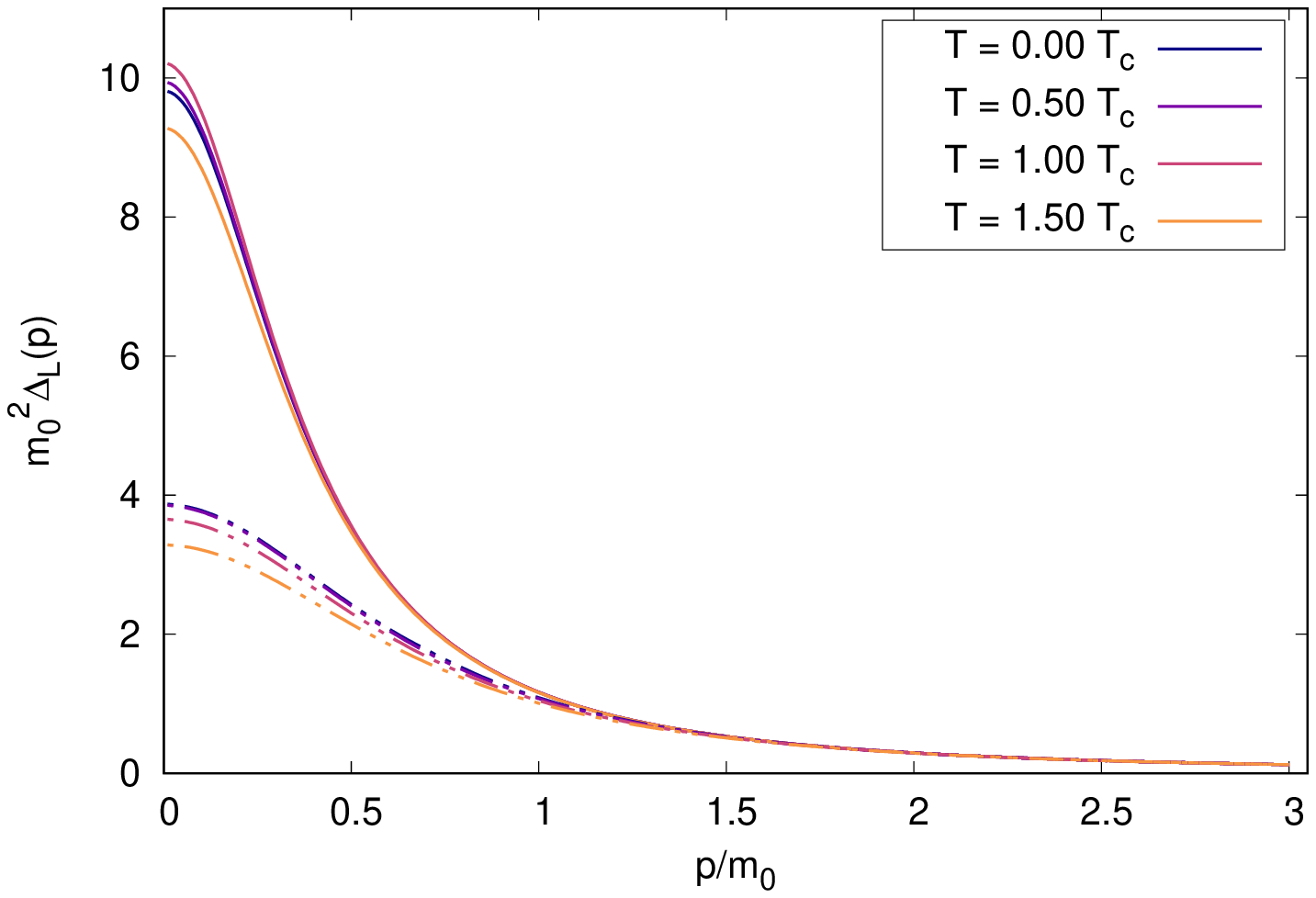}
    \caption{Transverse (left) and longitudinal (right) gluon propagator as a function of spatial momentum at finite $(T,\mu)$. Solid curves: $\mu=0$. Broken curves: $\mu=500$~MeV. Zero Matsubara frequency ($\omega_{n}=0$). $n_{f}=2+1$ with effective infrared quark masses $350,450$~MeV. Gluon mass parameter $m_{0}=656$~MeV. Top: $\pi_{0}=-1.5$. Bottom: $\pi_{0}=3.0$.}
    \label{qcdpropstabilityf0}
\end{figure*}

This analysis is confirmed by Fig.~\ref{qcdpropstabilityf0}, which displays the propagators computed for $\pi_{0}=-1.5$ (top) and $\pi_{0}=3.0$ (bottom) using the same masses as Sec.~IIIC. The figure, in which as before $T_{c}\approx 0.117\, m_{0}\approx 77$~MeV, shows that the qualitative behavior of the propagators with respect to changes in temperature and chemical potential is also independent of $\pi_{0}$. The only notable difference with our previous determinations is the fact that, as $\pi_{0}$ decreases to larger negative values, the propagators develop a maximum as functions of momentum. This happens because a large negative $\pi_{0}$ in the denominator of Eq.~\eqref{qcdglupropadim} can become of the same magnitude of the (positive) adimensional polarization functions to which it is summed, to the point of making the denominator vanish for some value of momentum when too large in absolute value. Clearly, $\pi_{0}$ must be chosen greater than the threshold value -- dependent on the quark and gluon masses -- at which the Euclidean propagator develops a spurious pole.

To summarize, the overall qualitative features of the gluon propagator, as computed at finite temperature and density using the screened massive expansion of QCD and modeling the quark masses as described in Sec.~IIIA, do not depend on the free parameters of the expansion and are thus genuine predictions of the approach. In the next section we will show that the same holds for the phase diagram of QCD, as determined from the behavior of the longitudinal gluon propagator with respect to changes in temperature.
\newpage

\section{The phase diagram of QCD}

\subsection{Maxima of the longitudinal gluon propagator and the critical temperature}

\begin{figure}[h]
    \includegraphics[width=0.32\textwidth,angle=270]{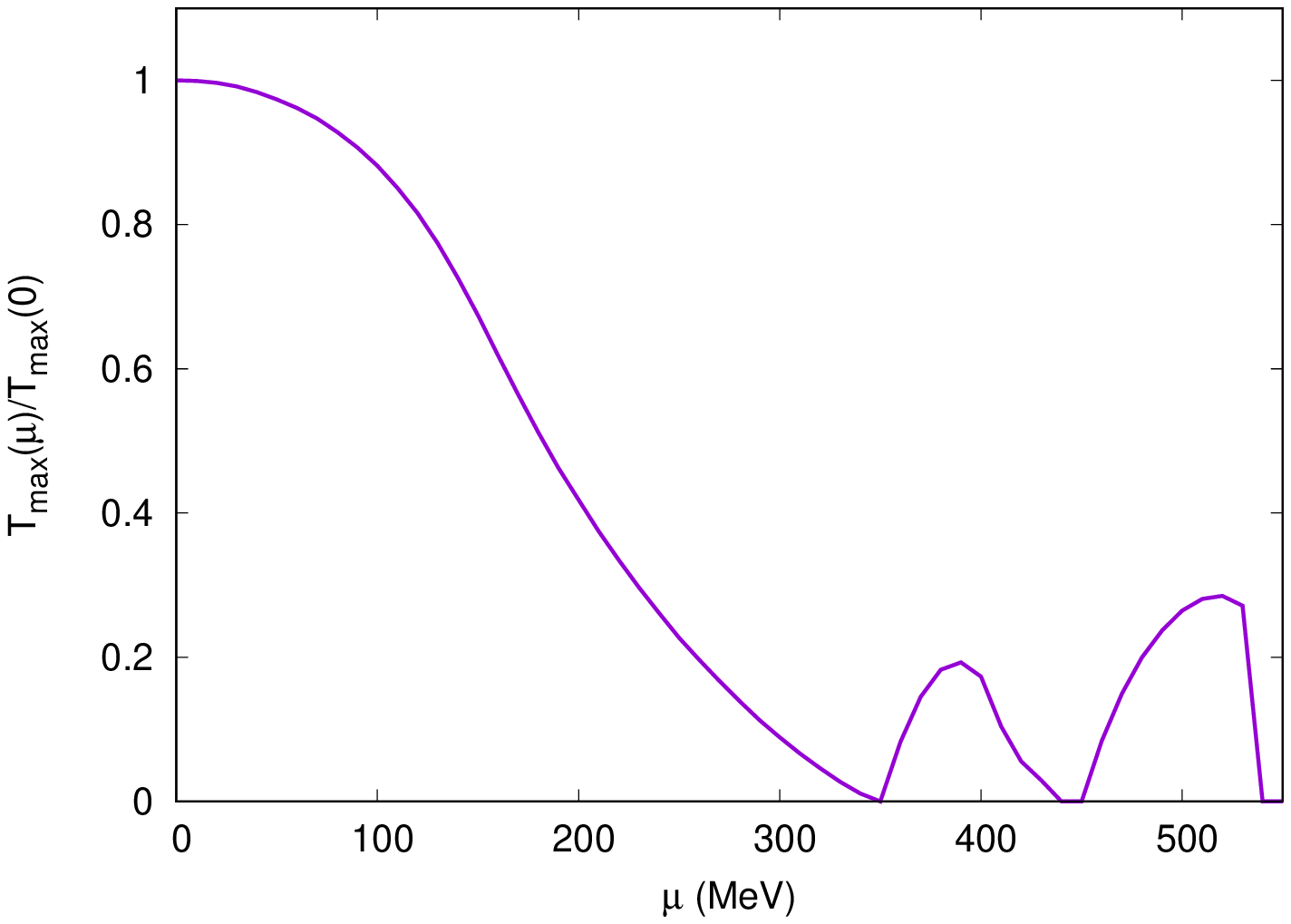}
    \caption{Temperature $T_{\text{max}}$ at which the zero-frequency ($\omega=0$) longitudinal gluon propagator $\Delta_{L}(\omega, {\bf p})$ attains a maximum in the $|{\bf p}|\to 0$ limit, as a function of chemical potential. $T_{\text{max}}(0)\approx0.117\,m_{0}\approx77$~MeV.}
    \label{qcdtmax}
\end{figure}

In Secs.~IIB, IIIC and IIID we saw that, both in pure Yang-Mills theory and in full QCD, the longitudinal component of the zero-frequency gluon propagator first increases with temperature for $T$'s smaller than a certain threshold value $T_{\text{max}}$, then decreases with it like the transverse one. In finite-density full QCD, at large enough chemical potentials and momenta, this threshold value is zero, meaning that the longitudinal propagator is a decreasing function of $T$. Thanks to quenched lattice calculations \cite{SOBC14}, we know that, in pure Yang-Mills theory, the temperature at which the longitudinal propagator attains its maximum is the critical temperature marking the transition from the confined to the deconfined phase. It is then natural to interpret an analogous change in full QCD as marking either a true phase transition, or a smooth crossover between phases. Motivated and guided by this interpretation, in this section we study the maxima of the longitudinal gluon propagator with respect to temperature and use them to gather information on the phase diagram of QCD.\\

In Fig.~\ref{qcdtmax} we display $T_{\text{max}}$ as a function of the chemical potential, as computed at $|{\bf p}| = 0$ using the mass configuration and parameters of Sec.~IIIC\footnote{Namely, $(M_{1},M_{2})=(350,450)$~MeV, $m=m_{0}=656$~MeV, $\pi_{0}=-0.876$.}. $T_{\text{max}}$ first decreases from $T_{\text{max}}(0)\approx0.117\,m_{0}\approx77$~MeV to zero as $\mu$ approaches $M_{1}=350$~MeV, then develops two humps that start at $M_{1}$ and $M_{2}$, and finally it decreases to zero again above $\mu\approx550$~MeV. The humps are the graphical expression of a feature that we have already highlighted in Sec.~IIIC (see Fig.~\ref{qcdlongprop}): over a finite interval of chemical potentials between $M_{1}$ and $M_{2}$, for small momenta including $|{\bf p}| = 0$, the propagators with $T\approx0$ are slightly suppressed with respect to those of slightly higher temperature. As a result, for these chemical potentials, $T_{\text{max}}$ is larger than zero when measured at $|{\bf p}| = 0$. The same happens at the $\mu=M_{2}$ threshold, causing the second hump to appear in Fig.~\ref{qcdtmax}.

The suppression observed at low momenta, $T\approx 0$ and $\mu\gtrapprox M_{1},M_{2}$ is due to the finite-density contribution of the quark loop to the gluon polarization. Going back to Eq.~\eqref{quarklongpol} and taking its $p\to 0$ limit first with respect to $\omega$ and then with respect to $|{\bf p}|$, we find that\\
\begin{align}
    &\lim_{|{\bf p}|\to0}\lim_{\omega\to0}\ [\Pi_{L}^{(f)}(\omega,{\bf p})]_{\text{M}}=\notag\\
    &=-\frac{1}{2\pi^{2}}\int_{0}^{+\infty}dq\ \frac{n_{F}(\varepsilon_{q};T,\mu)}{\varepsilon_{q}}\ (q^{2}+\varepsilon_{q}^{2})\ .
\end{align}
For $\mu$ just above the $M_{f}$ threshold and $T\approx 0$, the above expression increases with temperature; as $\mu$ is increased further, on the other hand, it becomes a strictly decreasing function of temperature. Since $[\Pi_{L}^{(f)}(\omega,{\bf p})]_{\text{M}}$ appears with a minus sign in the denominator of $\Delta_{L}(\omega,{\bf p})$, the same behavior is inherited at small spatial momenta by the zero-frequency gluon propagator\footnote{Note that the gluon and ghost thermal contributions to the gluon polarization are suppressed as $T\to 0$; in this limit, at finite chemical potential, the leading contribution to the gluon propagator is provided by the finite-density term of the quark loop.}.

If we assume the maxima of the longitudinal gluon propagator with respect to temperature to mark the transition between the confined and the deconfined phase, then -- at least for $\mu < M_{1}$ -- the $T_{\text{max}}(\mu)$ curve in Fig.~\ref{qcdtmax} can be identified with the critical temperature $T_{c}(\mu)$ of this transition. As for $\mu > M_{1}$, the interpretation of the regions below the two $T_{\text{max}}(\mu)$ humps in the figure is far from obvious. While our model does predict that the longitudinal propagator is not strictly decreasing with temperature at all momenta, at such large chemical potentials it is unclear whether the simple criterion of non-monotonicity versus strict decrease is sufficient to characterize the existence of a phase transition. Moreover, a few elements suggest that the prediction itself may actually be an artifact of our approximations: first of all, the suppression at low momenta is so small that it may disappear by going to higher orders in perturbation theory; second, it may disappear by giving an even small dependence on temperature and chemical potential to the free parameter $\pi_{0}$. Perhaps most importantly, as we noted above, the humps are caused by the chemical potential crossing a quark mass threshold at $T\approx 0$. It then follows that if the effective quark masses drop from $(M_{1},M_{2})$ to smaller values just above the $T_{\text{max}}(\mu)$ line, the suppression may disappear entirely. Of course, such a drop is expected in the deconfined phase due to the restoration of chiral symmetry. Here the effective quark masses are understood to originate from thermal and finite-density effects, rather than from condensates.

\begin{figure}[h]
    \includegraphics[width=0.32\textwidth,angle=270]{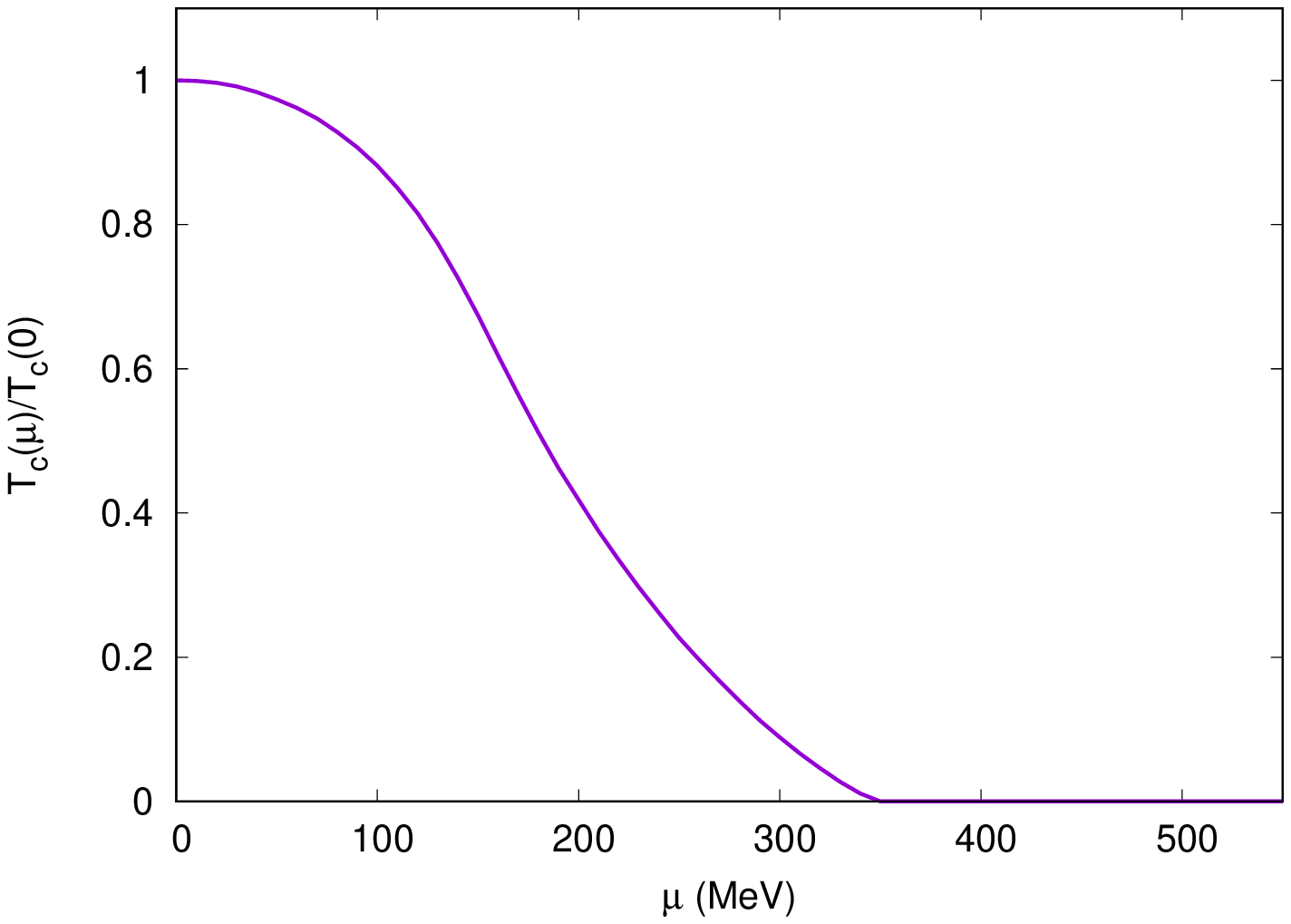}
    \caption{Phase diagram of QCD obtained by decreasing the quark masses from $(M_{1},M_{2})=(350, 450)$~MeV to $(M_{1}^{\prime},M_{2}^{\prime})=(125, 225)$~MeV beyond the first portion of the $T_{\text{max}}(\mu)$ curve. See the text for details.}
    \label{qcdphasediag}
\end{figure}

Within our approach, we calculated that if the quark masses were to abruptly decrease to less than about $250$-$280$~MeV beyond the first segment of the $T_{\text{max}}(\mu)$ curve -- that is, for $T > T_{\text{max}}(\mu)$ at $\mu < M_{1}$, and for all $T$ at $\mu > M_{1}$ --, then the two humps in Fig.~\ref{qcdtmax} would in fact disappear. This is shown in Fig.~\ref{qcdphasediag}, where the recomputed maxima are plotted as a function of chemical potential and marked as the $\mu$-dependent critical temperature of the system $T_{c}(\mu)$. In more detail, to obtain the curve in the figure, we first used our knowledge on $T_{\text{max}}(\mu)$ -- as evaluated under the assumption of constant quark masses over the whole $(T,\mu)$ plane -- to identify the region of the phase diagram which we expect to be the confined phase. Then we decreased the quark masses from $(M_{1},M_{2})=(350, 450)$~MeV to about half their value -- $(M_{1}^{\prime},M_{2}^{\prime})=(125, 225)$~MeV -- outside of that region and recomputed the propagators and their maxima as a function of chemical potential. As we will show in the next section, we found that for $\mu < M_{1}$ the maxima remain unchanged. On the other hand, for $\mu > M_{1}$, since at that point the quark masses $M_{1}^{\prime}, M_{2}^{\prime} < M_{1} < \mu$, the chemical potential does not cross mass thresholds anymore, and the propagator is no longer suppressed at $|{\bf p}|=0$ and $T\approx 0$: the temperature at which the longitudinal propagator attains a maximum becomes zero (meaning that the curve is strictly decreasing) over the whole momentum range, just like in the deconfined phase.

Since our model does not predict the values of the effective quark masses, we cannot determine which of the two behaviors -- low-momentum slight suppression for $\mu$ crossing the mass thresholds at $T\approx0$, or strictly decreasing propagators -- may be closer to the exact result. We note that there is also a third possibility: as suggested by \cite{RWB05,BFG05}, the lightest quark mass may drop substantially, while the heavier one may not do so enough to make the second hump in Fig.~\ref{qcdtmax} disappear, leading to suppression at one single mass threshold -- $\mu\gtrapprox M_{2}^{\prime}$ -- instead of two. To gain some insight into these matters, it would be interesting to extend the present study to the quark sector in such a way that the infared quark effective masses are determined dynamically from the interactions, instead of being free parameters as we are using them here. In any case, as stated before, it is unclear whether these issues ultimately have any influence on the phase diagram of QCD.\\

The non-monotonicity criterion is able to provide us with unambiguous information on the topology of the QCD phase diagram for $\mu < M_{1}$. The shape of the phase boundary -- whether marking a sharp transition or a smooth crossover between separate phases -- is the one expected from numerous studies carried out by employing diverse techniques \cite{GUE21}. Because of the qualitative nature of our results on the gluon propagator -- due to having used a one-loop approximation in which the free parameters do not depend on temperature and chemical potential -- we do not expect the actual values of the curve $T_{c}(\mu)$ to constitute a reliable physical prediction. This is already clear from our computed $T_{c}(0)\approx 77$~MeV, which is around twice as small as the current determinations, that set it at around $155$~MeV when measured from chiral and quark density susceptibilities or at around $175$~MeV when measured from the Polyakov loop \cite{Aoki2006a,Aoki2006b,Steinbrecher2019,GUE21}. On the other hand, this result is not dissimilar to that obtained in pure Yang-Mills theory -- see Sec.~II, where the predicted $T_{c}\approx 121$~MeV is also around twice as small as the expected $270$~MeV when computed using temperature-independent parameters. Indeed, the present approach provides a good estimate of the ratio between pure Yang-Mills' and full QCD's critical temperature at $\mu=0$ -- namely, $T_{c}^{(\text{YM})}\approx1.74\,T_{c}^{(\text{QCD})}(0)$ (susceptibilities) or $T_{c}^{(\text{YM})}\approx1.54T_{c}^{(\text{QCD})}(0)$ (Polyakov loop), against our $T_{c}^{(\text{YM})}\approx1.57\,T_{c}^{(\text{QCD})}(0)$. That our estimate is closer to the latter can be easily understood by observing that the Polyakov loop is both a gluonic observable and the actual order parameter of the deconfinement transition in pure Yang-Mills theory.

\subsection{Quark mass effects on the gluon propagator and stability of the phase diagram}

In the last section we stated that decreasing the quark masses from $(M_{1},M_{2})=(350, 450)$~MeV to $(M_{1}^{\prime},M_{2}^{\prime})=(125, 225)$~MeV beyond the first segment of the $T_{\text{max}}(\mu)$ curve computed at constant quark masses yields a longitudinal gluon propagator which is strictly decreasing with temperature for $\mu > M_{1}$ at all spatial momenta. In this section we present such results and compare the phase diagram obtained with the parameters of Secs.~IIIC and IVA with those obtained with the parameters of Sec.~IIID.\\

\begin{figure*}[t]
    \includegraphics[width=0.34\textwidth,angle=270]{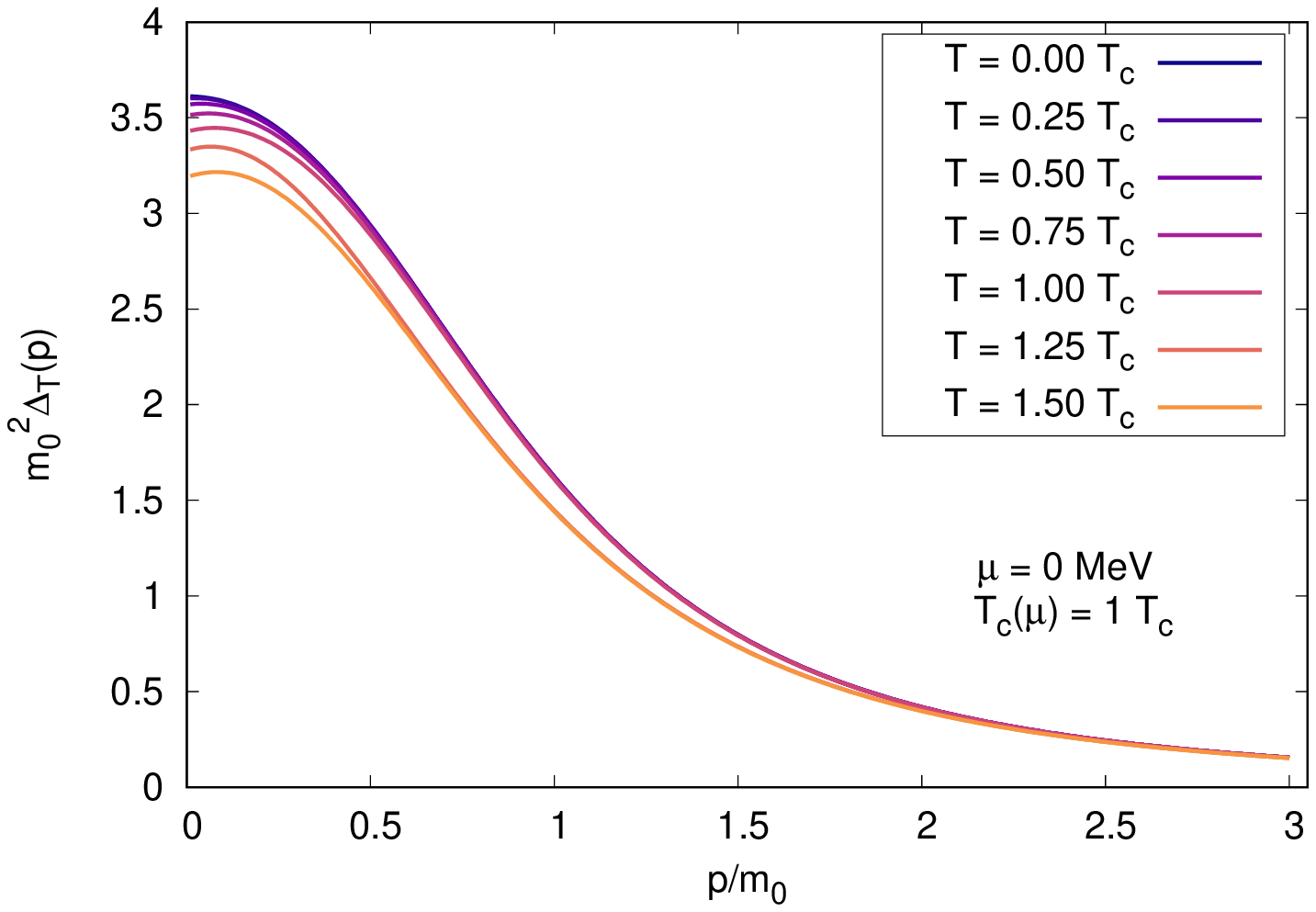}\hspace{5pt}%
    \includegraphics[width=0.34\textwidth,angle=270]{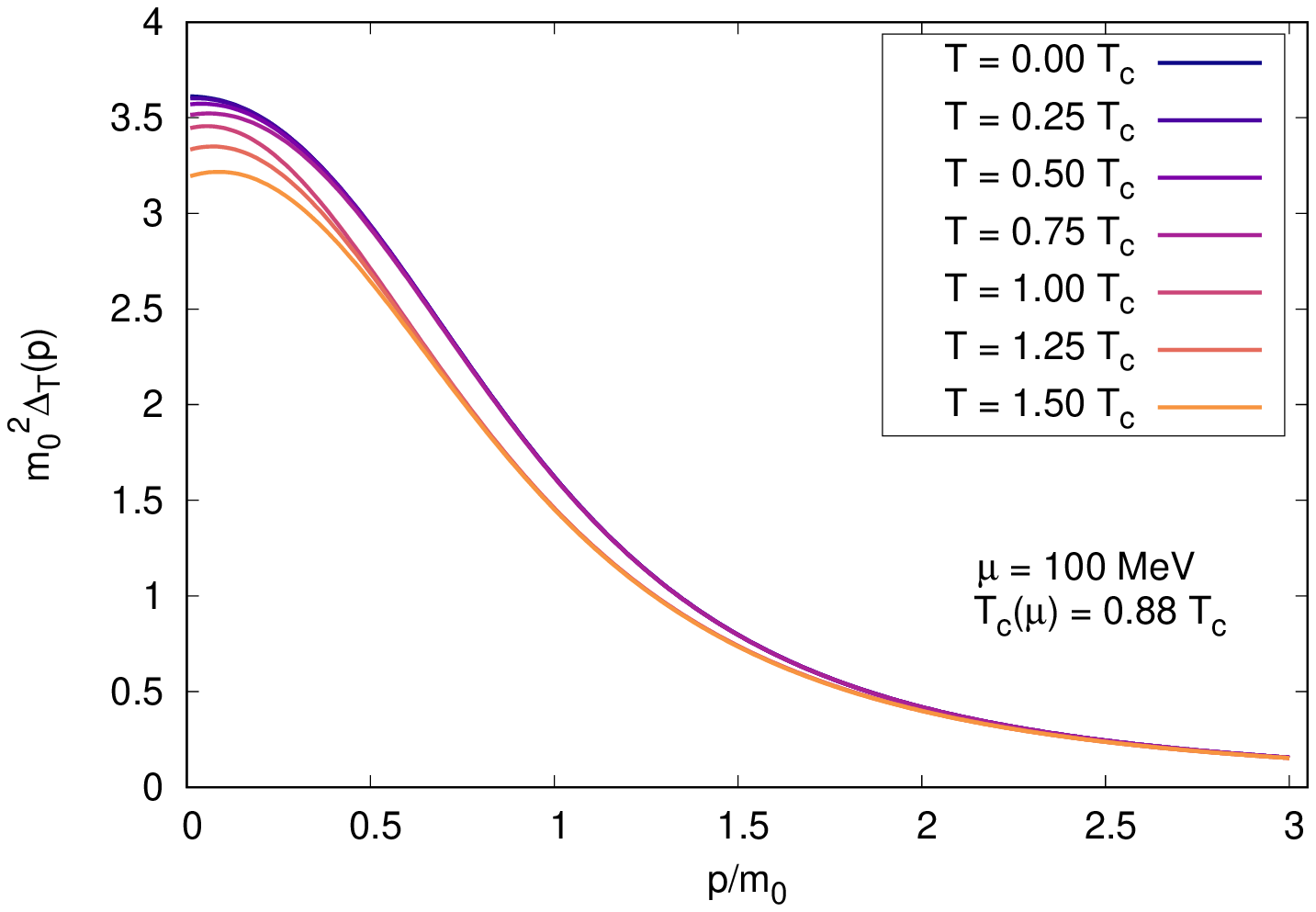}
    \includegraphics[width=0.34\textwidth,angle=270]{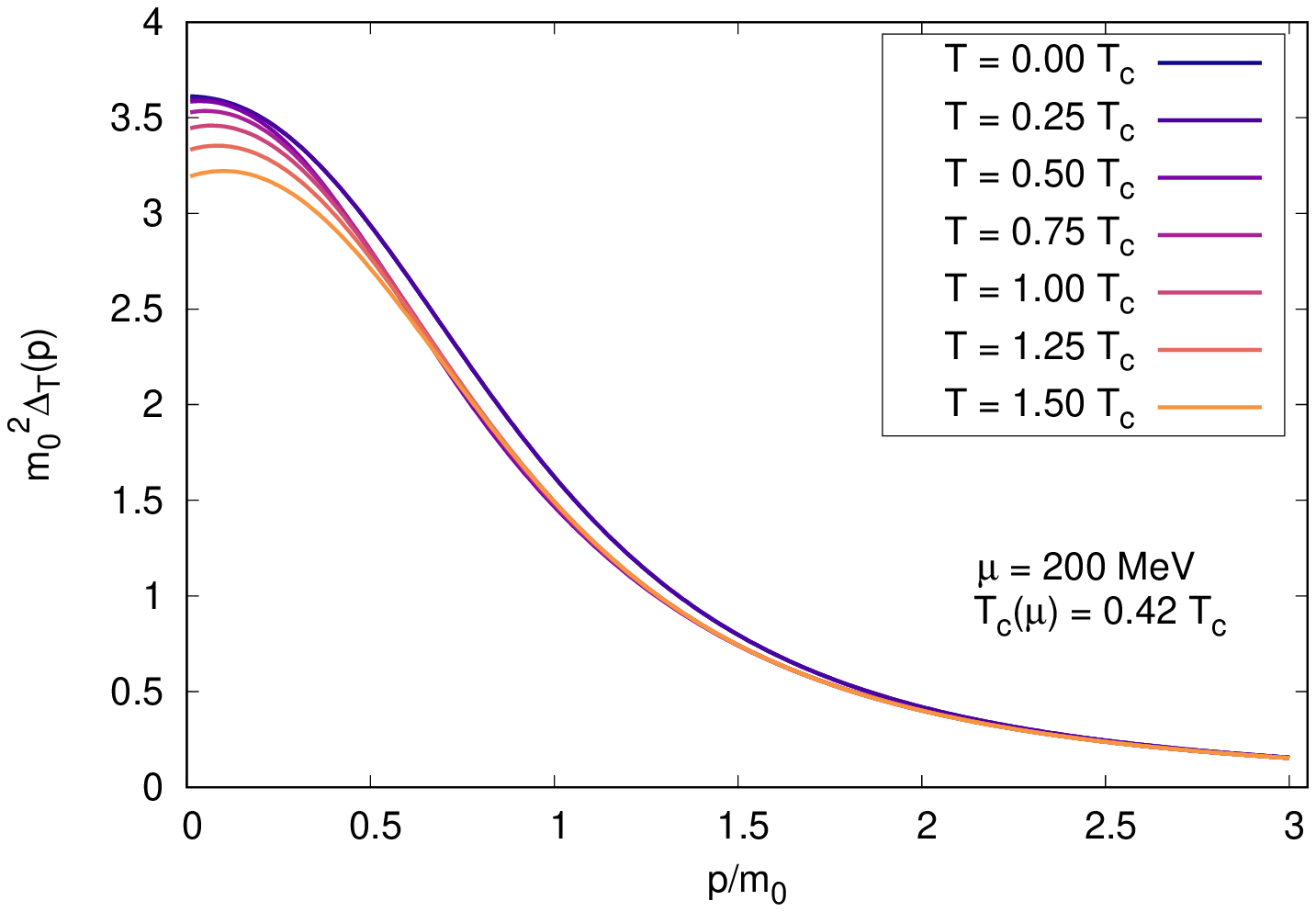}\hspace{5pt}%
    \includegraphics[width=0.34\textwidth,angle=270]{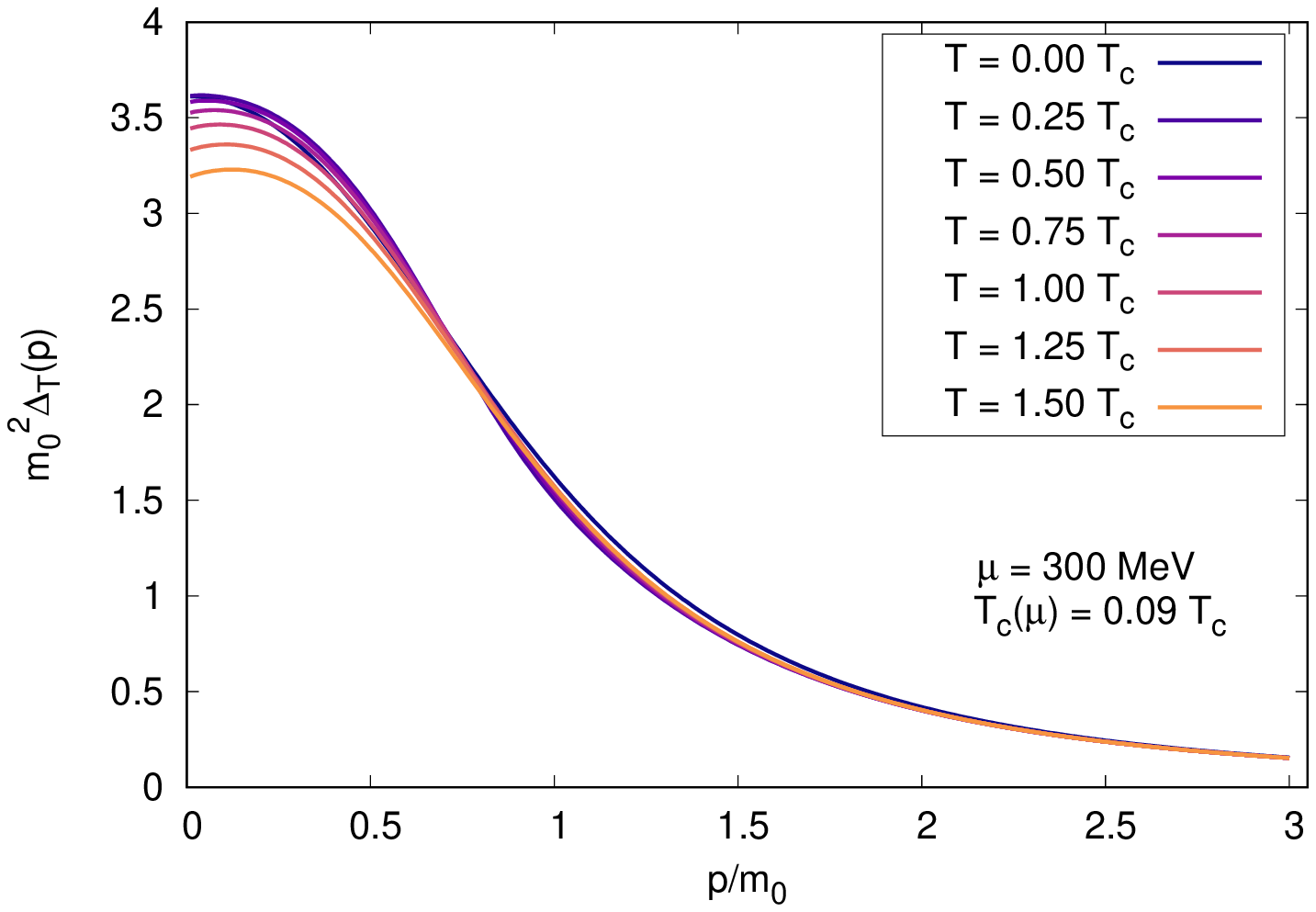}
    \includegraphics[width=0.34\textwidth,angle=270]{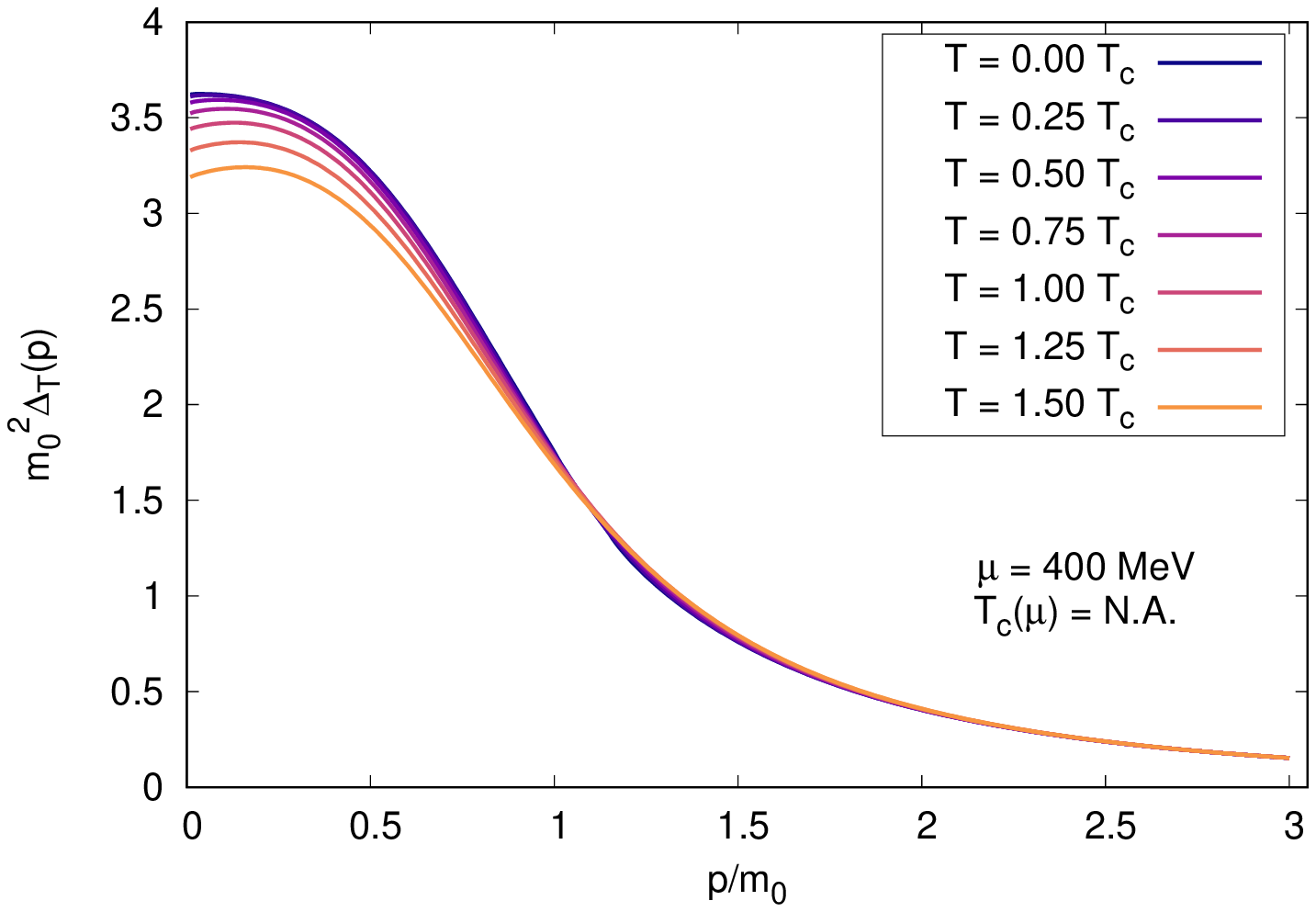}\hspace{5pt}%
    \includegraphics[width=0.34\textwidth,angle=270]{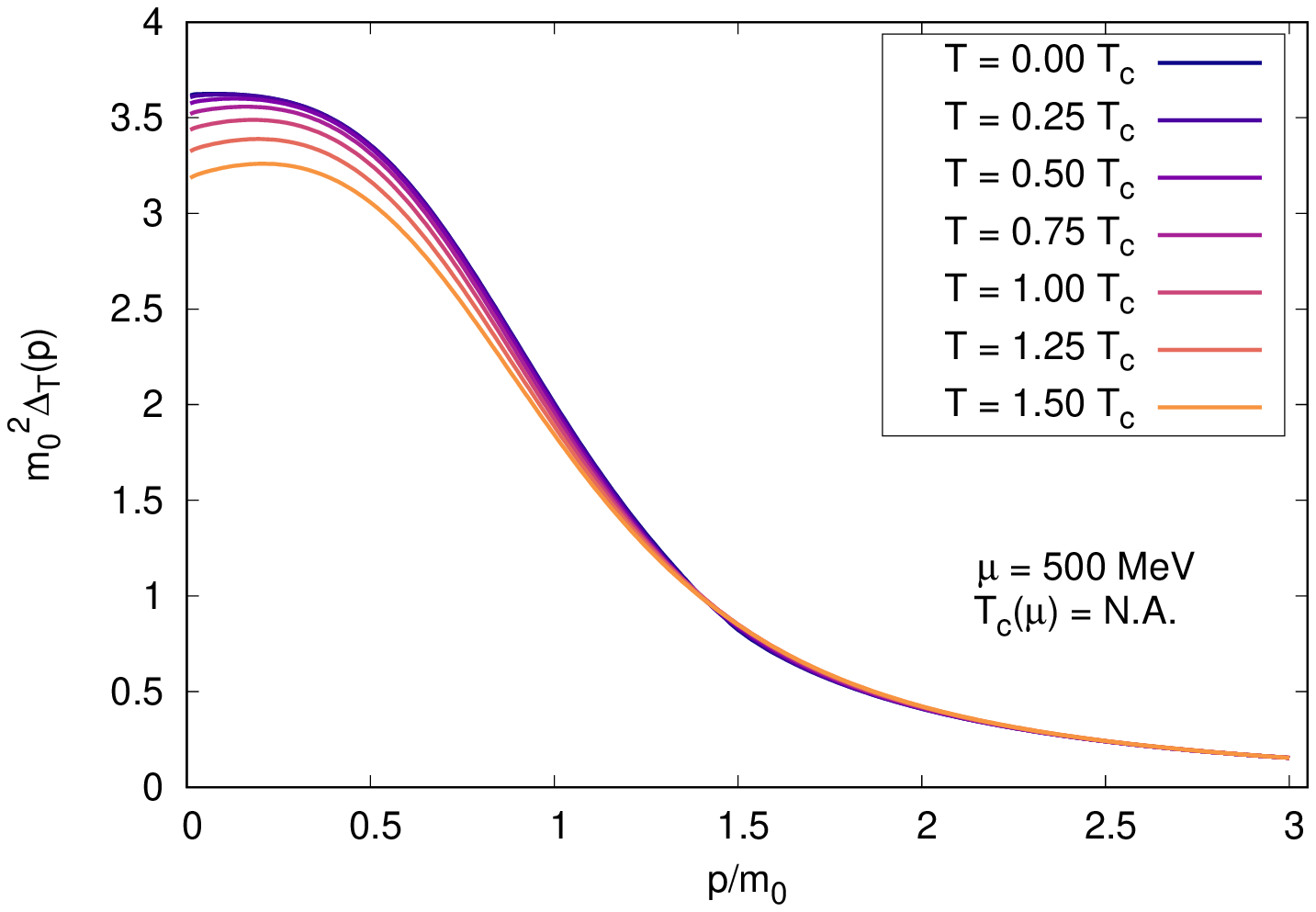}
    \caption{Transverse gluon propagator as a function of spatial momentum at finite $(T,\mu)$. Zero Matsubara frequency ($\omega_{n}=0$). $n_{f}=2+1$ with effective infrared quark masses $350,450$~MeV below $T_{c}(\mu)$ and $125,225$~MeV above $T_{c}(\mu)$. Gluon mass parameter $m_{0}=656$~MeV. $T_{c}=T_{c}(\mu=0)$.}
    \label{qcdtranspropcorr}
\end{figure*}

\begin{figure*}[t]
    \includegraphics[width=0.34\textwidth,angle=270]{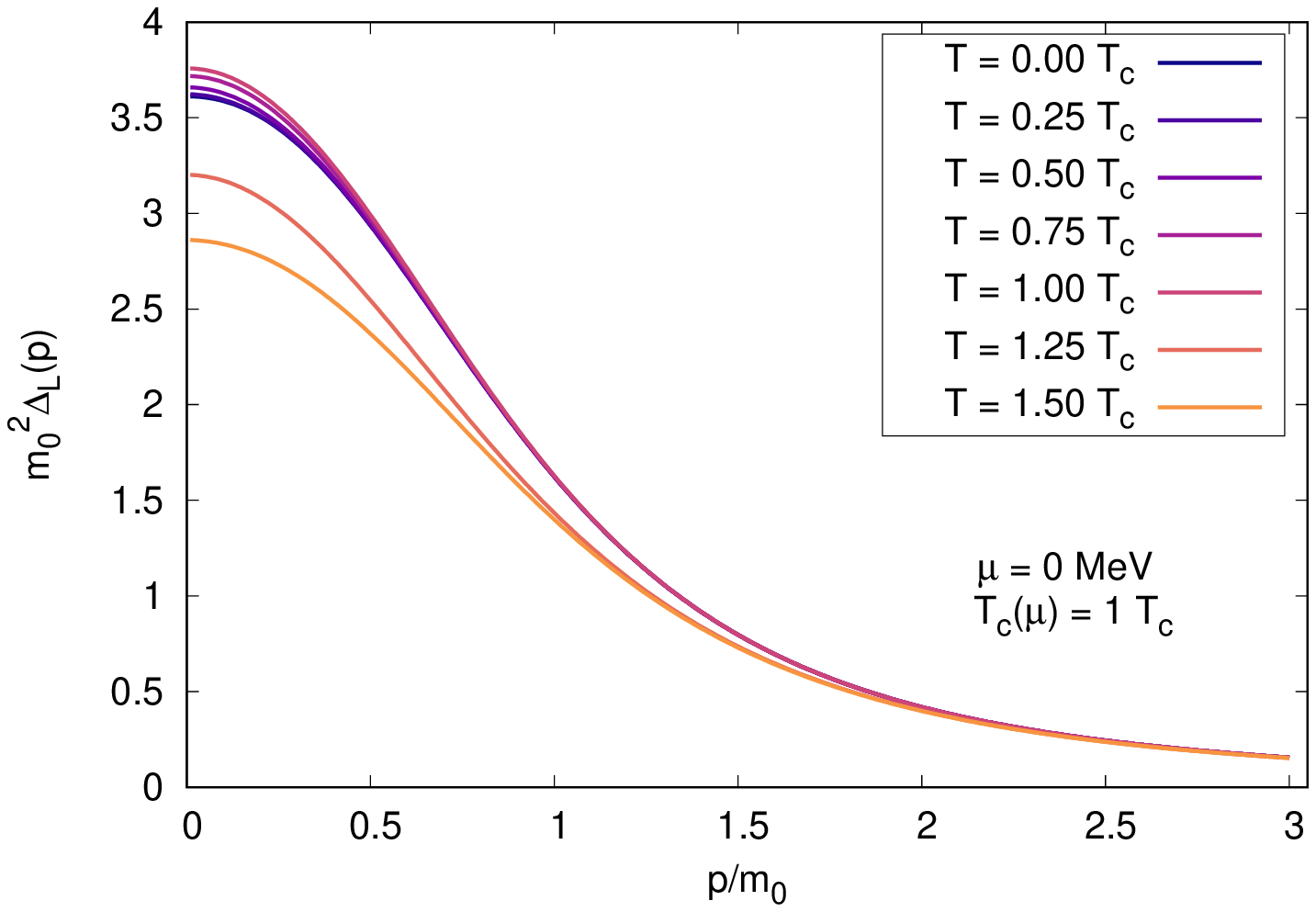}\hspace{5pt}%
    \includegraphics[width=0.34\textwidth,angle=270]{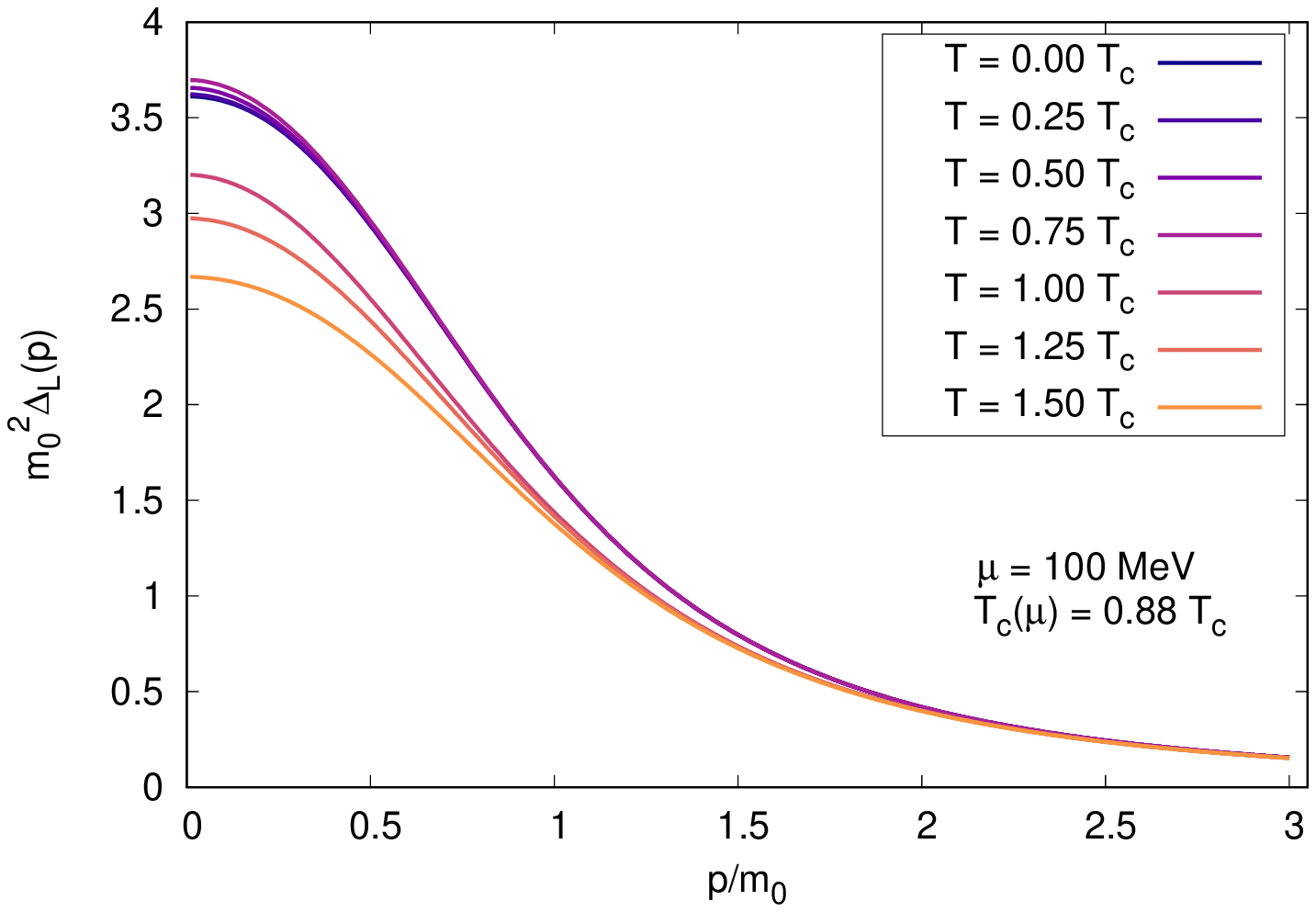}
    \includegraphics[width=0.34\textwidth,angle=270]{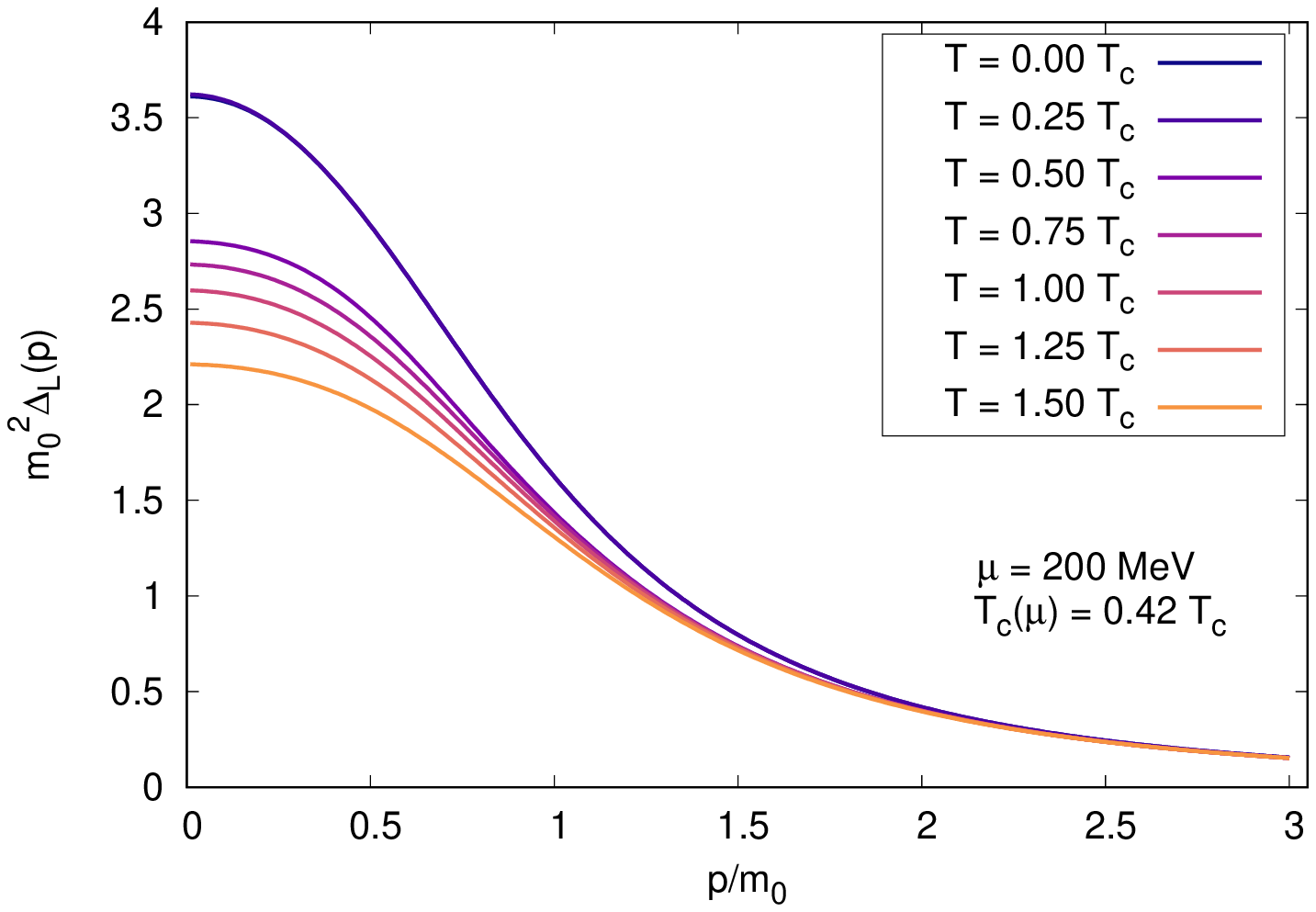}\hspace{5pt}%
    \includegraphics[width=0.34\textwidth,angle=270]{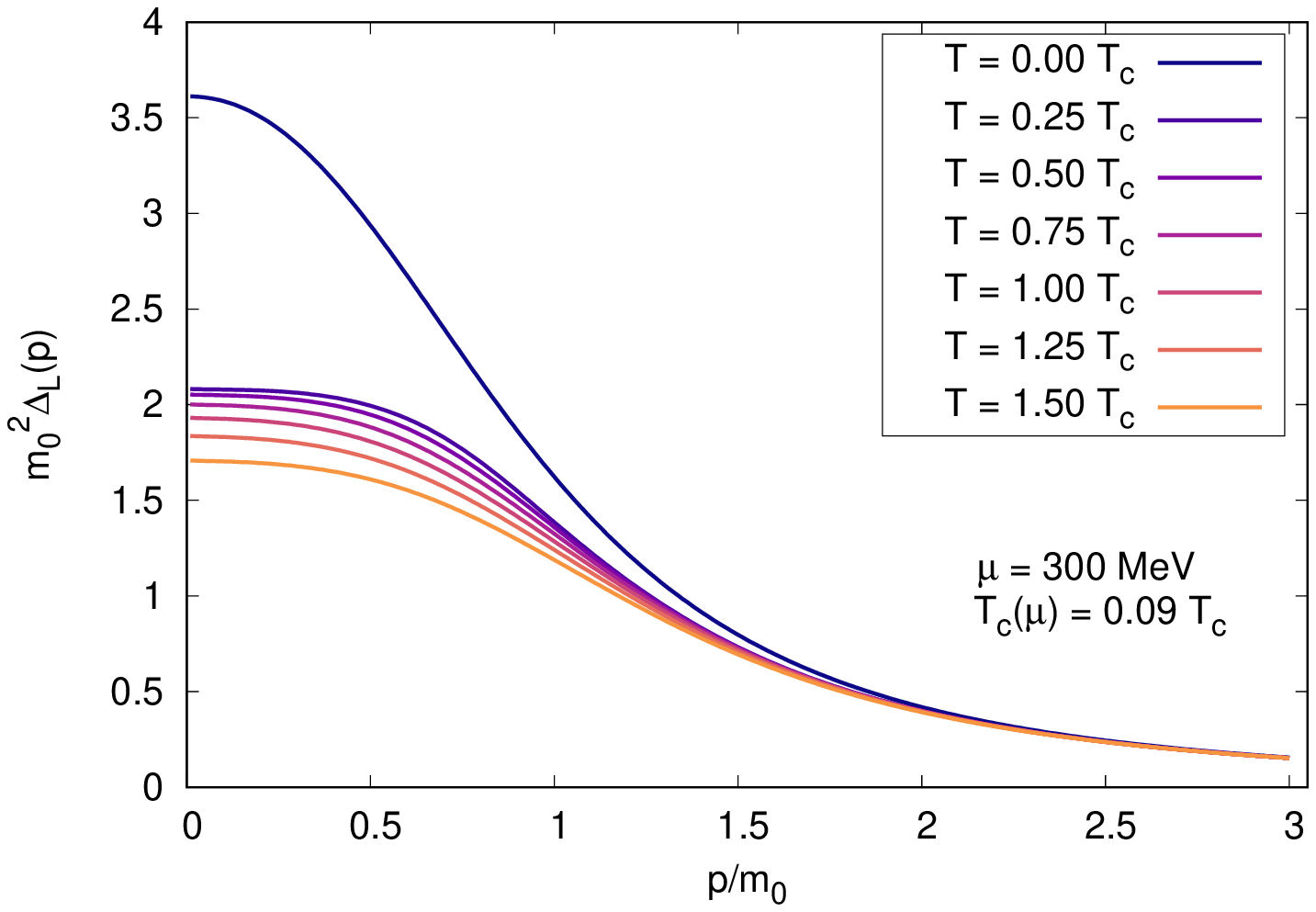}
    \includegraphics[width=0.34\textwidth,angle=270]{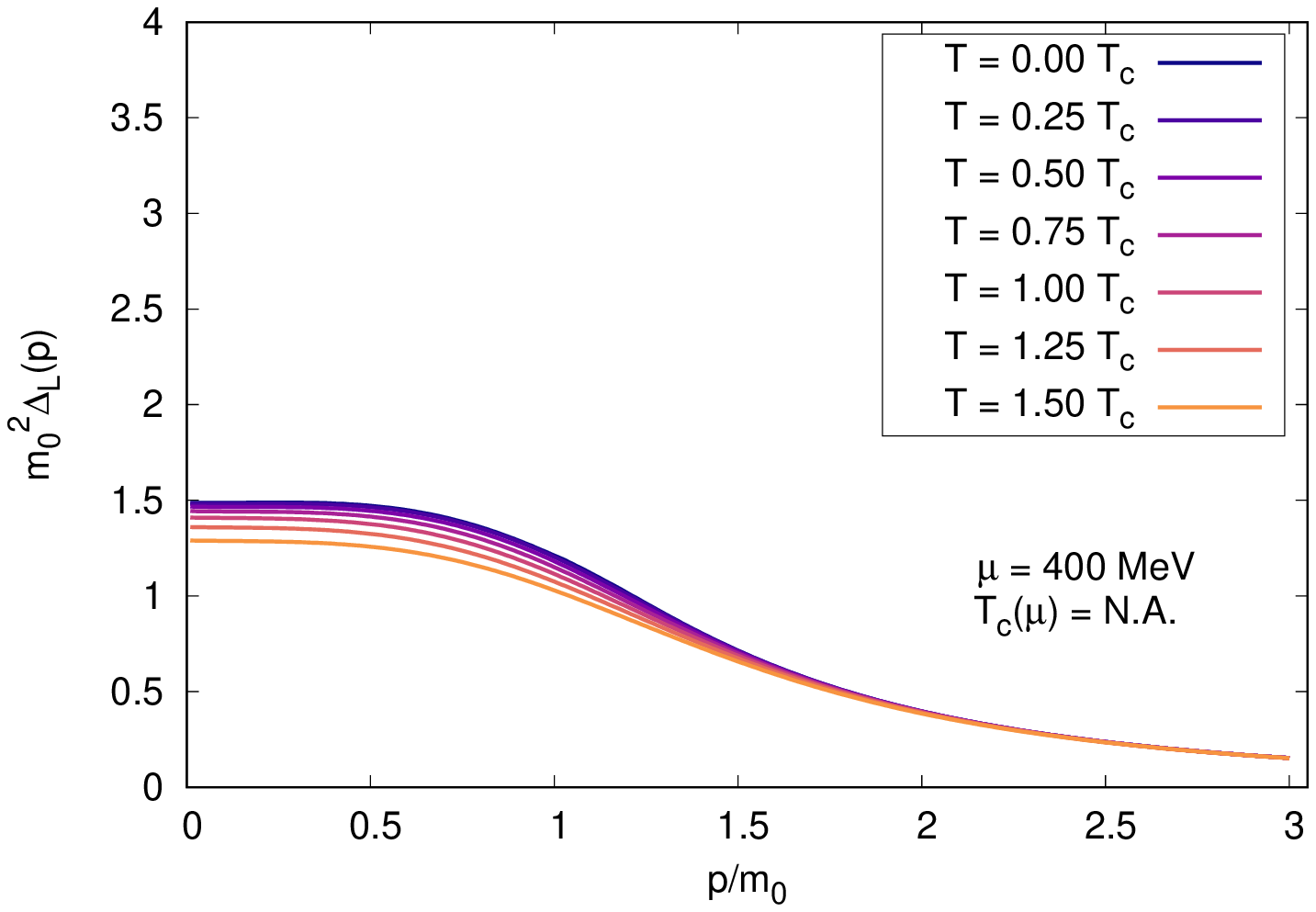}\hspace{5pt}%
    \includegraphics[width=0.34\textwidth,angle=270]{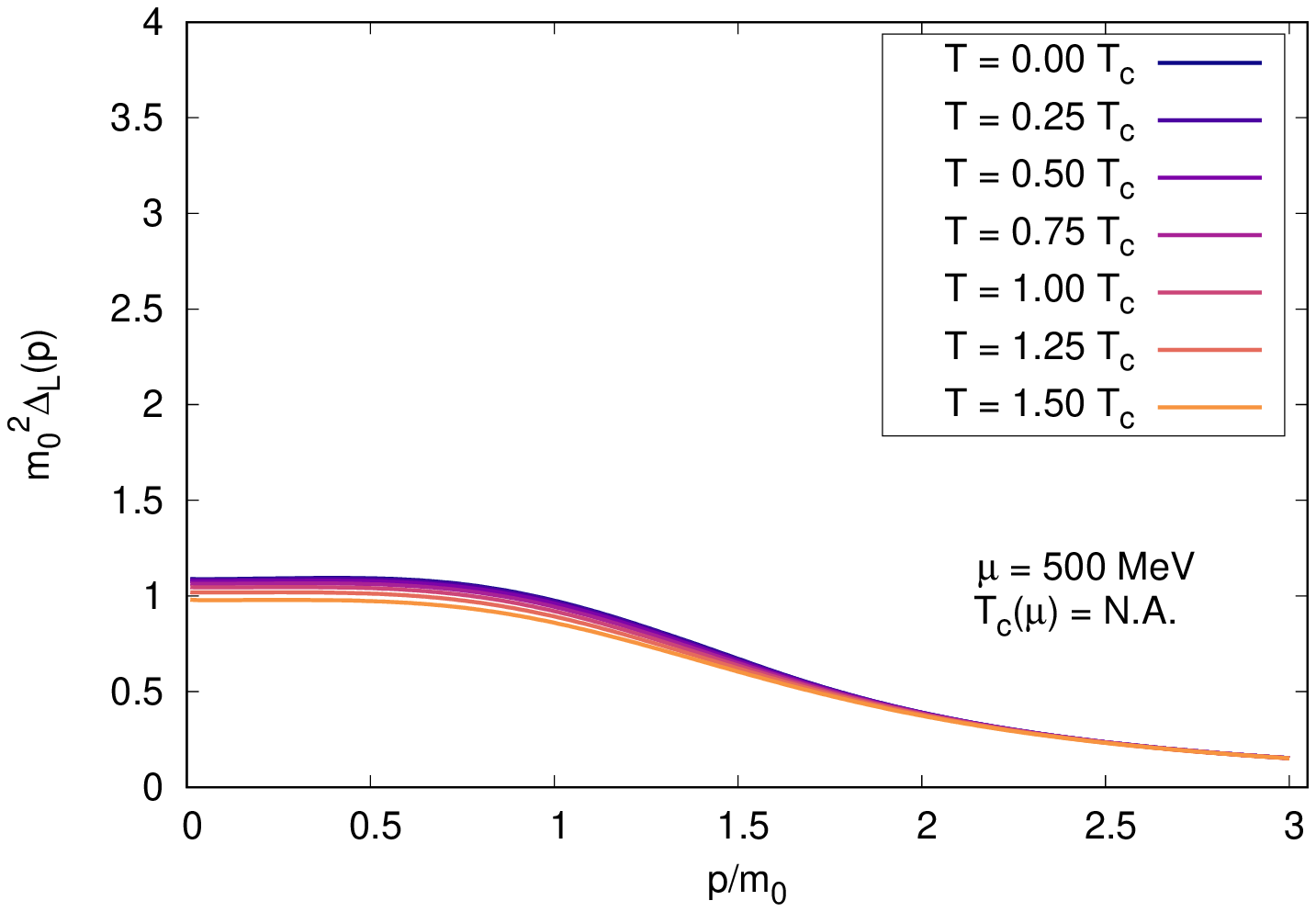}
    \caption{Longitudinal gluon propagator as a function of spatial momentum at finite $(T,\mu)$. Zero Matsubara frequency ($\omega_{n}=0$). Parameters and notation as in Fig.~\ref{qcdtranspropcorr}.}
    \label{qcdlongpropcorr}
\end{figure*}

In Figs.~\ref{qcdtranspropcorr} and \ref{qcdlongpropcorr} we display the transverse and the longitudinal zero-frequency gluon propagators as functions of spatial momentum for different values of $(T,\mu)$. The plots were obtained by first computing $T_{\text{max}}(\mu)$ -- Fig.~\ref{qcdtmax} -- assuming constant quark masses for all $(T,\mu)$, and then changing the value of the quark masses from $(M_{1},M_{2})=(350, 450)$~MeV to $(M_{1}^{\prime},M_{2}^{\prime})=(125, 225)$~MeV beyond the first segment of $T_{\text{max}}(\mu)$ -- which in what follows we will call $T_{c}(\mu)$ for simplicity, adopting its most obvious interpretation. As discussed in Sec.~IVA, this procedure is motivated by the expected restoration of (the remnants of) chiral symmetry, and is carried out to test whether an abrupt decrease in the quark masses outside the region of the phase diagram which can be identified with the confined phase would affect the behavior observed at the $\mu \gtrapprox M_{1,2}$ thresholds for small momenta and $T\approx 0$.

While in Sec.~IVA our main focus was on the longitudinal component of the propagator, studying the effects of such decreased quark masses on the transverse propagator is interesting in its own respect. As we can see from Fig.~\ref{qcdtranspropcorr}, at fixed $\mu < M_{1}$, the drop from $(M_{1},M_{2})$ to $(M_{1}^{\prime},M_{2}^{\prime})$ produces the separation of the transverse propagator in two sets of curves depending on whether $T < T_{c}(\mu)$ or $T > T_{c}(\mu)$\footnote{For the sake of definiteness, our plots use $(M_{1}, M_{2})$ as the quark masses at $T = T_{c}(\mu)$.}: the finite-temperature and density contribution of the quark loop to the polarization is larger for smaller quark masses, translating to a suppression of $\Delta_{T}$ for $T > T_{c}(\mu)$. As $\mu$ approaches $M_{1}$, the dependence of the transverse propagator on temperature becomes more involved, being the by-product of two competing effects: on the one hand, the curves with $T < T_{c}(\mu)$ are nearly insensitive to chemical potential and almost undisplaced from their position at lower $\mu$'s; on the other hand, the curves with $T > T_{c}(\mu)$, despite being initially suppressed, start to expand to larger values of momentum. The net effect of these two different evolutions is an inhomogeneous dependence of the propagator on temperature at fixed $\mu$ and spatial momentum -- see e.g. the $\mu=300$~MeV curves. For chemical potentials larger than $M_{1}$ we observe no qualitative differences with respect to the constant-mass propagators of Sec.~IIIC. This could be anticipated by our analysis of Sec.~IIID, which showed that -- modulo mass threshold effects which play no role here -- our results do not depend qualitatively on the quark masses.

As for the zero-frequency longitudinal gluon propagator, a sudden decrease in the quark masses beyond $T_{c}(\mu)$ does not change the qualitative behavior of $\Delta_{L}$ for $\mu < M_{1}$. Quantitatively, it produces a separation between the $T < T_{c}(\mu)$ and $T > T_{c}(\mu)$ curves similar to the one observed for $\Delta_{T}$, with the $T > T_{c}(\mu)$ ones being suppressed for analogous reasons. For $\mu > M_{1}$, we see that having decreased the quark masses to values smaller than $M_{1}$ made the mass threshold effects disappear. In particular, there are no near-discontinuities in the $T\approx 0$ propagator for $\mu \gtrapprox M_{1},M_{2}$, and no suppression is observed at small momenta with respect to curves of slightly higher temperature: for $\mu > M_{1}$, the longitudinal propagator is strictly decreasing with temperature at all fixed momenta. The temperature at which $\Delta_{L}$ attains a maximum as $|{\bf p}| \to 0$ is displayed in Fig.~\ref{qcdphasediag} as a function of chemical potential and presents no humps for $\mu > M_{1}$.\\

\begin{figure}[h]
    \includegraphics[width=0.32\textwidth,angle=270]{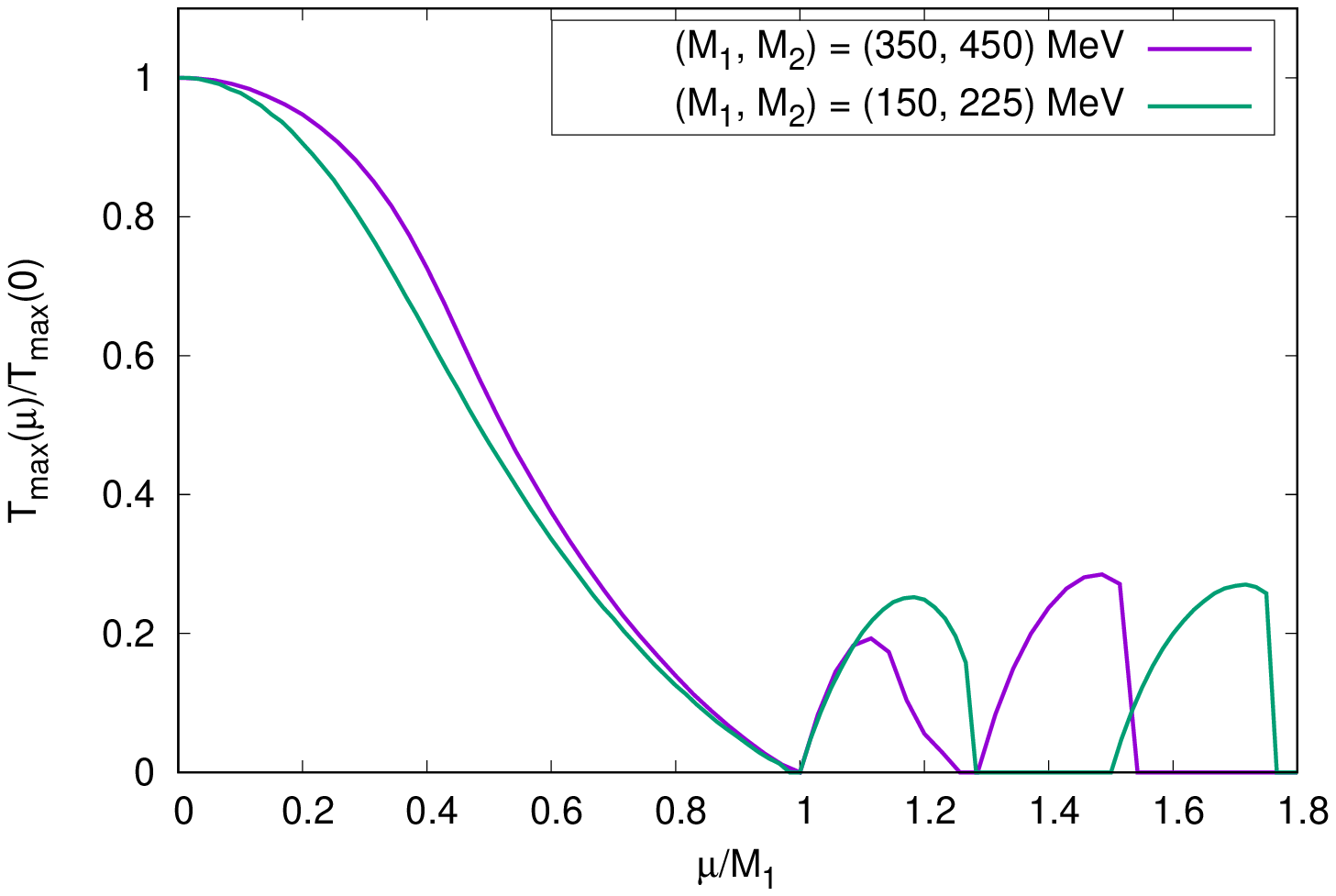}
    \caption{Zero-frequency longitudinal propagator's $T_{\text{max}}$ at $|{\bf p}|=0$ as a function of chemical potential for two constant-mass configurations, $(M_{1},M_{2})=(350, 450)$~MeV and $(M_{1},M_{2})=(150, 225)$~MeV. $m=656$~MeV.}
    \label{qcdtmaxstability}
\end{figure}

\begin{figure}[h]
    \includegraphics[width=0.32\textwidth,angle=270]{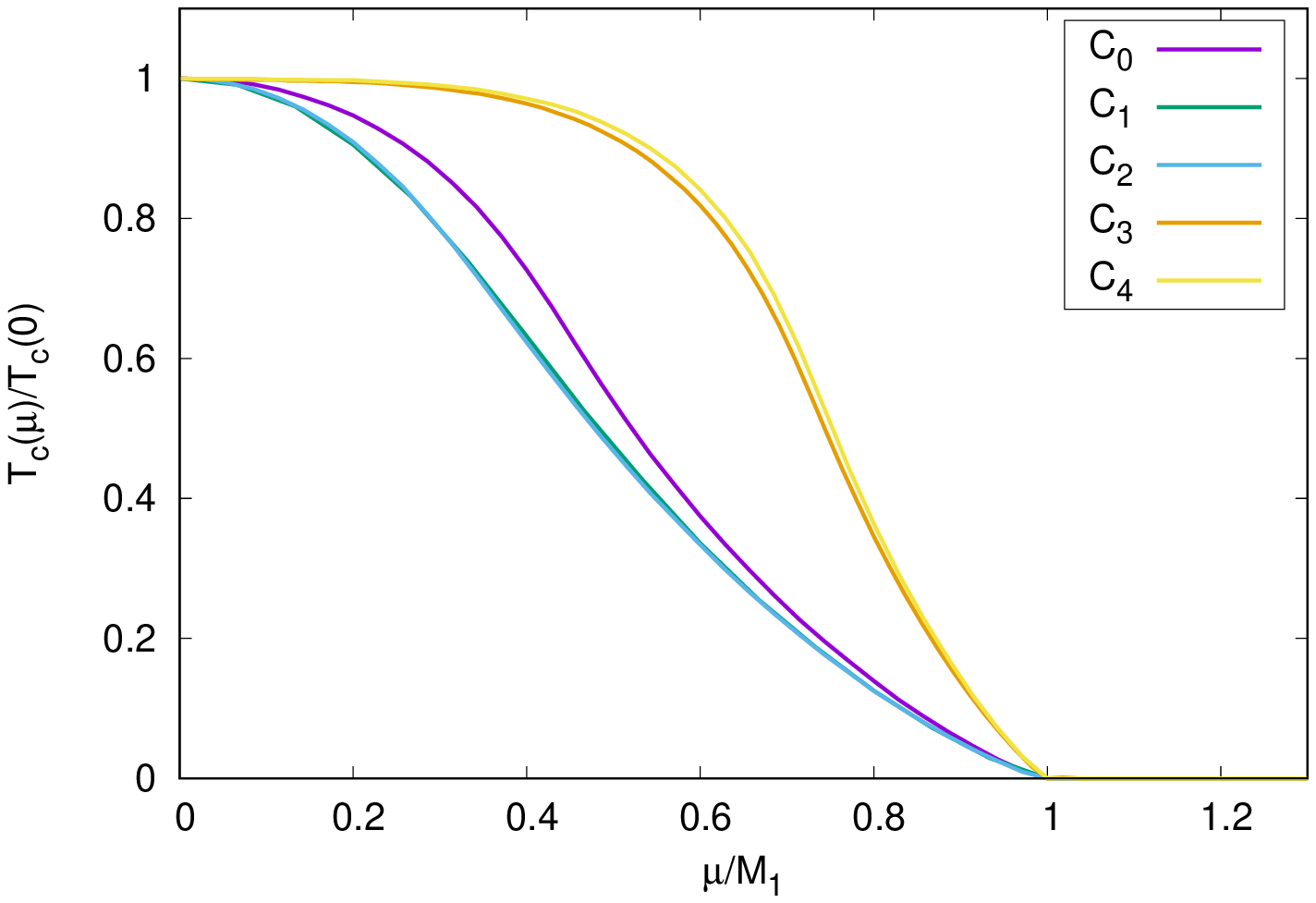}
    \caption{First portion of $T_{\text{max}}(\mu)$, $T_{c}(\mu)$, for different mass configurations. See Tab.~II.}
    \label{qcdphasediagstability}
\end{figure}

In Fig.~\ref{qcdtmaxstability} we display the full $T_{\text{max}}(\mu)$ curve computed for two constant-mass configurations -- namely, $(M_{1}, M_{2})=(350,450)$~MeV like in Sec.~IIIC and $(M_{1}, M_{2})=(150,225)$~MeV, both with $m=656$~MeV. The shape of the curve does not change with the masses, the only notable difference being an increased separation between the two humps in the second configuration. Fig.~\ref{qcdphasediagstability} shows the first portion of $T_{\text{max}}(\mu)$ -- which again we denote with $T_{c}(\mu)$ -- for a number of different mass configurations, reported in Tab.~II. As we can see, normalizing the chemical potential by $M_{1}$ and $T_{c}(\mu)$ by $T_{c}(0)$ yields curves whose dependence mostly comes from the ratio $m/M_{1}$. This is expected on the basis that, since the heavier quark only has a small influence on $T_{\text{max}}(\mu)$ for $\mu < M_{1} < M_{2}$, and since $T_{c}(0)$ and $M_{1}$ set the scale for $T_{c}(\mu)$ and $\mu_{c}$ respectively, $m/M_{1}$ is the only free parameter which can sensibly affect the curves when the latter are reported in adimensionalized units.

\begin{table}
    \begin{tabular}{|c|c|c|c|}
        \hline
        Configuration&$M_{1}$ (MeV)&$M_{2}$ (MeV)&$m$ (MeV)\\
        \hline
        $C_{0}$&$350$&$450$&$656$\\
        $C_{1}$&$150$&$225$&$656$\\
        $C_{2}$&$350$&$450$&$1000$\\
        $C_{3}$&$1100$&$1300$&$656$\\
        $C_{4}$&$350$&$225$&$200$\\
        \hline
    \end{tabular}
    \caption{Mass configurations used for Fig.~\ref{qcdphasediagstability}.}
    \end{table}

\section{Discussion}

In the present paper we have studied the zero-frequency Landau-gauge Euclidean gluon propagator at finite temperature and chemical potential as computed to one loop in full QCD for $n_{f}=2+1$ within the framework of the screened massive expansion. Our main objective was to obtain information on the deconfinement transition and on the phase diagram of QCD based on the observation that, in pure Yang-Mills theory, the longitudinal component of the propagator has distinct behaviors with respect to the temperature $T$ depending on whether $T<T_{c}$ (confined phase) or $T>T_{c}$ (deconfined phase), where $T_{c}$ is the critical temperature of the deconfinement transition. 
Thanks to lattice simulations \cite{SOBC14} we know that, below the critical temperature and for any given fixed spatial momentum, the longitudinal gluon propagator evaluated at zero Matsubara frequency experiences a sudden increase with temperature which peaks at $T_{c}\approx 270$~MeV, to then become a decreasing function of $T$ for $T>T_{c}$ just like the transverse propagator. In terms of a Debye mass defined as $m_{D}=\Delta_{L}^{-1/2}(\omega=0,|{\bf p}|\to 0)$, this translates to $m_{D}$ first decreasing to very small values as $T\to T_{c}$, and then growing with $T$ for $T>T_{c}$. When applied to pure Yang-Mills theory, the screened massive expansion \cite{SC21} is able to reproduce the qualitative behavior of the lattice propagator if its free parameters are kept fixed to their $T=0$ values, and to provide a good semi-quantitative estimate of the lattice data when the parameters are tuned independently and at each temperature for the two components of the propagator -- except, in the latter case, for the longitudinal one at $T\approx T_{c}$ and spatial momenta below $\approx 0.5$-$0.7$~GeV, where the one-loop approximation proves to be insufficient.

In order to extend our approach to full QCD and finite baryonic densities, we employed a $n_{f}=2+1$ model that treats the quarks as dynamical fields propagating with effective masses of the order of the QCD scale $\Lambda_{\text{QCD}}$ -- the consituent mass generated in the infrared as a consequence of chiral symmetry violation -- instead of their current masses. The model is inspired by previous work on dynamical mass generation in the quark sector \cite{CRBS21} and motivated by well-known zero-temperature and density results that see the light and chiral quarks acquiring masses of $350$-$450$~MeV in the infrared regime \cite{KBLW05}. For chemical potentials $\mu < M_{1}$, where $M_{1}$ is the mass of the lightest quarks -- set to $350$~MeV throughout most of our analysis -- we found that the gluon propagator has the same behavior as in pure Yang-Mills theory: its transverse component is a strictly decreasing function of the temperature $T$, whereas the longitudinal one first increases with $T$ for $T<T_{\text{max}}(\mu)$ and then decreases with it for $T>T_{\text{max}}(\mu)$, $T_{\text{max}}(\mu)$ being the (chemical-potential-dependent) temperature at which the propagator attains a maximum. Additionally, the transverse propagator is observed to be essentially independent of chemical potential. For $\mu > M_{1}$, the transverse propagator acquires a dependence on chemical potential that makes it expand toward larger values of momentum, whereas the longitudinal one becomes a strictly decreasing function of temperature for all but the lowest values of momenta and around the mass thresholds $\mu\gtrapprox M_{1},M_{2}$, where $M_{2}$ -- taken to be $450$~MeV throughout most of our analysis -- is the effective mass of the heavier quark. Around these thresholds, at low momenta, the $T\approx 0$ longitudinal propagator is slightly more suppressed than those of slightly larger temperatures, implying that $T_{\text{max}}(\mu)$ is not equal to zero for those values of chemical potential when measured at $|{\bf p}|=0$. We showed that this effect may be an artifact of our constant-mass approximation, since it disappears if the quark masses are made to abruptly decrease below around 50-70\% of their value beyond the first portion of the $T_{\text{max}}(\mu)$ curve. Such a decrease may be more realistic \cite{RWB05,BFG05} for the lighter quarks -- of mass $M_{1}$, representing the up and down quarks -- than for the heavier one -- of mass $M_{2}$, representing the strange --, although even for the latter it cannot be entirely ruled out. Overall, at variance with the transverse component of the propagator, for $\mu > M_{1}$ the longitudinal one is suppressed at larger chemical potentials.

If we assume that, in full QCD like in pure Yang-Mills theory, the longitudinal gluon propagator changes its behavior with temperature in passing from the confined to the deconfined phase, then our one-loop approach provides both a qualitative prediction of the phase boundary and a clear picture of the main factors affecting its shape. At small chemical potentials $\mu < M_{1} < M_{2}$, as the temperature $T$ increases from $T=0$ to larger values, the quark contribution to the gluon polarization is made up of two terms, one of which -- the vacuum part -- does not depend on temperature, and the other one -- the medium part -- which is exponentially suppressed by the Fermi distribution. In this regime, the dependence of the longitudinal propagator on temperature largely comes from the gauge-field self-interaction, and thus reflects that of pure Yang-Mills theory with its non-monotonic behavior. As chemical potential is increased, the medium part of the quark loops is activated by lower values of the temperature and becomes larger, contributing a positive term to the inverse propagator -- equivalently, at $|{\bf p}|=0$, to the Debye mass. This contribution counterbalances the increasing behavior brought by the gauge-field self-interaction, the net effect being that of decreasing the temperature $T_{\text{max}}(\mu)$ at which the longitudinal propagator attains its maximum -- and thus of what may interpreted to be the critical temperature $T_{c}(\mu)$ of the deconfinement transition. Since the Fermi distribution suppresses the quark loops the more they are massive, this effect is dominated by the lightest quark loops, i.e. by those of mass $M_{1}$. In this respect, we have verified that, for $\mu<M_{1}$, an analogous calculation carried out at $n_{f}=2$ -- that is, by excluding the quark loop of mass $M_{2}$ -- shows no significant differences with respect to the $n_{f}=2+1$ case studied in this paper except for a $4$-$9$\% increase in the values of $T_{c}(\mu)$. Similarly, we checked that adding a fourth quark loop of mass $M_{3}\sim1.2$~GeV to represent the charm quark's contribution in a $n_{f}=2+1+1$ model has virtually no effect on $T_{c}(\mu)$ for $\mu< M_{1}$. By the time the chemical potential has reached the first mass threshold $\mu=M_{1}$ -- threshold at which the lightest quark loop is active even for $T=0$ --, the fermionic contribution to the longitudinal polarization has become dominant with respect to the pure-gauge terms and has pushed $T_{c}(\mu)$ to zero.

The simple characterization of the confined (resp. deconfined) phase as the region of the phase diagram in which -- at least for $\mu < M_{1}$ -- the longitudinal gluon propagator is an increasing (resp. decreasing) function of temperature is inherently incapable of providing us with a criterion to determine whether the change from one phase to the other happens via a crossover or via a true phase transition -- and, in the latter case, to establish what is the order of the transition. As for the existence and position of a critical endpoint, such information remains out of reach as well, in the absence of any knowledge on the kind of transition. Still, we note that, regardless of the values of the masses and of the free parameters of the expansion, the curves $T_{c}(\mu)$ computed within the screened expansion display a change in concavity at $\mu=0.4$-$0.8\ M_{1}$ and $T=0.4$-$0.6\ T_{c}$. For $(M_{1},M_{2})=(350,450)$~MeV, the change happens at the point $(\mu,T)=(0.46 M_{1},0.62 T_{c})$ -- or $(\mu_{B},T)=(0.48\,\text{GeV},0.62 T_{c})$. While we have no indication that a change in concavity actually entails any new behavior for the observables of the system, it is tempting to interpret such a feature as a hint that the nature of the transition may change in a neighborhood of that point. 

For chemical potentials larger than $M_{1}$, the parallel with pure Yang-Mills theory becomes weaker, and it is unclear how a non-vanishing $T_{\text{max}}(\mu)$, if any, should be interpreted. In this region of the phase diagram, at low temperatures, we expect full QCD to display new phases of matter, like the two-flavour superconductive (2SC) and color-flavor locked (CFL) phases \cite{ARW98,RSS98,ARW99}, which have no analogue in pure Yang-Mills theory. While our approach provides some evidence for $T_{\text{max}}(\mu)\neq 0$ when measured at $|{\bf p}|=0$, as discussed above, this evidence is fragile, especially as far as the $\mu\gtrapprox M_{1}$ lightest-quark threshold is concerned. On the other hand, if the heavier quark mass $M_{2}$ did not decrease enough to make the $\mu\gtrapprox M_{2}$ hump in $T_{\text{max}}(\mu)$ disappear, the latter could hint to the influence of the strange quark on this region of the phase diagram -- which is actually expected, e.g., in the CFL phase. We do however need to be cautious in this respect, since, first of all, as stated before, this effect is very small and not at all assured to survive higher-order corrections, and second of all, because at very large chemical potentials, in the absence of a parallel with pure Yang-Mills theory, a non-zero $T_{\text{max}}(\mu)$ measured at $|{\bf p}|=0$ may simply not be a useful criterion for investigating the properties of the phase diagram.

Finally, we observe that the present study was carried out solely in the Landau gauge, whereas the thermodynamic properties of strongly interacting matter should obviously be gauge independent. In this respect, it would be interesting to extend our results to other covariant gauges, especially since to date -- to our knowledge -- no such results have been obtained outside of the Landau gauge on the lattice or using functional methods. The fact that pure Yang-Mills theory undergoes a true phase transition makes it clear that, in the absence of quarks, the critical temperature is gauge invariant. As a consequence, if we assume that the criterium of the maximum of the longitudinal gluon propagator holds in any gauge, then we should expect such maximum to be attained at the same temperature in all covariant gauges when computed in pure Yang-Mills theory. In the presence of quarks, on the other hand, the issue of gauge independence becomes more involved. For small enough chemical potentials and for physical quark masses, as we noted above, the transition is expected to be a crossover. Since the critical temperature, in this regime, depends on the observable used for its definition, measuring a gauge-dependent critical temperature from the (gauge-dependent) gluon propagator would not contradict any physical principle. At larger chemical potentials, however, the transition is expected to be a sharp (first-order) one, with a well-defined critical temperature. In this regime, if the $T_{\text{max}}$ criterion still holds outside of the Landau gauge, then the temperature at which the longitudinal gluon propagator attains a maximum should indeed be gauge independent when computed in full QCD as well. We will leave this interesting topic for future investigation.

\acknowledgments

This research was supported in part by PIACERI ``Linea di intervento 1'' (M@uRHIC) of the University of Catania and by PRIN2022 (project code 2022SM5YAS) within Next Generation EU fundings. G.C. thanks F. Murgana and M. Ruggieri for useful discussions on the QCD phase diagram at low temperatures and large densities. Numerical calculations were performed using the qcd-sme library \cite{qcd-sme-rs} powered by the Peroxide Rust library \cite{peroxide} for numerical integrations.

\bibliography{ThermalQCD}

\end{document}